\documentclass[a4paper,11pt]{article}
\pdfoutput=1 
\usepackage{jcappub}
\usepackage{amsmath}
\usepackage{graphicx}
\usepackage{latexsym}
\usepackage{xspace}
\usepackage{color}
\usepackage{hyperref} 
\usepackage{bm}
\usepackage{relsize}
\usepackage{tabularx}
\usepackage{multirow}
\usepackage{amssymb}
\usepackage[table]{xcolor}
\usepackage{braket}
\usepackage{comment}
\usepackage[normalem]{ulem} 
\usepackage{slashed}
\usepackage{cleveref}

\makeatletter
\gdef\@fpheader{}
\g@addto@macro\bfseries{\boldmath}
\makeatother

\newcommand{\Dirac}{\delta_{\mathrm{D}}}

\newcommand{\ie}{\textsl{i.e.~}}

\newcommand{\eg}{\textsl{e.g.~}}

\newcommand{\etc}{\textsl{etc.~}}

\DeclareMathOperator{\sinc}{sinc}

\newcommand{\dd}{\mathrm{d}}
\newcommand{\ee}{e}

\newcommand{\sss}[1]{{\scriptscriptstyle{#1}}}

\newcommand{\uPl}{\mathrm{Pl}}
\newcommand{\uin}{\mathrm{in}}

\newcommand{\uend}{\mathrm{end}}

\newcommand{\uc}{\mathrm{c}}

\newcommand{\usssPl}{\sss{\uPl}}

\newcommand{\Rea}{\Re \mathrm{e}\,}

\newcommand{\Mp}{M_\usssPl}

\newcommand{\Ni}{N_{\mathrm{in}}}

\newcommand{\beq}{\begin{equation}}
\newcommand{\eeq}{\end{equation}}
\newcommand{\bea}{\begin{equation}\begin{aligned}}
\newcommand{\eea}{\end{aligned}\end{equation}}

\newlength{\wsingfig}
\newlength{\wdblefig}
\newlength{\wquadfig}
\newlength{\wtriplefig}
\newcommand{\Fig}[1]{fig.~{\ref{#1}}}

\newcommand{\Refa}[1]{ref.~{\cite{#1}}}
\newcommand{\Refs}[1]{refs.~{\cite{#1}}}

\newcommand{\f}{\phi}

\newcommand{\field}{\phi}
\newcommand{\momentum}{\pi_\phi}
\newcommand{\cgfield}{\phi^{\mathrm{IR}}}
\newcommand{\cgmomentum}{\pi_\phi^{\mathrm{IR}}}

\newcommand{\fieldh}{
{\delta\phi}^{\rm IR}}
\newcommand{\momentumh}{
{\delta\pi}^{\rm IR}}

\newcommand{\Fieldh}{
{\delta\Phi}^{\rm IR}}
\newcommand{\Momentumh}{
{\delta\Pi}^{\rm IR}}

\newcommand{\ba}{\begin{eqnarray}}
\newcommand{\ea}{\end{eqnarray}}

\subheader{}

\title{Stochastic inflation with gradient interactions}

\author[a]{Vadim Briaud,}
\emailAdd{vadim.briaud@phys.ens.fr}

\author[b]{Ryodai Kawaguchi,}
\emailAdd{ryodai0602@fuji.waseda.jp}

\author[a]{Vincent Vennin}
\emailAdd{vincent.vennin@phys.ens.fr}

\affiliation[a]{Laboratoire de Physique de l'Ecole Normale Sup\'erieure, ENS, CNRS, Universit\'e PSL, Sorbonne Universit\'e, Universit\'e Paris Cit\'e, 75005 Paris, France}

\affiliation[b]{Department of Physics, Waseda University, 3-4-1 Okubo, Shinjuku, Tokyo 169-8555, Japan} 

\date{today}

\begin{document}

\begin{flushright}
WUCG-25-10\\
\end{flushright}

\sloppy

\abstract{Stochastic inflation rests on the separate-universe approximation, \ie the ability to describe long-wavelength fluctuations in an inflating universe as homogeneous perturbations of its background dynamics. Although this approximation is valid in most cases, it has been recently pointed out that it breaks down during transition periods between attractor and non-attractor phases. Such transitions are ubiquitous in single-field models giving rise to enhanced perturbations at small scales, that are required to form primordial black holes. The current inability to apply the stochastic-inflation program in such models is therefore one of the main obstacles to investigating the role of backreaction in primordial-black-hole scenarios. In this work, we show how gradient interactions can be incorporated in stochastic inflation, via a set of Langevin equations of higher dimension. We apply our formalism to a few cases of interest, including one with a sharp transition. In all cases, in the classical limit we show that gradient corrections as predicted from cosmological perturbation theory are properly recovered. We uncover the existence of a ``pullback'' effect by which the tails of the first-passage-time distributions are dampened by gradient interactions. We finally discuss the role of backreaction in the presence of gradient interactions.}

\arxivnumber{}

\maketitle

\section{Introduction}
\label{sec:intro}
While cosmic-microwave-background and large-scale-structure observations are consistent with primordial perturbations being scale invariant, they do not exclude the possibility for large fluctuations to exist at small scales, which could lead to the formation of primordial black holes (PBH)~\cite{Hawking:1971ei, Carr:1974nx, Carr:1975qj}. These objects are natural candidates for dark matter~\cite{Carr:2009jm,Niikura:2017zjd, Katz:2018zrn, Montero-Camacho:2019jte,Carr:2020gox} and could serve as seeds for the formation of supermassive black holes observed at high redshifts~\cite{1975A&A....38....5M, Duechting:2004dk, Kawasaki:2012kn, Clesse:2015wea, Carr:2018rid, Liu:2022bvr, Hutsi:2022fzw}.

In order to model such rare, large perturbations during inflation, non-perturbative methods are required. One such technique is the separate-universe approach~\cite{Starobinsky:1982ee, Starobinsky:1985ibc, Sasaki:1995aw, Sasaki:1998ug, Wands:2000dp}, by which long-wavelength fluctuations can be described by an ensemble of independent background-universe patches, evolving non linearly. Formally, this can be interpreted as the leading-order behaviour in a gradient expansion~\cite{Salopek:1990jq, Rigopoulos:2003ak, Tanaka:2007gh}. This gives rise to the $\delta N$ formalism~\cite{Sasaki:1998ug, Lyth:2003im, Lyth:2004gb}, where the curvature perturbation is identified with the amount by which these patches inflate.

As inflation proceeds, quantum fluctuations are stretched beyond the Hubble radius and source the dynamics of the background patches. This backreaction effect can be described by the stochastic-inflation formalism~\cite{Starobinsky:1982ee, Starobinsky:1986fx}, where the background fields follow random processes driven by Langevin (or, equivalently, Fokker-Planck) equations. Combined with the separate-universe approach, it leads to the stochastic-$\delta N$ formalism~\cite{Enqvist:2008kt,Fujita:2013cna,Fujita:2014tja,Vennin:2015hra}, where the curvature perturbations are related to first-passage times through the end-of-inflation hypersurface. Primordial black holes arise from those realisations of the Langevin process that inflate for an anomalously long time, and studying them thus requires to investigate the tail of first-passage-time distributions.

Both the classical- and stochastic-$\delta N$ formalisms have been applied to a wide range of models, see \eg \Refs{Pattison:2017mbe, Biagetti:2018pjj, Panagopoulos:2019ail, Figueroa:2020jkf, Pattison:2021oen, Ezquiaga:2019ftu, Vennin:2020kng, Ando:2020fjm, Tada:2021zzj, Achucarro:2021pdh, Kitajima:2021fpq, Hooshangi:2021ubn, Gow:2022jfb, Cai:2022erk, Animali:2022otk, Jackson:2022unc, Ezquiaga:2022qpw, Rigopoulos:2022gso, Briaud:2023eae, Hooshangi:2023kss, Kawaguchi:2023mgk, Raatikainen:2023bzk, Launay:2024qsm, Vennin:2024yzl, Inui:2024sce, Animali:2024jiz, Mizuguchi:2024kbl, Choudhury:2025kxg, Murata:2025onc, Kuroda:2025coa, Animali:2025pyf, Cruces:2025typ, Miyamoto:2025qqm}, where non-linear effects have been shown to induce heavy tails that strongly enhance the expected abundance of PBHs. Quantum diffusion thus seems to play a major role in shaping the statistics of extreme objects such as PBHs~\cite{Ezquiaga:2022qpw}, and to reduce the amount of fine tuning that is required to produce them.

There is, however, a cloud on the horizon: while the separate-universe approach has been shown to be valid both in slow roll (SR) and ultra-slow roll (USR) phases~\cite{Pattison:2019hef, Firouzjahi:2020jrj, Mishra:2023lhe}, it was recently pointed out~\cite{Jackson:2023obv, Artigas:2024ajh, Raveendran:2025pnz} that it fails on a finite range of super-Hubble scales at a sudden transition between these two regimes. Unfortunately, such transitions are commonplace in single-field models yielding enhanced perturbations at small scales~\cite{Garcia-Bellido:2017mdw, Ballesteros:2017fsr, Karam:2022nym}, as required for PBH formation. In order to apply the stochastic-$\delta N$ program to PBH models across all scales, it thus remains to include gradient interactions in the stochastic formalism of inflation. The goal of this paper is to fill this gap. 

After reviewing the description of perturbations during inflation in \cref{sec:setup}, using both linear perturbation theory and stochastic inflation, in \cref{sec:memory effect} we show how gradient interactions between nearby patches can be recast as a memory effect inside each patch individually. In practice, this means that the random processes driving the field dynamics are subject to coloured (\ie correlated over time) noises. This makes the stochastic dynamics non-Markovian, which would raise several technical challenges, if it were not for the existence of an equivalent formulation in terms of a set of Langevin equations with white noises only, albeit of higher dimension. This reinstates the ability to describe long-wavelengths perturbations in terms of separate universes, and to rely on standard techniques for solving Markovian stochastic differential equations. 

We then consider in \cref{sec:example} three cases of interest: SR, USR, and the Starobinsky piecewise-linear potential model that features a transition between these two regimes. In all three cases, we find that gradient interactions are properly accounted for by our improved stochastic formalism. This verification is performed at leading order in perturbation theory, since gradient corrections can be processed to all orders only in the perturbative framework. However, we also comment on non-linear effects, and the role of backreaction, in the presence of gradient interactions. We highlight the existence of a ``pullback effect'', by which the rare realisations that inflate for an anomalously long time are pulled back towards the average behaviour by their neighbour patches. This results in a reduction of the tail of the first-passage-time distributions (hence of the abundance of PBHs, to an extent that remains to be investigated). 

We summarise our main results in \cref{sec:conclusion} and end this article with several appendices to which some of the technical details are deferred. 

\section{Stochastic inflation and the separate-universe approach}
\label{sec:setup}

In this section, we review the construction of the stochastic-inflation formalism and its connection to the separate-universe approach. Our goal is to highlight that gradient interactions, which are neglected in this setup, may become important after sudden transitions away from the slow-roll attractor during inflation. The reader already familiar with these concepts may wish to skip this section and go directly to \cref{sec:memory effect}.

For simplicity, we focus on single-field inflationary models with canonical kinetic term and minimal coupling to gravity. The action of the inflaton $\phi$ is thus given by
\bea
\label{eq:action}
\mathcal{S}=\int  \dd ^4 x \sqrt{-g} \left[\frac{\Mp^{2}}{2}R-\frac{1}{2}g^{\mu\nu}\partial_{\mu}\phi\partial_{\nu}\phi-V(\phi)\right] ,
\eea
where $V(\phi)$ is the potential energy stored in the scalar field $\phi$, $g_{\mu\nu}$ is the metric tensor, $g$ is its determinant and $R$ is its Ricci scalar.

\subsection{Background}
\label{subsec:background}
On a homogeneous and isotropic background, described by the Friedmann-Lema\^itre-Robertson-Walker metric, the inflaton $\bar{\phi}$ and its velocity $\bar{\pi}_\phi\equiv \dd \phi/ \dd N$, where $N=\ln(a)$ is the number of e-folds and $a$ is the scale factor, obey the Klein-Gordon equation
\bea 
\label{eq:KG splitted}
\frac{\dd\bar{\f}}{\dd N}&=\bar{\pi}_\phi\, ,\\
\frac{\dd\bar{\pi}_\phi}{\dd N}&=-\left(3-\epsilon_1\right)\bar{\pi}_\phi- \frac{V_{,\f}(\bar{\phi})}{H^2\left(\bar{\phi},\bar{\pi}_\phi\right)}\, .
\eea
Here, a bar is used to denote background quantities, $V_{,\f}$ is the derivative of the potential function with respect to the field, $H=\dot{a}/a$ is the Hubble parameter (where a dot denotes derivative with respect to cosmic time $t$), and 
\bea
\epsilon_1 = -\frac{\dot{H}}{H^2}
\eea
is the first Hubble-flow parameter. The Friedmann equation relates the Hubble parameter to the field and its velocity according to
\bea
\label{eq:Friedmann:phibar:pibar}
H^2\left(\bar{\phi},\bar{\pi}_\phi\right) = \dfrac{V\left(\bar{\phi}\right)}{3\Mp^2-\dfrac{\bar{\pi}_\phi^2}{2}}\,,
\eea
which also leads to
\bea
\label{eq:epsilon1:pi}
\epsilon_1 = \frac{\bar{\pi}_\phi^2}{2\Mp^2}\,.
\eea
Inflation ($\ddot{a}>0$) takes place when $\epsilon_1<1$, \ie when $\vert \bar{\pi}_\phi\vert <\sqrt{2}\Mp$.

\subsection{Linear perturbation theory}
\label{subsec:linear_perturbation}

In single-field inflationary models, the scalar sector of cosmological fluctuations can be described by a single gauge-invariant degree of freedom, such as the Mukhanov-Sasaki variable $u(\bm{x},\eta)$~\cite{Mukhanov:1981xt,Sasaki:1986hm}. Its dynamics follows from the action
\bea
\label{eq:S:MS}
{\cal{S}}_{\mathrm{MS}}=\frac{1}{2}\int \dd \eta \dd ^3\bm{x}\left[(u')^2-(\partial_i u)^2+\frac{Z''}{Z}u^2\right]
\eea
where $\eta$ is the conformal time defined by $\dd \eta=\dd t/a$, a prime denotes derivative with respect to $\eta$, and $Z=a\vert \bar{\pi}_\phi\vert $. In Fourier space, the quantised Mukhanov-Sasaki variable $\hat{u}$ can be expanded on plane-wave solutions according to
\bea
\hat{u}(\bm{x},\eta)=\int\frac{ \dd ^3\bm{k}}{(2\pi)^{3/2}}\left[ \ee^{-i \bm{k}\cdot\bm{x}}u_{k}(\eta) \hat{a}_{\bm{k}}+\ee^{i \bm{k}\cdot\bm{x}}u_{k}^*(\eta) \hat{a}^\dagger_{\bm{k}}\right],
\eea
where $\hat{a}_{\bm{k}}$ and $\hat{a}_{\bm{k}}^\dagger$ are quantum annihilation and creation operators, which satisfy the usual commutation relation $[\hat{a}_{\bm{k}},\hat{a}_{\bm{k}'}^\dagger]=\delta^{(3)}(\bm{k}-\bm{k}')$. The action~\eqref{eq:S:MS} thus describes a set of independent harmonic oscillators, one per Fourier mode, with time-dependent squared frequencies $k^2-Z''/Z$. The mode functions $u_k$ satisfy the Mukhanov-Sasaki equation
\bea
\label{eq:Mukhanov-Sasaki}
    u''_k + \left( k^2 - \frac{Z''}{Z} \right) u_k = 0\,,
\eea
where the normalisation condition $u_k(u_k^*)'-u_k'u_k^*=i$ ensures that $\hat{u}$ and its conjugate momentum satisfy canonical commutation relations. The Mukhanov-Sasaki equation~\eqref{eq:Mukhanov-Sasaki} can be solved in a given background $Z(\eta)$ if an initial condition is specified, and a common choice is the Bunch-Davis vacuum~\cite{Bunch:1978yq}, which corresponds to the prescription $u_k(\eta) \to e^{-i k\eta}/\sqrt{2k}$ when $\eta\to -\infty$.

In stochastic inflation, linear perturbation theory is employed to describe cosmological fluctuations at small scales, \ie within the coarse-graining scale of the theory. In practice, the statistics of the noises are determined by the two-point functions of the scalar field fluctuation $\delta\phi$ and its velocity $\delta\pi_\phi = \dd \delta\phi/ \dd N$, in the uniform-expansion gauge. For the cases of interest below, these coincide with the same quantities in the spatially-flat gauge~\cite{Pattison:2019hef, Artigas:2021zdk, Artigas:2023kyo} where $\delta\phi_k=-u_k/a$ and $\delta\pi_k=(u_k- \dd u_k/ \dd N)/a$, up to negligible gauge corrections.\footnote{More precisely, in \Refa{Pattison:2019hef} it is shown that the gauge correction is suppressed by $\epsilon_1 \sigma^2$ in slow roll and by $\epsilon_1 \sigma^6$ in ultra-slow roll, where $\sigma$ is the coarse-graining parameter introduced in \cref{subsec:stochastic_inflation} (the gauge correction is also found to be negligible during the transition between these two regimes, as it occurs in the Starobinsky model). As will be made clear below, our goal is to incorporate gradient contributions up to order $\sigma^2$ in the stochastic formalism, hence such contributions are slow-roll suppressed in the slow-roll gauge corrections and absent from the ultra-slow-roll gauge corrections. This does not preclude the existence of regimes where they might play a role, but their inclusion would then be straightforward since it would have to be done at the perturbative level, \ie independently of the techniques developed below.} The two-point functions of $\delta\phi$ and $\delta\pi_\phi$ are then described by the reduced power spectra
\bea
\label{eq:powerspectra:field}
{\cal{P}}_{\f \f}(k,N)&=\frac{k^3}{2\pi^2}|\delta\phi_k(N)|^2\,,\\
{\cal{P}}_{\f \pi}(k,N)&=\frac{k^3}{2\pi^2}\mathrm{Re}\left[\delta\phi_k(N)\delta\pi_k^*(N)\right]\,,\\
{\cal{P}}_{\pi \pi}(k,N)&=\frac{k^3}{2\pi^2}|\delta\pi_k(N)|^2\,.
\eea

Below we will also consider the comoving curvature perturbation ${\cal{R}}=u/Z$, whose mode function obeys the following equation of motion
\bea
\label{eq:linear_curvature_pert_EOM}
{\cal{R}}_k''+2\frac{Z'}{Z}{\cal{R}}_k'+k^2{\cal{R}}_k=0\, ,
\eea
and whose reduced power spectrum is given by
\bea
\label{eq:powerspectrum:R}
{\cal{P}}_{\cal{R}}(k,N)&=\frac{k^3}{2\pi^2}|{\cal{R}}_k(N)|^2\, .
\eea
%

\subsection{Stochastic inflation}
\label{subsec:stochastic_inflation}

At long wavelengths, which are relevant for most cosmological observables when evaluated at the end of inflation, an effective description of quantum fields living on an inflating cosmological background is provided by the stochastic-inflation formalism~\cite{Starobinsky:1982ee, Starobinsky:1986fx,Nambu:1987ef,Nambu:1988je,Kandrup:1988sc,Nakao:1988yi,Nambu:1989uf,Mollerach:1990zf,Linde:1993xx,Starobinsky_1994,Finelli:2008zg,Finelli:2010sh}. In this approach, fields are coarse-grained at the fixed physical length scale $(\sigma H)^{-1}$ above the Hubble radius, where $\sigma \ll 1$ is a fixed parameter, and degrees of freedom at small scales are integrated out. They act as a source for the infrared (IR) part of the theory since more and more modes cross out the coarse-graining radius to join the IR sector as time goes on, as an effect of the accelerated expansion.

In practice, the IR parts of the fields are defined in Fourier space according to
\bea
\label{eq:coarse-graining}
\hat{\f}^{\mathrm{IR}}(\bm{x},N)&=\int\frac{ \dd ^3\bm{k}}{(2\pi)^{3/2}} W\left(\frac{k}{\sigma a H}\right)\left[ \ee^{-i \bm{k}\cdot\bm{x}}\f_{k}(N) \hat{a}_{\bm{k}}+\ee^{i \bm{k}\cdot\bm{x}}\f_{k}^*(N) \hat{a}^\dagger_{\bm{k}}\right]\, ,\\
\hat{\pi}_\phi^{\mathrm{IR}}(\bm{x},N)&=\int\frac{ \dd ^3 \bm{k}}{(2\pi)^{3/2}} W\left(\frac{k}{\sigma a H}\right) \left[ \ee^{-i \bm{k}\cdot\bm{x}}\pi_{k}(N) \hat{a}_{\bm{k}}+\ee^{i \bm{k}\cdot\bm{x}}\pi_{k}^*(N) \hat{a}^\dagger_{\bm{k}}\right]\, ,
\eea
where $W$ is a window function that selects modes $k$ with wavelength larger than the coarse-graining scale, \ie $W\simeq 1$ for $k\ll \sigma a H$ and $0$ for $k\gg \sigma a H$. The ultra-violet (UV) parts of the fields are defined as the complement of the IR parts, \ie $\hat{\f}^{\mathrm{UV}}=\hat{\f}-\hat{\f}^{\mathrm{IR}}$ and $\hat{\pi}^{\mathrm{UV}}=\hat{\pi}-\hat{\pi}^{\mathrm{IR}}$.

The effective dynamics of the long-wavelength degrees of freedom can be obtained by splitting the fields between their IR and UV parts according to \cref{eq:coarse-graining}, and inserting this decomposition in the field equations of motion. This leads to~\cite{Grain:2017dqa}
\bea 
\label{eq:Langevin general}
\frac{\dd\hat{\f}^{\mathrm{IR}}(\bm{x},N)}{\dd N}&=\hat{\pi}_\phi^{\mathrm{IR}}(\bm{x},N)+\hat{\xi}_\f(\bm{x},N)\, ,\\
\frac{\dd\hat{\pi}_\phi^{\mathrm{IR}}(\bm{x},N)}{\dd N}&=-\left(3-\epsilon_1\right)\hat{\pi}_\phi^{\mathrm{IR}}(\bm{x},N)- \frac{V_{,\f}}{H^2}+\hat{\xi}_{\pi_\phi}(\bm{x},N)+\frac{\Delta\hat{\f}^{\mathrm{IR}}(\bm{x},N)}{\left(a H\right)^2}\, ,
\eea 
where $\Delta$ denotes the Laplace operator, $V_{,\f}/H^2$ and $\epsilon_1$ are functions of $\hat{\f}^{\mathrm{IR}}(\bm{x},N)$ and $\hat{\pi}^{\mathrm{IR}}_\phi(\bm{x},N)$ through \cref{eq:Friedmann:phibar:pibar,eq:epsilon1:pi}, and the source functions $\hat{\xi}_\f$ and $\hat{\xi}_{\pi_\phi}$ read
\bea 
\label{eq:expression noises}
\hat{\xi}_\f(\bm{x},N)&=\int\frac{ \dd ^3k}{(2\pi)^{3/2}} \left[ \ee^{-i \bm{k}\cdot\bm{x}}\f_{k}(N) \hat{a}_{\bm{k}}+\ee^{i \bm{k}\cdot\bm{x}}\f_{k}^*(N) \hat{a}^\dagger_{\bm{k}}\right]\frac{\dd }{\dd N}W\left(\frac{k}{\sigma a H }\right)\, ,\\
\hat{\xi}_{\pi_\phi}(\bm{x},N)&=\int\frac{ \dd ^3k}{(2\pi)^{3/2}} \left[ \ee^{-i \bm{k}\cdot\bm{x}}\pi_{k}(N) \hat{a}_{\bm{k}}+\ee^{i \bm{k}\cdot\bm{x}}\pi_{k}^*(N) \hat{a}^\dagger_{\bm{k}}\right]\frac{\dd }{\dd N}W\left(\frac{k}{\sigma a H }\right)\, .
\eea 
Here, as mentioned above, the mode functions $\f_{k}$ and $\pi_k$ have to be evaluated in the uniform-expansion gauge in which the integrated expansion and the shift vector are unperturbed~\cite{Pattison:2019hef}.

Let us note that, while gradient interactions have been kept in the matter field equations of motion~\eqref{eq:Langevin general}, they have been discarded from the constraint equations. For instance, the Friedmann equation~\eqref{eq:Friedmann:phibar:pibar} has been employed to derive \cref{eq:Langevin general}, although in the ADM formalism it corresponds to the energy-constraint equation only at the background level and should otherwise receive gradient corrections~\cite{Artigas:2021zdk}. The momentum-constraint equation is also discarded. The reason for such a simplification is that we implicitly work in the so-called decoupling limit~\cite{Cheung:2007st}, where metric fluctuations are sub-dominant. The generalisation of our approach beyond that regime is left to future work.

In the stochastic formalism, \cref{eq:Langevin general} is treated as stochastic, Langevin equations, where $\hat{\xi}_\f$ and $\hat{\xi}_{\pi_\phi}$ are replaced with stochastic noises. Their statistics are determined from their quantum expectation values, which can be computed using standard cosmological perturbation theory because the noises $\hat{\xi}_\f$ and $\hat{\xi}_{\pi_\phi}$ involve scales that are not larger than the coarse-graining radius. At linear order, the noises are thus centred Gaussian noises, with covariance given by
\bea
\label{eq:cov:interm}
\left\langle \hat{\xi}_f (\bm{x},N) \hat{\xi}_g^\dagger(\bm{y},N') \right\rangle = & \int_{-\infty}^\infty \dd\ln(k)\frac{\dd }{\dd N}W\left[\frac{k}{\sigma a H\left(N\right)}\right]\frac{\dd }{\dd N'}W\left[\frac{k}{\sigma a H\left(N'\right)}\right]
\\ & \ \
\times\frac{k^3}{2\pi^2} f_{k}(N) g^*_{k}(N')\sinc\left(k|\bm{x}-\bm{y}|\right)
\,,
\eea
where $f,g=\f$ or $\pi_\phi$. If $W$ is a Heaviside step function, \ie $W=1$ if $k<k_\sigma=\sigma a H$ and $0$ otherwise, the covariance matrix becomes
\bea
\label{eq:covariance window function}
\left\langle \xi_f (\bm{x},N) \xi_g(\bm{y},N') \right\rangle = \mathcal{P}_{fg}\left(k_\sigma,N\right)\Dirac(N-N')\left[1-\epsilon_1(N)\right]\sinc\left(k_\sigma|\bm{x}-\bm{y}|\right)\, ,
\eea
where $\mathcal{P}_{fg}$ are the reduced power spectra given in \cref{eq:powerspectra:field} and $\Dirac$ denotes the Dirac distribution.

The presence of the Dirac distribution $\Dirac(N-N')$ implies that the noises are white, \ie uncorrelated over time, while the cardinal sine function indicates that the noises are mostly uncorrelated when evaluated between two points that are distant by more than the coarse-graining radius. 

\subsection{Stochastic inflation without gradient interactions}
\label{sec:Stochastic:Inflation:Standard}

In the separate-universe approach, gradient interactions are neglected at large scales since they are suppressed by $k^2/(aH)^2$, hence by $\sigma^2\ll 1$. In this limit, the Laplacian term in \cref{eq:Langevin general} can be dropped, and the Langevin equation becomes
\bea 
\label{eq:Langevin:standard}
\frac{\dd}{\dd N} {\f}^{\mathrm{IR}}= &{\pi}_\phi^{\mathrm{IR}}+{\xi}_\f\, ,\\
\frac{\dd}{\dd N}{\pi}_\phi^{\mathrm{IR}}= & -\left[3 -\epsilon_1\left({\f}^{\mathrm{IR}},{\pi}_\phi^{\mathrm{IR}}\right)\right]{\pi}_\phi^{\mathrm{IR}}- \frac{V_{,\f}}{H^2}\left({\f}^{\mathrm{IR}},{\pi}_\phi^{\mathrm{IR}}\right)+{\xi}_{\pi_\phi}\, .
\eea 
The covariance of the noises is still given by \cref{eq:covariance window function}.
The statistical properties of the IR fields can thus be described by producing separate realisations of the Langevin equation, each corresponding to the time evolution of one Hubble patch. This is the so-called separate-universe approach~\cite{Salopek:1990jq, Sasaki:1995aw, Wands:2000dp, Pattison:2019hef, Artigas:2021zdk}.
The above scheme constitutes the ``standard'' stochastic-inflation formalism, to which the improved version derived in \cref{sec:memory effect} will be compared below. 

\subsection{Importance of gradient effects}
\label{sec:Importance_gradient}

Before showing how gradient interactions can be incorporated within the stochastic formalism, in this section we explain in which models they are expected to play an important role, and why it is sufficient to keep only leading gradient ``corrections'' when this is the case. The following discussion is framed in linear perturbation theory, and revisits results from~\cite{Leach:2001zf, Jackson:2023obv, Artigas:2024ajh}.

\subsubsection{Gradient expansion}
\label{sec:gradient:expansion}

The gradient expansion is an expansion in powers of $k^2$ of the solutions to \cref{eq:linear_curvature_pert_EOM}. At leading order, the last term of that equation can be neglected, and the two independent solutions read
\bea 
\mathcal{G}_0 = & C_k\, ,\\
\mathcal{D}_0 = & D_k \int^{\eta}_{\eta_{\mathcal{D}}^{(0)}}\frac{\dd\tilde{\eta}}{Z^2(\tilde{\eta})}\, .
\eea 
In this expression, $C_k$, $D_k$ and $\eta_{\mathcal{D}}^{(0)}$ are integration constants, which are not independent since a shift in $\eta_{\mathcal{D}}^{(0)}$ can be reabsorbed in $C_k$. The solutions $\mathcal{G}_0$ and $\mathcal{D}_0$ are usually called ``growing''- and ``decaying''-mode solutions, hence the notation.  They can be used to evaluate the third term of \cref{eq:linear_curvature_pert_EOM} and derive the solution at order $k^2$, so on and so forth, such that one formally obtains
\bea 
\mathcal{R}_k(\eta) = \mathcal{G}(\eta) + \mathcal{D}(\eta)
\quad\text{with}\quad
\mathcal{G}(\eta) = \sum_{n=0}^\infty k^{2n}\mathcal{G}_{n}(\eta)
\quad\text{and}\quad
\mathcal{D}(\eta) = \sum_{n=0}^\infty k^{2n}\mathcal{D}_{n}(\eta),
\eea 
where $\mathcal{G}_n''+2 (Z'/Z) \mathcal{G}_n'=-\mathcal{G}_{n-1}$ and $\mathcal{D}_n''+2 (Z'/Z) \mathcal{D}_n'=-\mathcal{D}_{n-1}$, which can be solved according to
\bea
\label{eq:Gn:Dn:iterative}
\mathcal{G}_n(\eta) = & \int_\eta^{\eta_{\mathcal{G}}^{(n)}} \frac{\dd\tilde{\eta}}{Z^2(\tilde{\eta})} \int_{\bar{\eta}_{\mathcal{G}}^{(n)}}^{\tilde{\eta}} \dd\bar{\eta} Z^2(\bar{\eta}) \mathcal{G}_{n-1}(\bar{\eta})\, ,\\
\mathcal{D}_n(\eta) = & \int_\eta^{\eta_{\mathcal{D}}^{(n)}} \frac{\dd\tilde{\eta}}{Z^2(\tilde{\eta})} \int_{\bar{\eta}_{\mathcal{D}}^{(n)}}^{\tilde{\eta}} \dd\bar{\eta} Z^2(\bar{\eta}) \mathcal{D}_{n-1}(\bar{\eta})\, .
\eea 

\subsubsection{Homogeneous-matching procedure}

In the separate-universe picture~\cite{Lifshitz:1960, Starobinsky:1982ee, Salopek:1990jq, Comer:1994np, Sasaki:1995aw, Sasaki:1998ug, Wands:2000dp, Lyth:2003im, Rigopoulos:2003ak, Lyth:2004gb, Lyth:2005fi, Cai:2018dkf, Pi:2022ysn}, cosmological perturbation theory (to all orders in $k$) is employed to evolve $\mathcal{R}_k$ until the time when the wavelength of the Fourier mode $k$ becomes sufficiently larger than the Hubble radius, \ie when $k$ crosses $\sigma a H$. Past this point, the ``homogeneous'' solution ($k=0$) is used to track the evolution of $\mathcal{R}_k$, since gradient effects can be argued to play a subdominant role at super-Hubble scales. Our goal is to make this statement more precise. 

Let us denote by $\eta_*$ the time at which $k=\sigma a H$, and let us assume that $\mathcal{R}_k(\eta_*)$ and $\mathcal{R}_k'(\eta_*)$ have been computed by solving \cref{eq:linear_curvature_pert_EOM} until $\eta_*$, starting from the Bunch-Davies vacuum. At zeroth order, $\mathcal{R}_k^{(0)}=\mathcal{G}_0+\mathcal{D}_0$ at times $\eta>\eta_*$, where the integration constants $C_k$, $D_k$ and $\eta_{\mathcal{D}}^{(0)}$ can be set by requiring that $\mathcal{R}_k$ and $\mathcal{R}_k'$ are continuous at $\eta_*$. This leads to
\bea
\label{eq:Rk0}
{\cal{R}}_k^{(0)}(\eta)=\mathcal{R}_k(\eta_*)+\mathcal{R}'_k(\eta_*)Z^2(\eta_*)\int^{\eta}_{\eta_{*}}\frac{\dd \tilde{\eta}}{Z^2(\tilde{\eta})}\,.
\eea
This is referred to as the ``homogeneous-matching'' procedure, which lies at the heart of the separate-universe approach.

The same matching procedure can be repeated at order $k^2$ and one finds that $\mathcal{R}_k={\cal{R}}_k^{(0)} + k^2 {\cal{R}}_k^{(2)}$ where ${\cal{R}}_k^{(0)}$ is given by \cref{eq:Rk0} and 
\bea
\label{eq:Rk2}
{\cal{R}}_k^{(2)}(\eta)=-\mathcal{R}_k(\eta_*)\int_{\eta_*}^\eta \frac{\dd\tilde{\eta}}{Z^2(\tilde{\eta})}\int_{\eta_*}^{\tilde{\eta}}Z^2(\bar{\eta})\dd\bar{\eta} -\mathcal{R}'_k(\eta_*)Z^2(\eta_*)
\int^{\eta}_{\eta_{*}}\frac{\dd\tilde{\eta}}{Z^2(\tilde{\eta})}\int_{\eta_*}^{\tilde{\eta}} \dd\bar{\eta} Z^2(\bar{\eta}) \int_{\eta_*}^{\bar{\eta}} \frac{\dd \bar{\bar{\eta}}}{Z^2(\bar{\bar{\eta}})}\,.
\eea
Notice that all integration constants, including the integral bounds, have been uniquely determined by the two matching conditions, which is obviously expected since $\mathcal{R}_k$ satisfies a second-order linear differential equation.

\subsubsection{Failure of the zeroth-order solution during dynamical transitions}
\label{sec:failure:zeroth:order:in:transitions}

In order to assess the validity of the homogeneous-matching procedure, let us now compare ${\cal{R}}_k^{(0)}$ and $k^2 {\cal{R}}_k^{(2)}$ in a few situations of interest. Only when the latter is negligible compared to the former can the homogeneous solution~\eqref{eq:Rk0} be trusted at times $\eta\geq\eta_*$.

\paragraph{Slow-roll attractor}

Along the slow-roll (SR) attractor, $\epsilon_1$ is almost constant, hence $Z  \propto a$. At late times, the integrals appearing in \cref{eq:Rk0,eq:Rk2} reduce to
\bea
    Z^2(\eta_*)\int^\eta_{\eta_*}\frac{ \dd \tilde{\eta}}{Z^2(\tilde{\eta})}&\xrightarrow[\eta\to0]{}\frac{1}{3a_* H}\,,
    \\
    \int^\eta_{\eta_*}\frac{ \dd \tilde{\eta}}{Z^2(\tilde{\eta})}\int^{\tilde{\eta}}_{\eta_*} \dd \bar{\eta} Z^2(\bar{\eta}) &\xrightarrow[\eta\to0]{}\frac{1}{6a_*^2 H^2}\,,
    \\
    Z^2(\eta_*)\int^\eta_{\eta_*}\frac{ \dd \tilde{\eta}}{Z^2(\tilde{\eta})}\int^{\tilde{\eta}}_{\eta_*} \dd \bar{\eta} Z^2(\bar{\eta})\int^{\bar{\eta}}_{\eta_*}\frac{ \dd \bar{\bar{\eta}}}{Z^2(\bar{\bar{\eta}})} &\xrightarrow[\eta\to0]{}\frac{1}{30a_*^3 H^3}\,,
\eea
 where $a_*=a(\eta_*)$ and $H$ has been approximated to a constant. If initial conditions are set by the Bunch-Davies vacuum, \cref{eq:linear_curvature_pert_EOM} leads to ${\cal{R}}_k'(\eta_*)/{\cal{R}}_k(\eta_*) =-\sigma^2 a_* H$ at leading order in $\sigma$, and this implies that
    \bea
    \label{eq:ratio_k2_correction_SR}
    \left|\frac{k^2{\cal{R}}_k^{(2)}}{{\cal{R}}_k^{(0)}}\right|_{\eta\rightarrow0}\simeq \frac{\sigma^2}{6}\,.
    \eea
 Therefore, as long as $\sigma$ is chosen to be sufficiently small, the $k^2$ correction is suppressed compared to the $k^0$ solution and the standard separate-universe approach, which consists in keeping the $k^0$ solution only, is well justified. It should be stressed that, in SR inflation, ${\cal{G}}_0$ always gives the main contribution to $\mathcal{R}_k^{(0)}$, and the curvature perturbation is conserved on super-Hubble scales.

\paragraph{Constant-roll non-attractor models}

Let us now consider models where $Z^2\propto a^n$, hence $\epsilon_2\equiv \dd\ln(\epsilon_1)/\dd N=n-2$ is constant. If $n\simeq 2$ one recovers slow roll, but for $n<-1$ the dynamics does not proceed along a dynamical attractor. A prototypical example is ultra-slow roll (USR), for which $\epsilon_2=-6$ and $n=-4$. Keeping $n<-1$ generic, and assuming that $H$ is a constant (hence $\epsilon_1\ll 1$), the integrals appearing in \cref{eq:Rk0,eq:Rk2} now reduce to
\bea
     Z^2(\eta_*)\int^\eta_{\eta_*}\frac{ \dd \tilde{\eta}}{Z^2(\tilde{\eta})}&\xrightarrow[\eta\to0]{}\frac{-1}{(n+1)a_* H}\left[\frac{a_*}{a(\eta)}\right]^{n+1}\,,
    \\
    \int^\eta_{\eta_*}\frac{ \dd \tilde{\eta}}{Z^2(\tilde{\eta})}\int^{\tilde{\eta}}_{\eta_*} \dd \bar{\eta} Z^2(\bar{\eta}) &\xrightarrow[\eta\to0]{}\frac{1}{(n^2-1)a_*^2 H^2}\left[\frac{a_*}{a(\eta)}\right]^{n+1}\,,
    \\
    Z^2(\eta_*)\int^\eta_{\eta_*}\frac{ \dd \tilde{\eta}}{Z^2(\tilde{\eta})}\int^{\tilde{\eta}}_{\eta_*} \dd \bar{\eta} Z^2(\bar{\eta})\int^{\bar{\eta}}_{\eta_*}\frac{ \dd \bar{\bar{\eta}}}{Z^2(\bar{\bar{\eta}})} &\xrightarrow[\eta\to0]{}\frac{1}{2(n^2-1)a_*^3 H^3}\left[\frac{a_*}{a(\eta)}\right]^{n+1}\, .
\eea
 Unlike in slow roll, the curvature perturbation grows on super-Hubble scales in these models. Setting initial conditions in the Bunch-Davies vacuum, \cref{eq:linear_curvature_pert_EOM} leads to ${\cal{R}}_k'(\eta_*)/{\cal{R}}_k(\eta_*)=-(n+1)a_*H$ at leading order in $\sigma$, hence
    \bea
    \left| \frac{k^2{\cal{R}}_k^{(2)}}{{\cal{R}}_k^{(0)}}\right|_{\eta\rightarrow0}\simeq\frac{-\sigma^2}{2(n+1)}\,.
    \eea
    Gradient corrections are again suppressed by $\sigma^2$, hence the separate-universe approach is still valid in these models. The only difference with the slow-roll case is that, here, $\mathcal{D}_0$ provides the main contribution to $\mathcal{R}_k^{(0)}$.
    
\paragraph{SR/USR transitions}  

In most models featuring a period of ultra-slow roll, it follows a phase of SR inflation, and relaxes back to SR at late time. A prototypical example is the Starobinsky's piecewise linear potential~\cite{Starobinsky:1992ts}, see \cref{subsec::Starobinsky_Linear_potential}, but this behaviour is encountered in most inflection-point potentials. 

\begin{figure}[t]
    \centering
    \includegraphics[width=0.49\textwidth,trim={0cm 0cm 0cm 0cm}, clip]{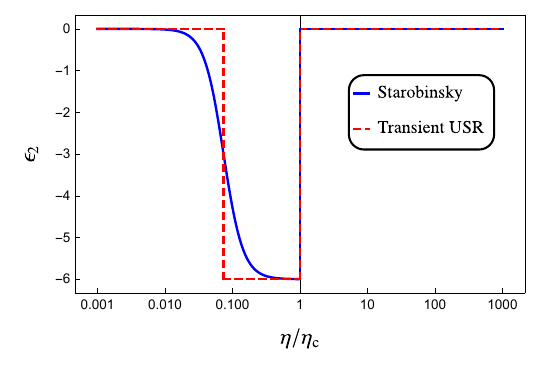}
    \includegraphics[width=0.49\textwidth,trim={0cm 0cm 0cm 0cm}, clip]{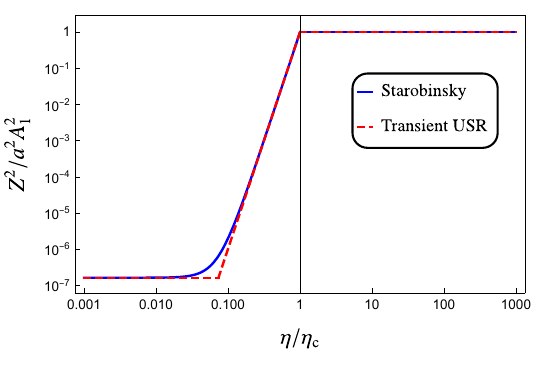}
    \caption{Second Hubble-flow parameter $\epsilon_2$, and $Z$ function normalised by the scale factor, in the piecewise Starobinsky model discussed in \cref{subsec::Starobinsky_Linear_potential}. The parameter values used in these figures are reported in \cref{table:Starobinsky}. The red dashed line corresponds to the transient-USR model~\eqref{eq:Z approximation transient}. Since $\eta<0$, time goes from right to left.}
      \label{fig:Z_and_eta}
    \end{figure}

Even though we have established the validity of the separate-universe approach within the SR and USR phases separately, we now examine its validity during the transitions in between these phases. In practice, we describe the three phases (SR, USR, SR) of the dynamics by approximating $\epsilon_2$ by a top-hat function ($\epsilon_2=0$, $\epsilon_2=-6$, $\epsilon_2=0$), which leads to  
    \bea
    \label{eq:Z approximation transient}
    Z^2(\eta)=
    \begin{cases}
    a^{2}(\eta)A_1^2 
    \quad\text{if}\quad
    \eta\le\eta_\uc\, \\ 
    a^{-4}(\eta)a_\uc^6 A_1^2
        \quad\text{if}\quad
    \eta_\uc<\eta\le\left({A_2}/{A_1}\right)^{\frac{1}{3}}\eta_\uc
    \, \\ 
    a^{2}(\eta)A_2^2
        \quad\text{if}\quad
    \left({A_2}/{A_1}\right)^{\frac{1}{3}}\eta_\uc<\eta\,
    \end{cases}\, .
    \eea
Here, $\eta_\uc$ is the conformal time at the SR-USR transition, $a_\uc=a(\eta_\uc)$, $A_1$ and $A_2$ are constant parameters that correspond to the velocity of the inflaton in the initial and final SR stages respectively, and we assume that $\epsilon_1\ll 1$ at all times hence $a(\eta)\simeq -1/(H \eta)$. The duration of the USR phase is set by the ratio $A_1/A_2$, which also determines the amount by which the power spectrum of curvature perturbations is amplified (more precisely, the power spectrum is enhanced at small scales by a factor $A_1^2/A_2^2$). In \Fig{fig:Z_and_eta}, we show the evolution of $Z$ and $\epsilon_2$ in the  Starobinsky's piecewise linear potential model discussed in depth in \cref{subsec::Starobinsky_Linear_potential}, and we superimpose \cref{eq:Z approximation transient} to show that it broadly captures the dynamics of the background in this kind of models.

 Fourier modes that exit the Hubble radius after $\eta_\uc$ are correctly described by the $k^0$ solution~\cite{Leach:2001zf,Jackson:2023obv}, hence we focus on scales $k< \sigma a_\uc H$, for which the matching is performed before the transition. In this regime, one obtains
    \bea
   Z^2(\eta_*)\int^\eta_{\eta_*}\frac{ \dd \tilde{\eta}}{Z^2(\tilde{\eta})} & \xrightarrow[\eta\to 0]{}\frac{1}{3a_* H}\left[1+\frac{2A_1}{A_2}\left(\frac{a_*}{a_\uc}\right)^3\right]\,,
    \\ 
    \int^\eta_{\eta_*}\frac{ \dd \tilde{\eta}}{Z^2(\tilde{\eta})}\int^{\tilde{\eta}}_{\eta_*} \dd \bar{\eta} Z^2(\bar{\eta}) & \xrightarrow[\eta\to0]{}\frac{1}{30a_*^2 H^2}\left[5+\frac{24A_1}{A_2}\left(\frac{a_*}{a_\uc}\right)^2-\frac{20A_1}{A_2}\left(\frac{a_*}{a_\uc}\right)^3\right]\,,
    \\
    &
    \hspace{-4cm}
    Z^2(\eta_*)\int^\eta_{\eta_*}\frac{ \dd \tilde{\eta}}{Z^2(\tilde{\eta})}\int^{\tilde{\eta}}_{\eta_*} \dd \bar{\eta} Z^2(\bar{\eta})\int^{\bar{\eta}}_{\eta_*}\frac{ \dd \bar{\bar{\eta}}}{Z^2(\bar{\bar{\eta}})}
    \\
    &
    \xrightarrow[\eta\to0]{}\frac{1}{30a_*^3 H^3}\left[1+\frac{8A_1}{A_2}\left(\frac{a_*}{a_\uc}\right)^2-\frac{10A_1}{A_2}\left(\frac{a_*}{a_\uc}\right)^3+\frac{4A_1}{A_2}\left(\frac{a_*}{a_\uc}\right)^5\right]\,,
    \eea
    where we dropped terms suppressed by $A_2/A_1$ that we assume to be small. Since the scales of interest exit the $\sigma$-Hubble radius during the first SR phase, one still has ${\cal{R}}_k'(\eta_*)/{\cal{R}}_k(\eta_*)=-\sigma^2 a_* H$ at leading order in $\sigma^2$, which gives rise to
    \bea
    \label{eq:ratio_k2_correction_transient_USR}
    \left|\frac{k^2{\cal{R}}_k^{(2)}}{{\cal{R}}_k^{(0)}}\right|_{\eta\rightarrow0}\simeq\frac{\sigma^2}{30}\left|\frac{5+24\frac{A_1}{A_2}\left(\frac{a_*}{a_\uc}\right)^2-20\frac{A_1}{A_2}\left(\frac{a_*}{a_\uc}\right)^3}{1-\frac{2}{3}\sigma^2\frac{A_1}{A_2}\left(\frac{a_*}{a_\uc}\right)^3}\right|\,.
    \eea
Two cases need to be distinguished. If $A_1/A_2\ll (a_\uc/a_*)^2$, \ie if $k\ll \sigma a_\uc H \sqrt{A_2/A_1}$, one recovers the slow-roll result~\eqref{eq:ratio_k2_correction_SR}. This is because, at those scales, the decaying mode has been sufficiently suppressed during the first SR phase such that it does not catch up with the growing mode during the USR phase, hence the curvature perturbation remains frozen during the transition. In contrast, if $k\gg \sigma a_\uc H \sqrt{A_2/A_1}$, the $k^2$ correction becomes large (unless $\sigma^2\ll A_2/A_1$), which signals a breakdown of the separate-universe approach.  

\subsubsection{Why the second-order solution is sufficient}
\label{sec:k2:is:enough}

In models with transitions, we have thus found that there are scales for which $k^2 \mathcal{R}_k^{(2)}$ is comparable to, and even possibly larger than, the zeroth-order solution $\mathcal{R}_k^{(0)}$. A natural concern is that the whole gradient expansion breaks down at these scales, and that $k^4 \mathcal{R}_k^{(4)}$, $k^6 \mathcal{R}_k^{(6)}$, \etc, may all become relevant. As we now show, this is in fact not the case, and including the $k^2$-``corrections'' (quotes are used since such ``corrections'' may be large) is enough.

Let us consider again the gradient expansion as introduced in \cref{sec:gradient:expansion}. At each order, new integration constants appear, in the form of integral bounds. However, since the curvature perturbation follows a second-order linear differential equation, there are only two independent integration constants, hence there exist degeneracies between the integral bounds. As mentioned in \cref{sec:gradient:expansion}, at zeroth order, a shift in $\eta_{\mathcal{D}}^{(0)}$ can be absorbed in a redefinition of $C_k$. At order $k^2$, from \cref{eq:Gn:Dn:iterative} it is clear that a shift in $\bar{\eta}_{\mathcal{G}}^{(1)}$ or $\bar{\eta}_{\mathcal{D}}^{(1)}$ generates a term proportional to $\mathcal{D}_0$, \ie it redefines the amplitude of the decaying mode at zeroth order. As a consequence, the effective amplitude of the zeroth-order decaying mode depends on the prescription used for these integration constants. %
In other words, what is classified as a $k^2$ correction in some prescription belongs to the zeroth-order sector for some other prescription~\cite{Artigas:2025nbm}. 

This ambiguity in power counting results in the failure of the zeroth order solution for some prescriptions for the integral bounds. However, by including the second-order contributions, one can be sure that, regardless of the prescription, all terms relevant at zeroth order are accounted for. In some sense, the gradient expansion is always valid, the failure of the separate-universe approach is simply due to an ambiguity in the definition of the zeroth-order contributions. When second-order contributions are also included, that ambiguity is benign.\footnote{This does not preclude the possibility of accidental cancellations between $k^0$ and $k^2$ terms, for which the leading non-vanishing contributions would appear at order $k^4$.}

This is the reason why, in~\cite{Jackson:2023obv}, it is found that the power spectrum of curvatures perturbations in the Starobinsky's piecewise  linear potential model as obtained from numerically solving \cref{eq:linear_curvature_pert_EOM}, \ie including all terms in the gradient expansion, is well reproduced by the second-order solution~\eqref{eq:Rk2}, while the zeroth-order result~\eqref{eq:Rk0} breaks down at scales mildly larger than the comoving Hubble radius at the SR-USR transition.

Let us finally note that, even in models with sharp transitions, the zeroth-order solution of the homogeneous-matching procedure is always accurate if one chooses $\sigma^2$ to be sufficiently small. However, the drawback of using a small value of $\sigma^2$ in the (classical or stochastic) $\delta N$ formalism is that non-linearities are not accounted for inside the $\sigma$-Hubble radius. By decreasing $\sigma$, one thus looses access to potentially relevant non-perturbative effects. This is why, in the next section, we show how $k^2$ ``corrections'' can be incorporated in the stochastic-inflation formalism, getting rid of the need to use small values for $\sigma^2$.

\section{Gradient interactions as a memory effect}
\label{sec:memory effect}

In the separate-universe picture, the universe is described as an ensemble of independent $\sigma$-Hubble patches. The fact that these patches evolve independently implies that gradient interactions between them are neglected. As discussed in \cref{sec:Importance_gradient}, although this approximation is expected to be valid in most cases, in models presenting a sharp transition between an attractor and a non-attractor phase, finite gradient effects need to be taken into account. In principle, this requires to describe interactions between patches, hence to give up the technical simplicity of the separate-universe picture. Nevertheless, in this section we show that, in the stochastic-inflation formalism, gradient interactions can be modelled as a memory effect, which allows us to retain a description in terms of separate universes. 

\subsection{Gradient interactions on a lattice}
\label{sec:discretised gradient}

\begin{figure}[t]
\centering
\includegraphics[height=0.5\textwidth,trim={0cm 0cm 0cm 0cm}, clip]{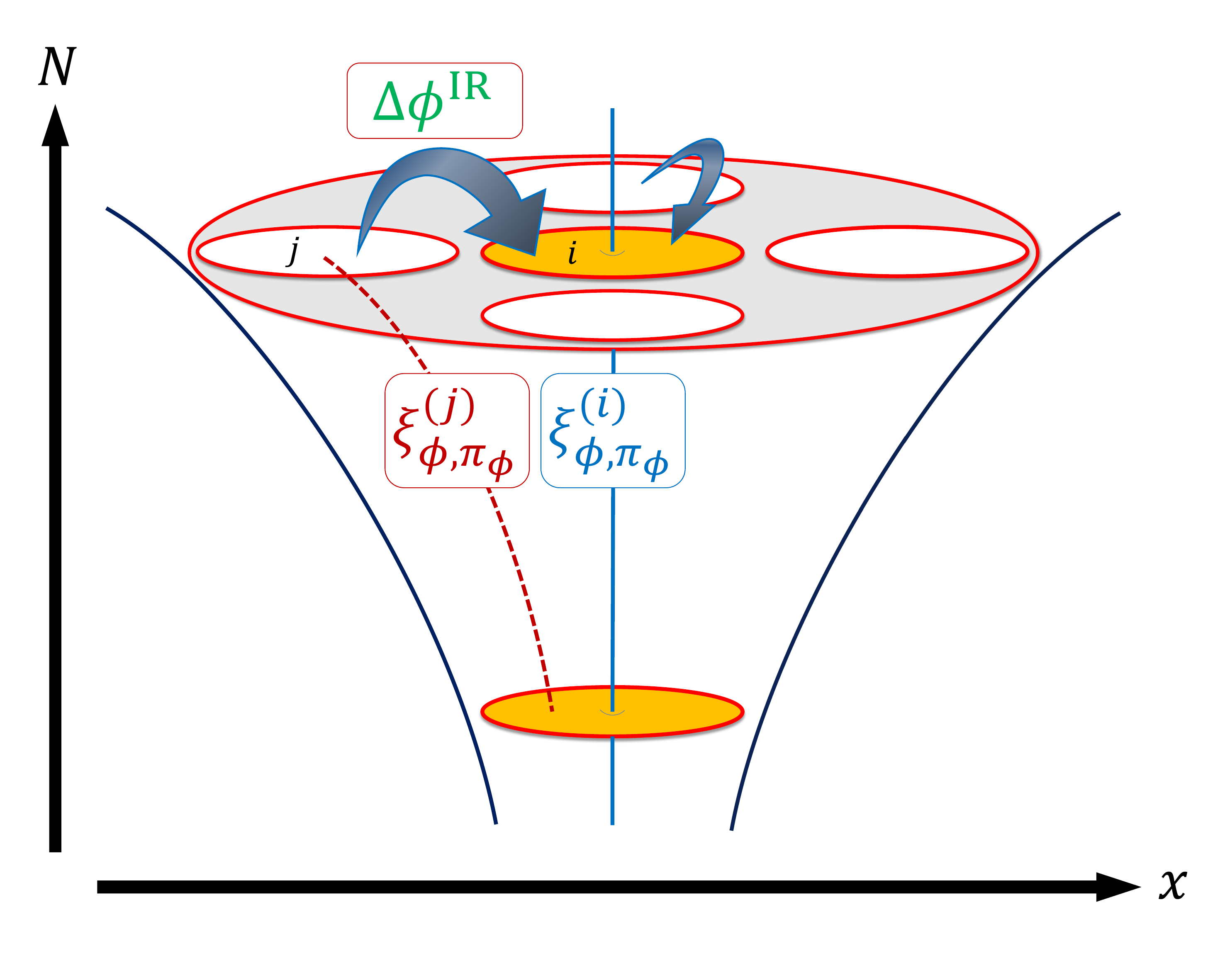}
\caption{Time evolution of a reference Hubble patch during inflation. As the expansion proceeds, this patch leads to several patches. The gradient interaction comes from the difference in the field value between the reference patch in orange and the neighbour patches in white. This difference is the result of the accumulation of distinct stochastic realisations of the noises along the two worldlines displayed in blue and red. Since these noises are perfectly correlated in the distant past, where they are evaluated within the same patch, gradient interactions can be interpreted as a local memory effect.}
\label{fig:memory effect}
\end{figure}

Before introducing gradient corrections explicitly in the stochastic-inflation formalism, we first consider gradient interactions as they would be implemented on a lattice, in order to gain some intuition on the nature of this type of interactions during inflation. The Laplace term $\Delta\cgfield/\left(aH\right)^2$ in the Langevin equation \eqref{eq:Langevin general} can be discretised on a cubic lattice with spacing $\left(\sigma aH\right)^{-1}$ using finite differences, see \cref{fig:memory effect}. At first order, at the site $i$ it can be evaluated according to
\bea
\label{eq:discretised Laplace}
\frac{\Delta\phi^{\mathrm{IR}}_i}{\left(aH\right)^2} = \sigma^2\sum_{j\in\mathrm{NN}(i)}\left(\phi^{\mathrm{IR}}_j-\phi^{\mathrm{IR}}_i\right)\,,
\eea
where the sum is over the nearest neighbours of the site $i$, denoted $\mathrm{NN}(i)$. This expression makes it clear that the gradient contributions to the Langevin equation is driven by the difference between the field value at position $i$ and at the neighbouring positions $j$.\footnote{Strictly speaking, if two spatial points $i$ and $j$ are distant by more than the Hubble radius, they do not interact, because of the causal structure of inflating space-times. However, the field value associated to each lattice node corresponds to its coarse-grained average within the cell that surrounds it, and two adjacent cells  $i$ and $j$ always share pairs of points that are closer to each other than the Hubble distance, hence they gradient interact. Moreover, in the picture of \cref{fig:memory effect}, the field is implicitly coarse-grained with top-hat window functions in real space, while in the stochastic formalism we use top-hat window functions in Fourier space. This is why, in real space, the window functions of two nearby patches overlap, and until the details of the resulting gradient interactions are derived below, \cref{fig:memory effect} should be understood as schematic only.} 

The way by which the field value at neighbouring positions $i$ and $j$ at time $N$ differ is determined by recent realisations of the noise around the site $i$: indeed, until the time $N-\ln(2)$, the comoving patches centred on $i$ and $j$ belong to the same Hubble sphere, hence they share the same field values. Differences accumulate between $N-\ln(2)$ and $N$, at which point the two comoving patches become adjacent Hubble patches, see \cref{fig:memory effect}. These differences are thus determined by recent and nearby realisations of the noise. In other words, the mean-field expression~\eqref{eq:discretised Laplace} is local in space by design, but in inflating space-times it is also local in time.
In what follows, we will show that $\cgfield_j-\cgfield_i$ can be expressed in terms of the noise realisation within the reference patch $i$ over its past evolution, hence that gradient interactions can be interpreted as a memory, \ie non-Markovian, effect within each Hubble patch individually.   

\subsection{Gradient interaction for a continuous field}
\label{sec:continuous expression}

The above considerations prompt us to rewrite $\Delta\cgfield$ in terms of the noises $\xi_\field$ and $\xi_{\momentum}$. In the present and next subsections, we derive that expression step by step. First, the Laplace term in the Langevin equation \eqref{eq:Langevin general} has the trivial representation
\bea
\label{eq:Delta:phiIR:start}
\Delta\cgfield(\bm{x},N)&= \int \dd ^3 \bm{y}\Delta_{\bm{y}}\left[\cgfield(\bm{y},N)\right]\Dirac(\bm{y}-\bm{x})\,.
\eea
Hereafter, when $\Delta$ appears without a subscript, the differentiation is performed with respect to the $\bm{x}$ variable. Introducing the field difference
\bea
\label{eq:delta:cgfield:def}
\delta\cgfield\left(\bm{x},\bm{y},N\right) = \cgfield\left(\bm{y},N\right) - \cgfield\left(\bm{x},N\right)\,,
\eea
this is equivalent to
\bea
\Delta\cgfield(\bm{x},N)&= \int \dd ^3 \bm{y}\Delta_{\bm{y}}\left[\delta\cgfield(\bm{x},\bm{y},N)\right]\Dirac(\bm{y}-\bm{x})\,.
\eea
Note that the field difference $\delta\cgfield$ is not equal to the perturbation around the background value, \ie $\delta\cgfield\neq\cgfield-\bar{\phi}$, as it rather compares the field at two different positions. Integrating by parts, one obtains
\bea
\label{eq:expression gradient delta_phi}
\Delta\cgfield(\bm{x},N)&= \int \dd ^3 \bm{y}\delta\phi^{\mathrm{IR}}(\bm{x},\bm{y},N)I\left(|\bm{x}-\bm{y}|\right)\, ,
\eea
where we have introduced the representation of the Laplacian operator
\bea
\label{eq:I:distrib:Delta}
I\left(|\bm{x}-\bm{y}|\right)=\Delta_{\bm{y}}\Dirac(\bm{y}-\bm{x})\, .
\eea
As shown in \cref{app:I}, this can be interpreted as a smoothing function that selects patches that are close enough to $\bm{x}$. Thus, \cref{eq:expression gradient delta_phi} may be seen as the continuous version of \cref{eq:discretised Laplace}.
Our next task is thus to express the field difference $\delta\phi^{\mathrm{IR}}$ in terms of the noises $\xi_\field$ and $\xi_{\momentum}$.

\subsection{Linear evolution of the local field difference}
\label{sec:linear_evolution_of_field_difference}

In the stochastic-inflation formalism, the source terms in the Langevin equation (\ie the noises) are computed at linear order in cosmological perturbation theory. The reason is that these terms are made of field fluctuations around the Hubble scale, hence their statistics is shaped by the details of the sub-Hubble dynamics while stochastic inflation only accounts for non-linearities at super-Hubble scales. As discussed around \cref{fig:memory effect}, $\Delta\cgfield$ also arises from fluctuations around the Hubble scale: since the field difference builds up from noise realisations within the recent past of a given patch, it does not contain scales that are much larger than the Hubble radius. For this reason, it can be evaluated in linear perturbation theory.
Moreover, as explained in \cref{sec:k2:is:enough}, only the leading gradient corrections are needed, hence it is enough to compute it at leading order in the gradient expansion.

By expanding \cref{eq:Langevin general} at linear order in $\delta\cgfield$, and discarding the Laplacian term that only arises at next-to-leading order in the gradient expansion, one obtains
\bea
\label{eq:Langevin deltaphi}
\frac{ \dd \delta\cgfield}{ \dd N}\left(\bm{x},\bm{y},N\right)&=\delta\cgmomentum\left(\bm{x},\bm{y},N\right)+\xi_\field\left(\bm{y},N\right)-\xi_\field\left(\bm{x},N\right)\,,
\\
\frac{ \dd \delta\cgmomentum}{ \dd N}\left(\bm{x},\bm{y},N\right)&=\alpha(\bm{x},N)\delta\cgfield\left(\bm{x},\bm{y},N\right)+\beta(\bm{x},N)\delta\cgmomentum\left(\bm{x},\bm{y},N\right)+\xi_{\momentum}\left(\bm{y},N\right)-\xi_{\momentum}\left(\bm{x},N\right)\,,
\eea
where
\bea 
\label{eq:alpha:beta:def}
\alpha(\bm{x},N) =& 
\left[\frac{1}{2}{(\cgmomentum)^2(\bm{x},N)}-3\Mp^2\right] \left(\frac{V_{,\f}}{V}\right)_{\!\!\!,\f}\left[\cgfield(\bm{x},N)\right]\, ,
\\
\beta(\bm{x},N) =& -3+\frac{1}{2\Mp^2}{(\cgmomentum)^2(\bm{x},N)}+\frac{V_{,\phi}}{V}\left[\cgfield(\bm{x},N)\right] \cgmomentum(\bm{x},N)\, ,
\eea 
are evaluated along the trajectory followed by the fields $\cgfield\left(\bm{x},N\right)$ and $\cgmomentum(\bm{x},N)$. 
The system of equations~\eqref{eq:Langevin deltaphi} is linear, hence it can be solved formally in terms of two homogeneous solutions $\{\fieldh_{(1)},\momentumh_{(1)}\}$ and $\{\fieldh_{(2)},\momentumh_{(2)}\}$ of the source-less system
\bea
\label{eq:linearised:sourceless:system}
\frac{ \dd \fieldh}{ \dd N}&=\momentumh_\phi\,,
\\
\frac{ \dd \momentumh_\phi}{ \dd N}&=\alpha\fieldh+\beta\momentumh_\phi\, .
\eea
Hereafter, since $\bm{x}$ is fixed, for conciseness it is not always stated explicitly in the arguments of external functions such as  $\alpha$, $\beta$, $\fieldh_{(i)}$ or $\momentumh_{(i)}$.
Using Green's function method, the general solution of \cref{eq:Langevin deltaphi} is
\bea
\label{eq:general solution field difference}
& \delta\cgfield\left(\bm{x},\bm{y},N\right) = 
\\ &\quad \int^N_{\Ni}\mathrm{d}N'\Biggl\{\left[\fieldh_{(2)}\left(N\right)\Momentumh_{(1)}\left(N'\right)-\fieldh_{(1)}\left(N\right)\Momentumh_{(2)}\left(N'\right)\right]\left[\xi_\field\left(\bm{y},N'\right)-\xi_\field\left(\bm{x},N'\right)\right]\,\\
&\quad\quad +\left[\fieldh_{(1)}\left(N\right)\Fieldh_{(2)}\left(N'\right)-\fieldh_{(2)}\left(N\right)\Fieldh_{(1)}\left(N'\right)\right]\left[\xi_{\momentum}\left(\bm{y},N'\right)-\xi_{\momentum}\left(\bm{x},N'\right)\right]\Biggr\}\,,
\eea
where the functions $\Fieldh_{(i)},\Momentumh_{(i)}$ are defined as
\bea
\label{eq:deltaPhi:deltaPi:def}
\Fieldh_{(i)} = \frac{\fieldh_{(i)}}{\fieldh_{(2)}\momentumh_{(1)}-\fieldh_{(1)}\momentumh_{(2)}} \, ,\\
\Momentumh_{(i)} = \frac{\momentumh_{(i)}}{\fieldh_{(2)}\momentumh_{(1)}-\fieldh_{(1)}\momentumh_{(2)}}\,.
\eea
and $\Ni$ is some initial time that should be taken in the asymptotic past. The above expressions are valid if $\delta\phi^{\mathrm{IR}}$ and $\delta\pi^{\mathrm{IR}}_\phi$ vanish initially, hence we dropped the homogeneous solutions in the general solution, since $\bm{x}$ and $\bm{y}$ belong to the same patch in the past (see the discussion around \cref{fig:memory effect}). In \cref{app:sol:local:field:difference}, we check explicitly that \cref{eq:general solution field difference} is indeed a solution of \cref{eq:Langevin deltaphi}.

We can now insert this result into \cref{eq:expression gradient delta_phi} to express the Laplacian term in terms of the noises $\xi_\field$ and $\xi_{\momentum}$. The result is
\bea
&\Delta\cgfield(\bm{x},N)=\\
&\quad \int^N_{\Ni}\dd N'\left\{\left[\fieldh_{(2)}\left(N\right)\Momentumh_{(1)}\left(N'\right)-\fieldh_{(1)}\left(N\right)\Momentumh_{(2)}\left(N'\right)\right]\int \dd ^3\bm{y}I\left(|\bm{x}-\bm{y}|\right)\xi_\field\left(\bm{y},N'\right)\,
\right. \\ & \quad\quad\left.
+\left[\fieldh_{(1)}\left(N\right)\Fieldh_{(2)}\left(N'\right)-\fieldh_{(2)}\left(N\right)\Fieldh_{(1)}\left(N'\right)\right]\int \dd ^3\bm{y}I\left(|\bm{x}-\bm{y}|\right)\xi_{\momentum}\left(\bm{y},N'\right)\right\}\,,
\eea
Notice that the terms involving $\xi_{\field}\left(\bm{x},N'\right)$ and $\xi_{\momentum}\left(\bm{x},N'\right)$ give vanishing contributions since the integral of the Laplacian distribution $I$ is trivially null, see \cref{eq:vanishing integral I} in \cref{app:I}. Using \cref{eq:I:distrib:Delta}, and integrating by parts, we obtain
\bea
\label{eq:gradient noise generic window function}
\Delta\cgfield(\bm{x},N)= \int^N_{\Ni}&\dd N'\left\{\left[\fieldh_{(2)}\left(N\right)\Momentumh_{(1)}\left(N'\right)-\fieldh_{(1)}\left(N\right)\Momentumh_{(2)}\left(N'\right)\right]\Delta\xi_\field\left(\bm{x},N'\right)\right. \\
&\left.+\left[\fieldh_{(1)}\left(N\right)\Fieldh_{(2)}\left(N'\right)-\fieldh_{(2)}\left(N\right)\Fieldh_{(1)}\left(N'\right)\right]\Delta\xi_{\momentum}\left(\bm{x},N'\right)\right\}\,,
\eea
where the action of the Laplace operator on the noises is given in Fourier space by
\bea
\label{eq:Deltaxiphi:Deltaxipi}
&\Delta\hat{\xi}_\field\left(\bm{x},N\right) = -\int\frac{ \dd ^3 \bm{k}}{(2\pi)^{3/2}}k^2\left[ \ee^{-i \bm{k}\cdot\bm{x}}\field_{k}(N) \hat{a}_{\bm{k}}+\ee^{i \bm{k}\cdot\bm{x}}\field_{k}^*(N) \hat{a}^\dagger_{\bm{k}}\right]\frac{\mathrm{d}}{\mathrm{d}N}W\left(\frac{k}{\sigma aH}\right)\, , \\
&\Delta\hat{\xi}_{\momentum}\left(\bm{x},N\right) = -\int\frac{ \dd ^3\bm{k}}{(2\pi)^{3/2}}k^2\left[ \ee^{-i \bm{k}\cdot\bm{x}}\pi_{k}(N) \hat{a}_{\bm{k}}+\ee^{i \bm{k}\cdot\bm{x}}\pi_{k}^*(N) \hat{a}^\dagger_{\bm{k}}\right]\frac{\mathrm{d}}{\mathrm{d}N}W\left(\frac{k}{\sigma aH}\right)\,.
\eea
This equation is written in terms of quantum operators to allow the derivation of the statistics of the corresponding random variables.
It is clear that \cref{eq:gradient noise generic window function} can be obtained more directly, by acting the Laplacian operator $\Delta_{\bm{y}}$ onto \cref{eq:general solution field difference}, and by noticing that, in \cref{eq:Delta:phiIR:start}, $\Delta_{\bm{y}}$ can be replaced by $\Delta_{\bm{x}}$. We nonetheless provided a detailed derivation of this result for the sake of mathematical rigour, and also to remain close to the intuition built from the discrete setup discussed around \cref{fig:memory effect}.

\subsection{Coloured noises}

We now specify the above formulas to the case where the window function employed to coarse-grain the fields in stochastic inflation is a top-hat function in Fourier space, \ie $W=1$ if $k<\sigma a H$ and $0$ otherwise. As explained in \cref{subsec:stochastic_inflation}, such a window function ensures that the noises are white in standard (\ie without gradient corrections) stochastic inflation, and it allows \cref{eq:Deltaxiphi:Deltaxipi} to reduce to
\bea
&\Delta\xi_\field\left(\bm{x},N\right) = -\left(\sigma aH\right)^2\xi_\field\left(\bm{x},N\right)\, , \\
&\Delta\xi_{\momentum}\left(\bm{x},N\right) = -\left(\sigma aH\right)^2\xi_{\momentum}\left(\bm{x},N\right)\,.
\eea
Inserting this result into \cref{eq:gradient noise generic window function}, the gradient interaction can be decomposed into four terms,
\bea
\label{eq:gradient noise Heaviside window function}
\frac{\Delta\cgfield(\bm{x},N)}{\left(aH\right)^2}= \sum_{i=1}^4 \xi_\Delta^{(i)}\left(\bm{x},N\right)\, ,
\eea
where
\bea
\label{eq:xiDelta:def}
\xi_\Delta^{(1)}\left(\bm{x},N\right) &= -\sigma^2\int^N_{\Ni} \dd N' e^{2\left(N'-N\right)}\fieldh_{(2)}\left(N\right)\Momentumh_{(1)}\left(N'\right)\xi_\field\left(\bm{x},N'\right) \, ,\\
\xi_\Delta^{(2)}\left(\bm{x},N\right) &= \sigma^2\int^N_{\Ni} \dd N' e^{2\left(N'-N\right)}\fieldh_{(1)}\left(N\right)\Momentumh_{(2)}\left(N'\right)\xi_\field\left(\bm{x},N'\right) \, ,\\
\xi_\Delta^{(3)}\left(\bm{x},N\right) &= -\sigma^2\int^N_{\Ni} \dd N' e^{2\left(N'-N\right)}\fieldh_{(1)}\left(N\right)\Fieldh_{(2)}\left(N'\right)\xi_{\momentum}\left(\bm{x},N'\right)\, , \\
\xi_\Delta^{(4)}\left(\bm{x},N\right) &= \sigma^2\int^N_{\Ni} \dd N' e^{2\left(N'-N\right)}\fieldh_{(2)}\left(N\right)\Fieldh_{(1)}\left(N'\right)\xi_{\momentum}\left(\bm{x},N'\right)\,,
\eea
These can be seen as four new noise terms in the Langevin equation~\eqref{eq:Langevin general}, which now contains six noises: $\xi_\f$, $\xi_{\pi_\phi}$, and the $\xi_\Delta^{(i)}$'s. The main difference with standard stochastic inflation is that these noises are coloured, since the $\xi_\Delta^{(i)}$'s are combinations of $\xi_\f$ and $\xi_{\pi_\phi}$ at previous times. For instance, making use of \cref{eq:covariance window function}, one has
\bea 
\label{eq:coloured:correlator:example}
\langle \xi_\f(N) \xi_\Delta^{(1)}(N') \rangle =
-\sigma^2 \fieldh_{(2)}\left(N'\right)\Momentumh_{(1)}\left(N\right) \mathcal{P}_{\f\f}\left[k_\sigma(N), N\right]
e^{-2\left(N'-N\right)}\theta(N'-N), 
\eea
where $\theta$ is the Heaviside function. This correlator vanishes when $N>N'$, and decays exponentially with $N'-N$ otherwise, but it does not involve a Dirac distribution of the time difference $N'-N$. As a consequence, the Langevin equation~\eqref{eq:Langevin general} does not describe a Markovian process anymore. Although methods are available to solve stochastic differential equations with coloured noises~\cite{Van_Kampen1989-gg, gardiner2004handbook}, the analysis of such systems is substantially more complicated than Markovian setups, so the price to pay for including gradient corrections in the stochastic formalism of inflation seems very high indeed. 

In passing, the fact that the correlator~\eqref{eq:coloured:correlator:example} decays exponentially with $N'-N$ is consistent with the intuition developed in \cref{sec:discretised gradient} that the memory effect should be somewhat local in time, \ie the realisation of the gradient-induced noises at time $N$ should only depend on the recent past history of the local patch. This also justifies the linear treatment of \cref{sec:linear_evolution_of_field_difference}, since the non-vanishing correlations arise from modes that have spent at most a few e-folds outside the Hubble radius, hence they can still be described within linear perturbation theory.    
\subsection{Higher-dimensional Langevin system}

Coloured noises raise a number of technical difficulties that may discourage us from proceeding any further. However, fundamentally, in the above setup there are only two noise fields, namely $\xi_\f$ and $\xi_{\pi_\phi}$, from which the $\xi_\Delta^{(i)}$'s are derived. As we now show, this allows one to recast the coloured stochastic differential equation as a system of Langevin equations with white noises, albeit of higher dimension.

Indeed, upon time differentiating \cref{eq:xiDelta:def}, one finds
\bea
\label{eq:Langevin gradient term}
\frac{\mathrm{d}}{\mathrm{d}N}\xi_\Delta^{(1)} &=\left(\frac{{\fieldh_{(2)}}'}{\fieldh_{(2)}}-2\right) \xi_\Delta^{(1)} -\sigma^2 \fieldh_{(2)}\left(N\right)\Momentumh_{(1)}\left(N\right)\xi_\field \, ,\\
\frac{\mathrm{d}}{\mathrm{d}N}\xi_\Delta^{(2)} &= \left(\frac{{\fieldh_{(1)}}'}{\fieldh_{(1)}}-2\right)\xi_\Delta^{(2)} +\sigma^2 \fieldh_{(1)}\left(N\right)\Momentumh_{(2)}\left(N\right)\xi_\field\, , \\
\frac{\mathrm{d}}{\mathrm{d}N}\xi_\Delta^{(3)} &= \left(\frac{{\fieldh_{(1)}}'}{\fieldh_{(1)}}-2\right)\xi_\Delta^{(3)} -\sigma^2 \fieldh_{(1)}\left(N\right)\Fieldh_{(2)}\left(N\right)\xi_{\momentum}\, , \\
\frac{\mathrm{d}}{\mathrm{d}N}\xi_\Delta^{(4)} &= \left(\frac{{\fieldh_{(2)}}'}{\fieldh_{(2)}}-2\right)\xi_\Delta^{(4)} +\sigma^2 \fieldh_{(2)}\left(N\right)\Fieldh_{(1)}\left(N\right)\xi_{\momentum} \,,
\eea
where a prime denotes a derivative with respect to $N$. Together with \cref{eq:Langevin general}, \ie 
\bea 
\label{eq:Langevin:extended}
\frac{\dd}{\dd N} {\f}^{\mathrm{IR}}= &{\pi}_\phi^{\mathrm{IR}}+{\xi}_\f\, ,\\
\frac{\dd}{\dd N}{\pi}_\phi^{\mathrm{IR}}= & -\left[3-\frac{\left({\pi}_\phi^{\mathrm{IR}}\right)^2}{2\Mp^2}\right]{\pi}_\phi^{\mathrm{IR}}- \frac{V_{,\f}}{H^2}\left({\f}^{\mathrm{IR}},{\pi}_\phi^{\mathrm{IR}}\right)+{\xi}_{\pi_\phi}+\sum_{i=1}^4\xi_\Delta^{(i)}
\, ,
\eea 
this may be seen as a system of six coupled Langevin equations, for the six stochastic fields $\{{\f}^{\mathrm{IR}},\,{\pi}_\phi^{\mathrm{IR}},\,\xi_\Delta^{(1)},\,\xi_\Delta^{(2)},\,\xi_\Delta^{(3)},\,\xi_\Delta^{(4)}\}$, sourced by the two white noises ${\xi}_\f$ and ${\xi}_{\pi_\phi}$, subject to the correlator~\eqref{eq:covariance window function}.

We have thus shown that gradient corrections can be incorporated in stochastic inflation by increasing the order of the differential system, thereby keeping track of the gradient-induced noises $\xi_\Delta^{(i)}$ while preserving the ability to cast the dynamics of the system in terms of Langevin equations subject to white noises (or equivalently, in terms of Fokker-Planck equations). Different realisations of these Langevin equations follow the worldlines of different Hubble patches that evolve independently, hence inflating space-times can still be described in terms of separate universes. This is the main result of this work.

\section{Examples}
\label{sec:example}

In the rest of this paper, we illustrate the formalism developed above with a few cases of interest. The goal is to make explicit its application to concrete setups, and to verify that gradient effects are properly accounted for. 

In order to check that gradient effects are indeed captured, we will compare the results of our extended stochastic formalism to linear perturbation theory, \ie to solutions of the Mukhanov-Sasaki equation~\eqref{eq:Mukhanov-Sasaki}, that include all orders in the gradient expansion. Stochastic inflation also accounts for non-linear evolution above the Hubble radius, hence for numerical applications we will work in regimes where these non-linear effects are subdominant and a fair comparison with perturbation theory can be performed. 

In numerical simulations of the Langevin equation, a quantity that can be readily extracted is the first-passage time $\mathcal{N}$ through the end-of-inflation condition $\vert\pi_\f^{\mathrm{IR}}\vert=\sqrt{2}\Mp$, starting from an initial configuration $\{\phi_\uin^{\mathrm{IR}},\pi_{\f,\uin}^{\mathrm{IR}}\}$~\cite{Fujita:2013cna,Vennin:2015hra}. Fluctuations in that first-passage time coincide with the curvature perturbation $\mathcal{R}$ at super-Hubble scales~\cite{Starobinsky:1982ee, Starobinsky:1986fxa, Sasaki1996, Sasaki:1998ug, Lyth:2004gb}, $\mathcal{R}=\delta \mathcal{N}$. Therefore, the second centred moment $\langle \delta {\cal{N}}^2\rangle$ receives an integrated contribution from all scales comprised between $k_\uin$ and $k_\uend$,
\bea
\left\langle \delta{\cal{N}}^2\right\rangle\left(\phi_\uin^{\mathrm{IR}},\pi_{\f,\uin}^{\mathrm{IR}}\right) =\int_{k_\uin}^{k_\uend}  {\cal{P}}_{\cal{R}}\left(k,N_\uend\right)\dd\ln k\, ,
\eea
where $N_\uend$ denotes the time at the end of inflation, $k_\uend = \sigma a_\uend H_\uend$ is the $\sigma$-Hubble comoving scale at the end of inflation, and $k_\uin = \sigma a_\uin H_\uin$ is the $\sigma$-Hubble comoving scale at initial time. In practice, $\ln(k_\uend/k_\uin) = N_\uend - N_\uin + \ln(H_\uend/H_\uin) = \langle\mathcal{N}\rangle + \delta\mathcal{N}+\ln(H_\uend/H_\uin)$, so at leading order in perturbation theory the above reduces to
\bea
\label{eq:variance as integrated power spectrum}
\left\langle \delta {\cal{N}}^2\right\rangle_{\mathrm{pert}}\left(\phi_\uin^{\mathrm{IR}},\pi_{\f,\uin}^{\mathrm{IR}}\right) = \int^{\langle {\cal{N}} \rangle\left(\phi_\uin^{\mathrm{IR}},\pi_{\f,\uin}^{\mathrm{IR}}\right)+\ln(H_\uend/H_\uin)}_0 \,{\cal{P}}_{\cal{R}}\left(k_\uend e^{-N},N_\uend\right)\dd N\, .
\eea
Here, $\mathcal{P}_{\mathcal{R}}$ needs to be evaluated as explained in \cref{subsec:linear_perturbation}.
This formula will be used below to make contact between the stochastic formalism, by which $\left\langle \delta {\cal{N}}^2\right\rangle$ will be computed over a large sample of realisations of the Langevin equations~\eqref{eq:Langevin gradient term} and \eqref{eq:Langevin:extended}, and linear perturbation theory, which will provide \cref{eq:variance as integrated power spectrum}.\footnote{One could differentiate \cref{eq:variance as integrated power spectrum} in order to reach an expression for the power spectrum alone, but differentiated quantities are subject to larger statistical noise so we will not follow this route.}

\subsection{Slow-roll inflation in a linear  potential}
\label{subsec:Linear_potential}

Let us first consider slow-roll inflation in a linear potential
\bea
\label{eq:pot:lin}
V(\phi) = V_0\left(1+\frac{A_1}{\Mp^2} \phi\right) .
\eea 
In the field regime $A_1\phi\ll \Mp^2$, where the potential function is dominated by its constant term,  $\epsilon_1 = (A_1/\Mp)^2/2$ is a constant, which we assume to be small.

\subsubsection{Background}
\label{sec:lin:background}
At the background level, see \cref{subsec:background}, at leading order in $\epsilon_1$ the equation of motion~\eqref{eq:KG splitted} reduces to
\bea 
\label{eq:KG homogeneous linear potential}
\frac{\dd\bar{\f}}{\dd N}&=\bar{\pi}_\phi\, ,\\
\frac{\dd\bar{\pi}_\phi}{\dd N}&=-3\bar{\pi}_\phi-3A_1\, ,
\eea
where we have replaced $V_{,\f}/H^2 = 3 A_1$ in the limit mentioned above. This can be solved as
\bea
\label{eq:background:sol:linear:pot}
\bar{\phi}(N) =& \bar{\phi}_{\uin}+A_1\left(N_\uin-N\right)+\frac{\bar{\pi}_{\phi,\uin}+A_1}{3}\left[1-e^{3(N_\uin-N)}\right]\, ,\\
\bar{\pi}_\phi(N) = &\left(\bar{\pi}_{\phi,\uin}+A_1\right)e^{3(N_\uin-N)}-A_1\, .
\eea
At late time $\bar{\pi}_\phi=-A_1$ is constant and $\bar{\phi}=\mathrm{constant}-A_1 N$, which corresponds to the slow-roll attractor where $\epsilon_1={A_1^2}/({2\Mp^2})$ is indeed constant, see \cref{eq:epsilon1:pi}.

\subsubsection{Linear perturbation theory}

Along the slow-roll attractor where $\epsilon_1$ is a constant and all higher slow-roll parameters vanish, $Z''/Z=(aH)^2(2-\epsilon_1)\simeq 2/\eta^2$ at leading order in $\epsilon_1$, hence the Bunch-Davies solution to \cref{eq:Mukhanov-Sasaki} reads $u_k(\eta)=[1-i/(k\eta)]e^{-ik\eta}/\sqrt{2k}$. When evaluated at the coarse-graining scale $k=k_\sigma(N)=\sigma a(N) H$, the power spectra~\eqref{eq:powerspectra:field} are thus given by
\bea 
\label{eq:noise_amplitudes_Linear}
{\cal{P}}_{\f \f}\left[k_\sigma(N),N\right]&=\frac{H^2}{4\pi^2}(1+\sigma^2)\, ,\\
{\cal{P}}_{\f\pi}\left[k_\sigma(N),N\right]&=-\frac{H^2}{4\pi^2}\sigma^2\, ,\\
{\cal{P}}_{\pi\pi}\left[k_\sigma(N),N\right]&=\frac{H^2}{4\pi^2}\sigma^4\, .
\eea 
These are constant, and they constitute the covariance matrix of the two noises $\xi_\phi$ and $\xi_{\pi_\f}$ in the Langevin equations~\eqref{eq:Langevin gradient term} and \eqref{eq:Langevin:extended}.

The power spectrum of the curvature perturbation can also be obtained from the above solution to the Mukhanov-Sasaki equation, see \cref{eq:powerspectrum:R}, and one finds
\bea 
\label{eq:PR:SR}
\mathcal{P}_{\mathcal{R}}(k,\eta) = \frac{H^2}{4\pi^2 A_1^2}\left[1+\left(k\eta\right)^2\right]\, .
\eea 
Inserted into \cref{eq:variance as integrated power spectrum}, it leads to
\bea 
\label{eq:variance as integrated power spectrum SR}
\left\langle \delta \mathcal{N}^2 \right\rangle_{\mathrm{pert}} = & \frac{H^2}{4\pi^2 A_1^2} \left[\left\langle\mathcal{N}\right\rangle+\frac{\sigma^2}{2}\left(1-e^{-2\langle\mathcal{N}\rangle}\right)\right]\, .
\eea 

\subsubsection{Stochastic inflation}
\label{sec:line:pot:stochastic:inflation}
The only quantities that remain to be evaluated in \cref{eq:Langevin gradient term,eq:Langevin:extended} are the functions $\delta\phi_{(i)}^{\mathrm{IR}}$, $\delta\Phi_{(i)}^{\mathrm{IR}}$ and $\delta\Pi_{(i)}^{\mathrm{IR}}$, which require to solve the sourceless linearised system~\eqref{eq:linearised:sourceless:system}. At leading order in slow roll, \cref{eq:alpha:beta:def} reduces to $\alpha=0$ and $\beta=-3$, hence \cref{eq:linearised:sourceless:system} reads
\bea
\label{eq:linearised:sourceless:system:SR}
\frac{ \dd \fieldh}{ \dd N}&=\momentumh_\phi\,,
\\
\frac{ \dd \momentumh_\phi}{ \dd N}&=-3\momentumh_\phi\, .
\eea
Two independent solutions to this system are given by
\bea
\label{eq:hom:field:fluctuations:sol:SR}
\delta\phi^{\mathrm{IR}}_{(1)}(N)&=1\,,
\quad
\delta\phi^{\mathrm{IR}}_{(2)}(N)=-\frac{1}{3}e^{-3N}\,,
\\
\delta\pi^{\mathrm{IR}}_{(1)}(N)&=0\,,
\quad
\delta\pi^{\mathrm{IR}}_{(2)}(N)=e^{-3N}\,,
\eea
hence \cref{eq:deltaPhi:deltaPi:def} gives rise to
\bea 
\delta\Phi_{(1)}^{\mathrm{IR}} = -e^{3N}\, , \quad
\delta\Phi_{(2)}^{\mathrm{IR}} = \frac{1}{3}\, ,\quad
\delta\Pi_{(1)}^{\mathrm{IR}} = 0\, ,\quad
\delta\Pi_{(2)}^{\mathrm{IR}} = -1\, .
\eea 
Inserted into \cref{eq:Langevin gradient term}, these expressions lead to
\bea
\label{eq:Langevin gradient term:linear:potential}
\frac{\mathrm{d}}{\mathrm{d}N}\xi_\Delta^{(1)} &=-5 \xi_\Delta^{(1)} \, ,\\
\frac{\mathrm{d}}{\mathrm{d}N}\xi_\Delta^{(2)} &= -2\xi_\Delta^{(2)} -\sigma^2 \xi_\field\, , \\
\frac{\mathrm{d}}{\mathrm{d}N}\xi_\Delta^{(3)} &= -2\xi_\Delta^{(3)} -\frac{\sigma^2}{3} \xi_{\momentum}\, , \\
\frac{\mathrm{d}}{\mathrm{d}N}\xi_\Delta^{(4)} &= -5\xi_\Delta^{(4)} +\frac{\sigma^2}{3} \xi_{\momentum} \, .
\eea
Since the gradient noises must vanish initially, the first equation leads to $\xi_\Delta^{(1)}=0$. The third and fourth equations imply that $\xi_\Delta^{(3)}$ and $\xi_\Delta^{(4)}$ are sourced by $\xi_{\momentum}$, which is of order $\sigma^2$ according to \cref{eq:noise_amplitudes_Linear}, hence they can be discarded as well. The only gradient noise that remains is $\xi_\Delta^{(2)}$, and the system of Langevin equations we need to solve is given by
\bea
\label{eq:Langevin equations linear potential with gradient}
&\frac{ \dd \cgfield}{ \dd N}=\cgmomentum+\xi_\phi\,,
\\
&\frac{ \dd \cgmomentum}{ \dd N}=-3\cgmomentum-3A_1+\xi_{\momentum}+\xi_\Delta^{(2)}\,, \\
&\frac{ \dd \xi_\Delta^{(2)}}{ \dd N}= -2\xi_\Delta^{(2)}-\sigma^2\xi_\phi\,.
\eea
%

\subsubsection[Stochastic inflation vs perturbation theory: field perturbations]{Stochastic inflation versus perturbation theory: field perturbations}
\label{sec:lin:stochastic:versus:CPT:field:pert}

The system of Langevin equations~\eqref{eq:Langevin equations linear potential with gradient} being linear, it can be solved analytically, and one finds
\bea
\label{eq:sol:lin}
\cgfield\left(N\right) =& \cgfield_\uin+\int_{N_\uin}^N \dd N' \left[\cgmomentum(N')+\xi_\f(N')\right] ,\\
\cgmomentum\left(N\right) =& \pi_{\f,\uin}^{\mathrm{IR}}e^{-3(N-N_\uin)}+A_1\left[e^{3(N_\uin-N)}-1\right]+\int_{N_\uin}^N \dd N' \left[\xi_{\momentum}(N')+\xi_\Delta^{(2)}(N')\right] e^{3(N'-N)}\, ,\\
\xi_\Delta^{(2)}(N)=&-\sigma^2\int_{N_\uin}^N \dd N' \xi_\f(N')e^{2(N'-N)}\, .
\eea 
From these expressions, it is clear that the fields $\cgfield$ and $\cgmomentum$ have Gaussian statistics, whose means $\langle \cgfield\rangle$ and $\langle \cgmomentum \rangle$ coincide with the background solution~\eqref{eq:background:sol:linear:pot}, and whose second centred moments $\langle (\delta\field)^2\rangle$, $\langle(\delta\momentum)^2\rangle$ and $\langle \delta\field \delta\momentum\rangle$, where $\delta\field=\cgfield-\langle \cgfield\rangle$ and $\delta\momentum=\cgmomentum-\langle \cgmomentum\rangle$ (not to be confused with $\delta\cgfield$ and $\delta\cgmomentum$ introduced in \cref{eq:delta:cgfield:def}), are computed below.
The goal is to compare these moments with the result expected from linear perturbation theory, in order to check that gradient corrections are properly included. 

As noted below \cref{eq:background:sol:linear:pot}, along the slow-roll attractor the terms involving $e^{-3(N-N_\uin)}$ can be discarded, hence one has
\bea 
\label{eq:sol:lin:SR}
\cgfield\left(N\right) =&\cgfield_\uin - A_1(N-N_\uin)
\\ & 
+\int_{N_\uin}^N\dd N' \left\lbrace\xi_\f(N') +\int_{N_\uin}^{N'}\dd N''\left[\xi_\Delta^{(2)}(N'')+\xi_{\momentum}(N'')\right]e^{3\left(N''-N'\right)}\right\rbrace\, , \\
\cgmomentum\left(N\right) =&-A_1+\int_{N_\uin}^N \dd N' \left[\xi_{\momentum}(N')+\xi_\Delta^{(2)}(N')\right] e^{3(N'-N)}\, .\\
\eea 
Let us first consider $\langle(\delta\field)^2\rangle$. Using \cref{eq:sol:lin:SR}, one has
\bea 
\langle (\delta\field)^2\rangle =&
\int_{N_\uin}^N \dd N_1\int_{N_\uin}^N \dd N_1' \left\langle \xi_\f(N_1)\xi_\f(N_1')\right\rangle 
\\ & 
+ 2\int_{N_\uin}^N \dd N_1\int_{N_\uin}^N \dd N_1'\int_{N_\uin}^{N_1'}\dd N_2'e^{3(N_2'-N_1')} \left\langle \xi_\f(N_1)\left[\xi_\Delta^{(2)}(N_2')+\xi_{\momentum}(N_2') \right]\right\rangle
\\ & 
+\int_{N_\uin}^N \dd N_1\int_{N_\uin}^{N_1} \dd N_2 \int_{N_\uin}^N \dd N_1'\int_{N_\uin}^{N_1'} \dd N_2' e^{3(N_2+N_2'-N_1-N_1')}
\\ & \hspace{1cm}
\times\left\langle \left[\xi_\Delta^{(2)}(N_2)+\xi_{\momentum}(N_2)\right] \left[\xi_\Delta^{(2)}(N_2')+\xi_{\momentum}(N_2')\right] \right\rangle\, .
\eea 
The last integral provides contributions of order $\sigma^4$, while the two first integrals require to compute
\bea 
\label{eq:linear:noise:correlators}
\left\langle \xi_{\f}(N_1)\xi_{\f}(N_2) \right\rangle =& \mathcal{P}_{\f\f}\left[k_\sigma(N_1),N_1\right]\delta(N_1-N_2)\, ,\\
\left\langle \xi_{\f}(N_1)\xi_{\momentum}(N_2) \right\rangle =& \mathcal{P}_{\f\pi}\left[k_\sigma(N_1),N_1\right]\delta(N_1-N_2)\, ,\\
\left\langle \xi_{\f}(N_1)\xi_\Delta^{(2)}(N_2) \right\rangle =&
-\sigma^2 e^{2(N_1-N_2)}\mathcal{P}_{\phi\phi}\left[k_\sigma(N_1),N_1\right]\theta(N_2-N_1)\, ,
\eea
where we have made used of the expression for $\xi_\Delta^{(2)}$ given in \cref{eq:sol:lin} as well as \cref{eq:covariance window function}. At order $\sigma^2$, one thus finds
\bea 
\label{eq:deltapi:squared:phi^2:interm}
\langle (\delta\field)^2\rangle =&
\int_{N_\uin}^N \dd N_1 \mathcal{P}_{\phi\phi}\left[k_{\sigma}(N_1),N_1\right] 
\\ & 
-2\sigma^2 \int_{N_\uin}^N \dd N_1\int_{N_\uin}^N \dd N_1'\int_{N_\uin}^{N_1'}\dd N_2'e^{3(N_2'-N_1')+2(N_1-N_2')}\mathcal{P}_{\phi\phi}\left[k_{\sigma}(N_1),N_1\right]\theta(N_2'-N_1)
\\ &
+2 \int_{N_\uin}^N \dd N_1\int_{N_\uin}^N \dd N_1'e^{3(N_1-N_1')}\mathcal{P}_{\phi\pi}\left[k_{\sigma}(N_1),N_1\right]\theta(N_1'-N_1)
+\mathcal{O}(\sigma^4)\, .
\eea 
In the linear potential considered here, the power spectra $\mathcal{P}_{fg}[k_\sigma(N),N]\equiv \mathcal{P}_{fg}$ do not depend on time, hence the integrals over $N_1$ and $N_2$ in \cref{eq:deltapi:squared:phi^2:interm} can be performed analytically, and one finds
\bea 
\langle (\delta\field)^2\rangle =&
\mathcal{P}_{\phi\phi}\left(N-N_\uin\right)
+\frac{\sigma^2}{18}\mathcal{P}_{\phi\phi}\left[5+4e^{3(N_\uin-N)}-9e^{2(N_\uin-N)}+6(N_\uin-N)\right]
\\ & 
-\frac{2}{9}\mathcal{P}_{\phi\pi}\left[1-e^{3(N_\uin-N)}+3(N_\uin-N)\right]+\mathcal{O}(\sigma^4)\, .
\eea 
By replacing the power spectra by their expressions given in \cref{eq:noise_amplitudes_Linear}, one finally obtains
\bea 
\label{eq:deltaphi^2:lin}
\langle (\delta\field)^2\rangle =&
\left(\frac{H}{2\pi}\right)^2\left\lbrace N-N_\uin+\frac{\sigma^2}{2}\left[1-e^{2(N_\uin-N)}\right]\right\rbrace
+\mathcal{O}(\sigma^4)\, .
\eea

In cosmological perturbation theory, the power spectrum of the inflaton's fluctuations reads $\mathcal{P_{\phi\phi}}(k,\eta)=[H/(2\pi)]^2[1+(k\eta)^2]$, which leads to
\bea 
\label{eq:field:moment:lin:phiphi}
\int_{k_\sigma(N_\uin)}^{k_\sigma(N)} \dd\ln k\, \mathcal{P}_{\phi\phi}\left(k,N\right) =\left(\frac{H}{2\pi}\right)^2\left\lbrace N-N_\uin+\frac{\sigma^2}{2}\left[1-e^{2(N_\uin-N)}\right]\right\rbrace\, .
\eea 
This coincides with the expression~\eqref{eq:deltaphi^2:lin} found above for $\langle (\delta\field)^2\rangle$, at order $\sigma^2$ (\ie at order $k^2$ in the gradient expansion). Similar calculations can be performed for $\langle \delta\field \delta\momentum\rangle$ and $\langle (\delta\momentum)^2\rangle$, the details of which can be found in \cref{app:moments:linear}. There we show that
\bea 
\label{eq:field:moment:lin:phiphi:phipi}
\langle \delta\field \delta\momentum\rangle = &- \left(\frac{H}{2\pi}\right)^2\frac{\sigma^2}{2}\left[1-e^{2(N_\uin-N)}\right]+\mathcal{O}(\sigma^4)
=
\int_{k_\sigma(N_\uin)}^{k_\sigma(N)} \dd\ln k\, \mathcal{P}_{\phi\pi}\left(k,N\right)+\mathcal{O}(\sigma^4) \, ,\\
\langle(\delta\momentum)^2\rangle =&\left(\frac{H}{2\pi}\right)^2\frac{\sigma^4}{4}\left[1-e^{4(N_\uin-N)}\right]+\mathcal{O}(\sigma^6)
=\int_{k_\sigma(N_\uin)}^{k_\sigma(N)} \dd\ln k\, \mathcal{P}_{\pi\pi}\left(k,N\right)+\mathcal{O}(\sigma^6)\, .
\eea 
This concludes the proof that leading gradient corrections are properly accounted for by our improved stochastic formalism, in the linear potential studied in this section.

\subsubsection[Stochastic inflation vs perturbation theory: curvature perturbations]{Stochastic inflation versus perturbation theory: curvature perturbations}
%
\begin{table}
    \renewcommand{\arraystretch}{1.3}
    \centering
    \begin{tabular}{c @{\quad} c @{\quad} c @{\quad} c @{\quad} c @{\quad} c} 
        \hline 
        Parameters  & $H/\Mp$ & $A_1/\Mp$ & $\phi_\uin/\Mp$ & $\pi_{\f,\uin}/\Mp$ & $\phi_\uend/\Mp$ \\ \hline 
        Values & $2.0\times10^{-2}$ &$0.01$ &$0.2$ &$-0.01$ &$0.0$ \\
        \hline 
    \end{tabular}
    \caption{Parameter values used for numerical simulations in the linear potential.}
    \label{table:linear}
\end{table}
We now proceed to compare stochastic inflation with perturbation theory at the level of the curvature perturbation $\mathcal{R}=\delta \mathcal{N}$. The reason is two-fold. First, this will require us to solve the Langevin equations numerically in order to compute the first-passage time, thereby illustrating the concrete use of our formalism in numerical applications. Second, as shown in \cref{sec:lin:stochastic:versus:CPT:field:pert}, the model considered here is fully linear at the level of the field perturbations, which therefore feature Gaussian statistics. The first-passage time is however non-Gaussian because of the presence of the absorbing boundary at the end-of-inflation hypersurface, and the tail of its distribution function is shaped by non-perturbative effects that only the stochastic formalism is able to capture. A comparison between the two formalisms, in the presence of gradient corrections, will also be interesting from that perspective.

\begin{figure}[t]
\centering
\includegraphics[width=0.48\textwidth,trim={0cm 0cm 0cm 0cm}, clip]{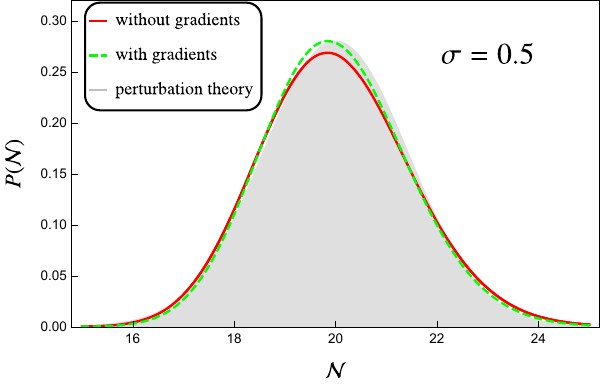}
    \hspace{0.2cm}
    \includegraphics[width=0.48\textwidth,trim={0cm 0cm 0cm 0cm}, clip]{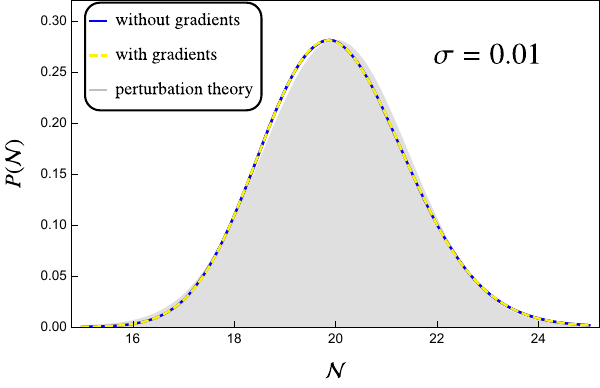}
    \includegraphics[width=0.48\textwidth,trim={0cm 0cm 0cm 0cm}, clip]{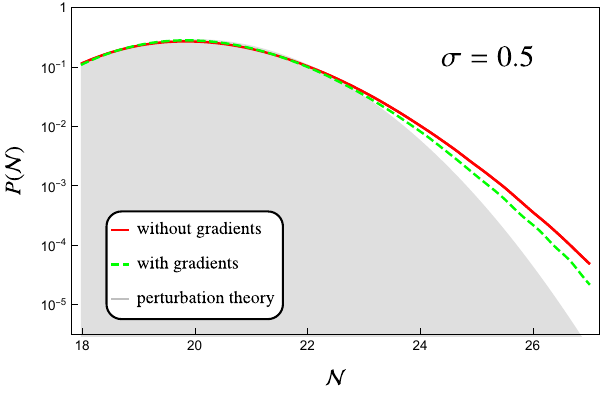}
    \hspace{0.2cm}
    \includegraphics[width=0.48\textwidth,trim={0cm 0cm 0cm 0cm}, clip]{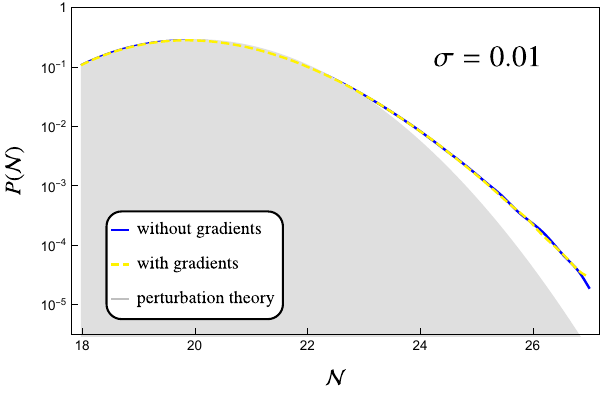}
\caption{First-passage-time distribution in the linear potential~\eqref{eq:pot:lin} for the parameters listed in \cref{table:linear}. The left panels correspond to $\sigma=0.5$ and the right panels to $\sigma=0.01$. Solid lines are obtained from $10^7$ realisations of the Langevin equations~\eqref{eq:Langevin:standard} without gradient interactions, \ie without the $\xi_\Delta^{(i)}$ noises, while dashed curves include gradient interactions and are drawn from \cref{eq:Langevin equations linear potential with gradient}. The grey-shaded curves are Gaussian distributions following the prediction~\eqref{eq:variance as integrated power spectrum} from linear perturbation theory. The bottom panels zoom in on the tails, with a logarithmic scale on the vertical axis.}
\label{fig:PDF_linear}
\end{figure}

The system~\eqref{eq:Langevin equations linear potential with gradient} is solved numerically for the parameter values listed in \cref{table:linear}. The distribution function for the first-passage time is displayed in \cref{fig:PDF_linear}, for $\sigma=0.5$ (left panels) and $\sigma=0.01$ (right panels). The solid lines are obtained from $10^7$ realisations of the standard (\ie without gradient corrections) Langevin equations~\eqref{eq:Langevin:standard}. The dashed lines correspond to the full system~\eqref{eq:Langevin equations linear potential with gradient}, and the grey-shaded curves are Gaussian distributions obtained from linear perturbation theory, \ie whose variance is given by \cref{eq:variance as integrated power spectrum SR}.

One can see that, with $\sigma=0.5$, if gradient corrections are discarded then the stochastic and perturbative results are substantially discrepant, since homogeneous matching is performed at a time when gradient effects are still important. However, when gradients corrections are included, the two fall in much better agreement near the maximum of the distribution function. With $\sigma=0.01$, gradient corrections play a more minor role since they are already negligible when homogeneous matching is performed.

\begin{figure}[t]
\includegraphics[width=0.49\textwidth,trim={0cm 0cm 0cm 0cm}, clip]{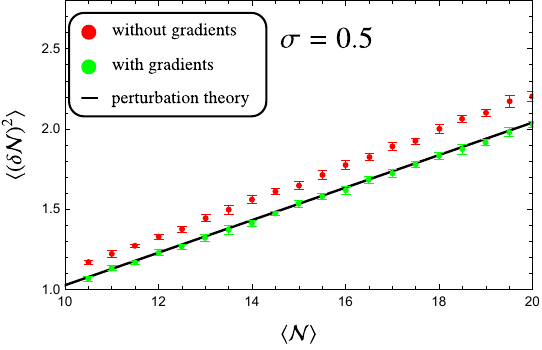}
\includegraphics[width=0.49\textwidth,trim={0cm 0cm 0cm 0cm}, clip]{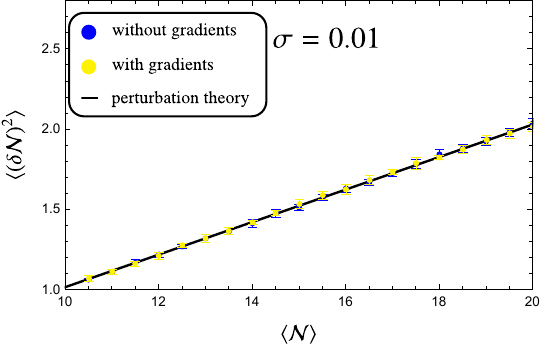}
\caption{Variance of the first-passage-time distribution, $\langle(\delta {\cal{N}})^2\rangle$, for different initial conditions along the slow-roll attractor, parametrised by its mean number of e-folds $\langle {\cal{N}}\rangle$. The colour code is the same as in \cref{fig:PDF_linear}, where perturbation theory, \ie \cref{eq:variance as integrated power spectrum}, is displayed with black solid lines. $10^4$ realisations have been used for each point, whose error bars have been evaluated using Jackknife resampling.
}
\label{fig:Integrated_power_linear}
\end{figure}

This is confirmed in \cref{fig:Integrated_power_linear}, where the variance of the first-passage time is displayed for different initial conditions chosen along the slow-roll attractor. It is rather remarkable that, if gradient corrections are included by means of our improved stochastic formalism, gradient effects are properly accounted for even when $\sigma$ is of order one, \ie even if coarse-graining is performed at scales where gradients are not subdominant. 

Finally, we note in \cref{fig:PDF_linear} that small differences persist between the improved stochastic formalism and perturbation theory, that are more pronounced in the tails. This is because, as mentioned above, the stochastic formalism allows one to capture non-Gaussianities coming from the presence of a finite absorbing boundary, and non-perturbative effects mostly affect the tail. In principle, these non-Gaussian features depend on the value used for $\sigma$, since larger values of $\sigma$ allow us to include backreaction over a wider range of scales, and we further comment on this point in \cref{sec:conclusion}. Here, we notice that gradient interactions reduce the amplitude of curvature perturbations, both at the level of its variance and along its upper tail (this is more visible with $\sigma=0.5$ for which gradient interactions are more prominent). This can be interpreted as originating from the fact that spatial gradients act like a spring force in the field equations, which pulls back the rarer regions closer to the mean behaviour. This is analogous to the pullback effect found in \Refs{Clough:2016ymm, Caravano:2024tlp, Caravano:2024moy}. It reduces the probability for large excursions, thereby softening the upper tail.

\subsection{Ultra-slow roll}
\label{subsec::pure_USR}
Next, let us consider the ultra-slow roll limit~\cite{Inoue:2001zt, Kinney:2005vj, Dimopoulos:2017ged, Pattison:2018bct} where the inflaton's potential is completely flat, $V_{,\phi}=0$. We follow the same steps as in \cref{subsec:Linear_potential} and since they unfold similarly we shall proceed at a slightly increased pace.

At the background level, the equations of motion for the inflaton and its velocity are given by 
\bea 
\label{eq:KG_homogeneous_pure_USR}
\frac{\dd\bar{\f}}{\dd N}&=\bar{\pi}_\phi\, ,\\
\frac{\dd\bar{\pi}_\phi}{\dd N}&=-3\bar{\pi}_\phi\, ,
\eea
whose solution reads
\bea
\bar{\phi} (N) = & \bar{\phi}_\uin -\frac{1}{3} \bar{\pi}_{\phi,\uin} \left[e^{-3(N-N_\uin)}-1\right]\, ,\\
\bar{\pi}_\phi (N) = & \bar{\pi}_{\phi,\uin} e^{-3(N-N_\uin)}\, .
\eea
From these expressions, the first Hubble-flow parameter is such that $\epsilon_1\propto e^{-6N}$, see \cref{eq:epsilon1:pi}, hence $\epsilon_2=-6$.

At the perturbative level, since $Z=a\bar{\pi}_\phi\propto e^{-2N}$, one has $Z''/Z=2/\eta^2$, as in the slow-roll case. Therefore, the field power spectra are still given by \cref{eq:noise_amplitudes_Linear}, and in the stochastic formalism the noise correlators are the same as in slow roll. For the curvature perturbation however, since $Z$ behaves differently, one finds a different expression, namely
\bea
{\cal{P}}_{\cal{R}}(k,\eta)=\frac{H^2}{4\pi^2\bar{\pi}_\phi^2(\eta)}\left[1+(k\eta)^2\right]\,.
\eea
In USR, the curvature perturbation is not conserved at super-Hubble scales, contrary to what was found in \cref{eq:PR:SR}.
Together with \cref{eq:variance as integrated power spectrum}, this leads to the following variance for the first-passage time, 
\bea
\label{eq:NvsdeltaN2_pure_USR}
\langle(\delta {\cal{N}})^2\rangle =\frac{H^2}{4\pi^2\bar{\pi}_\phi^2(N_\uin+\langle {\cal{N}} \rangle)}\left[\langle {\cal{N}} \rangle+\frac{\sigma^2}{2}\left(1-e^{-2\langle {\cal{N}} \rangle}\right)\right]\,.
\eea

Let us now derive the Langevin equations of the stochastic-inflation formalism. From \cref{eq:alpha:beta:def}, one has $\alpha=0$ and $\beta=-3$, which coincides with what was found in the linear potential case, see \cref{sec:line:pot:stochastic:inflation}. As a consequence, the Langevin equations are the same as \cref{eq:Langevin equations linear potential with gradient} with $A_1=0$, namely
\bea
\label{eq:Langevin equations pure USR with gradient}
&\frac{ \dd \cgfield}{ \dd N}=\cgmomentum+\xi_\phi\,,
\\
&\frac{ \dd \cgmomentum}{ \dd N}=-3\cgmomentum+\xi_{\momentum}+\xi_\Delta^{(2)}\,, \\
&\frac{ \dd \xi_\Delta^{(2)}}{ \dd N}= -2\xi_\Delta^{(2)}-\sigma^2\xi_\phi\,,
\eea
where the noise covariance is still given by \cref{eq:noise_amplitudes_Linear}.

\subsubsection[Stochastic inflation vs perturbation theory: field perturbations]{Stochastic inflation versus perturbation theory: field perturbations}
\label{sec:USR:stochastic:versus:CPT:field:pert}

The same solution as in \cref{eq:sol:lin} is found,
\bea
\label{eq:sol:USR}
\cgfield\left(N\right) =& \cgfield_\uin+\int_{N_\uin}^N \dd N' \left[\cgmomentum(N')+\xi_\f(N')\right] ,\\
\cgmomentum\left(N\right) =& \pi_{\f,\uin}^{\mathrm{IR}}e^{-3(N-N_\uin)}+\int_{N_\uin}^N \dd N' \left[\xi_{\momentum}(N')+\xi_\Delta^{(2)}(N')\right] e^{3(N'-N)}\, ,\\
\xi_\Delta^{(2)}(N)=&-\sigma^2\int_{N_\uin}^N \dd N' \xi_\f(N')e^{2(N'-N)}\, ,
\eea 
where we have set $A_1=0$. The main difference with \cref{eq:sol:lin} is that the slow-roll attractor is not attained at late times, hence the above expression cannot be further simplified. However, the terms that were dropped in \cref{eq:sol:lin:SR} belong to the background, and do not affect the moments of the field fluctuations. This is why these moments are still given by the same expressions, namely \cref{eq:field:moment:lin:phiphi,eq:field:moment:lin:phiphi:phipi}. Moreover, as mentioned above the field power spectra as obtained from linear perturbation theory are also the same as in slow roll. This confirms that leading gradient corrections are properly included.

\subsubsection[Stochastic inflation vs perturbation theory: curvature perturbations]{Stochastic inflation versus perturbation theory: curvature perturbations}

%
\begin{table}
    \renewcommand{\arraystretch}{1.3}
    \centering
    \begin{tabular}{c @{\quad} c @{\quad} c @{\quad} c @{\quad} c @{\quad} c} 
        \hline 
        Parameters  & $H/\Mp$ & $\phi_\uin/\Mp$ & $\pi_{\phi,\uin}/\Mp$ & $\phi_\uend/\Mp$ \\ \hline 
        Values & $1.0\times10^{-3}$ & $0.3325$ &$-1.0$ &$0.0$ \\
        \hline 
    \end{tabular}
    \caption{Parameter values used for numerical simulations in the USR model.}
    \label{table:pure_USR}
\end{table}

Let us then compute the statistics of the first-passage time by solving the Langevin equations~\eqref{eq:Langevin equations pure USR with gradient} numerically. In practice, we use the parameter values listed in \cref{table:pure_USR}, which lie far from the validity regime of the classical limit. Indeed, in the absence of stochastic noise, the initial velocity that is required to cross the USR well is $\pi_{\phi,\uc}=-3(\phi_\uin-\phi_\uend)$. As shown in \cite{Pattison:2021oen}, the relative importance of stochastic effects can be assessed with the parameter $y\equiv \pi_{\phi,\uin}/\pi_{\phi,\uc}$. The classical regime corresponds to $y\gg 1$, while the quantum-well limit is recovered when $y\ll 1$. Here, $y\simeq 1$, which implies that we probe an intermediate regime where neither limit applies. The reason is that, since we have already established that gradient interactions are properly included in the perturbative regime, we want to further characterise their role in non-perturbative processes. 

    \begin{figure}[t]
    \centering
    \includegraphics[width=0.48\textwidth,trim={0cm 0cm 0cm 0cm}, clip]{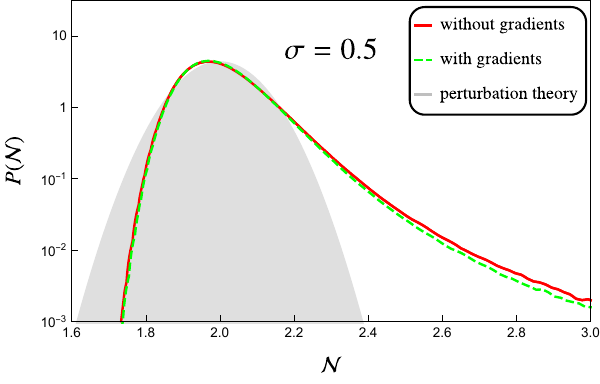}
    \hspace{0.2cm}
    \includegraphics[width=0.48\textwidth,trim={0cm 0cm 0cm 0cm}, clip]{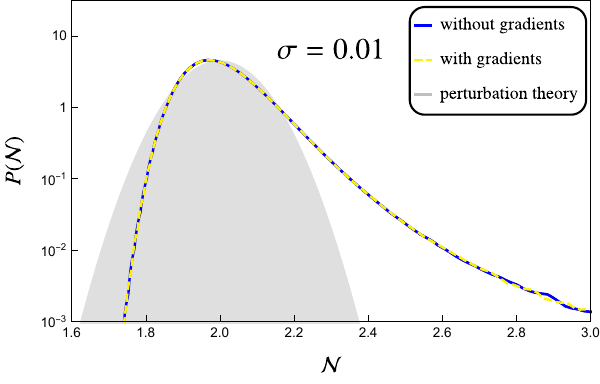}
    \caption{First-passage-time distribution in the USR model for the parameters listed in \cref{table:pure_USR}, for $\sigma=0.5$ (left panel) and $\sigma=0.01$ (right panel). Solid lines are obtained from $10^7$ realisations of the Langevin equations~\eqref{eq:Langevin:standard} without gradient interactions, while dashed curves include gradient corrections and are drawn from \cref{eq:Langevin equations pure USR with gradient}. The grey-shaded curves are Gaussian distributions following the prediction~\eqref{eq:NvsdeltaN2_pure_USR} from linear perturbation theory.}
    \label{fig:PDF_Pure_USR}
    \end{figure}
    \begin{figure}[t]
    \centering
    \includegraphics[width=0.48\textwidth,trim={0cm 0cm 0cm 0cm}, clip]{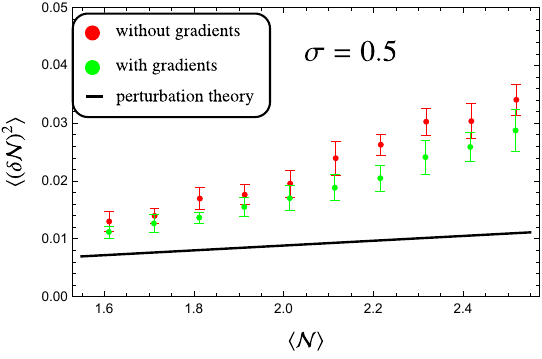}
    \hspace{0.2cm}
    \includegraphics[width=0.48\textwidth,trim={0cm 0cm 0cm 0cm}, clip]{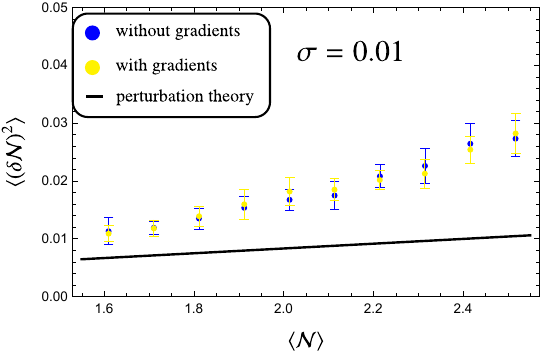}
    \caption{Variance of the first-passage-time distribution, $\langle(\delta {\cal{N}})^2\rangle$, for different initial field values, parametrised by their mean number of e-folds $\langle {\cal{N}}\rangle$. The colour code is the same as in \cref{fig:PDF_Pure_USR}, where perturbation theory, \ie \cref{eq:NvsdeltaN2_pure_USR}, is displayed with black solid lines. $10^4$ realisations have been used for each point, whose error bars have been evaluated using Jackknife resampling.}
    \label{fig:Integrated_power_Pure_USR}
    \end{figure}

In \cref{fig:PDF_Pure_USR}, the probability density function of the first-passage time is displayed, for $\sigma=0.5$ in the left panel and $\sigma=0.01$ in the right panel. The result from perturbation theory is displayed in grey for indication, although the stochastic prediction substantially deviates from it for the reason mentioned above. Non-perturbative effects are responsible for the heavy tail at large $\mathcal{N}$ (\ie large curvature perturbation), while they do not alter much the lower tail of the distribution function. This is in agreement with the results of \cite{Pattison:2021oen}.

One can also note that the inclusion of gradient interactions does not modify much the result when $\sigma=0.01$. This is not only the case for the second moment of $\mathcal{N}$ displayed in \cref{fig:Integrated_power_Pure_USR} (and for which gradient effects are indeed expected to provide a small correction only, at least at the perturbative level), but also for the far tail of the distribution. For $\sigma=0.5$ however, gradient interactions reduce the amplitude of curvature perturbations, both at the level of its variance and along its upper tail, as was also the case in slow roll, see \cref{fig:PDF_linear,fig:Integrated_power_linear}. This can again be interpreted as originating from the pullback effect mentioned above.

\subsection{Starobinsky's piecewise linear potential}
\label{subsec::Starobinsky_Linear_potential}
Finally, let us apply our formalism to a model featuring a dynamical transition between SR and USR, since gradient interactions are expected to play an important role in such scenarios, as explained in \cref{sec:Importance_gradient}. In practice, we consider Starobinsky's piecewise linear potential~\cite{Starobinsky:1992ts}, 
\bea 
V(\phi) =& 
\begin{cases}
V_0\left(1+\frac{A_1}{\Mp^2} \phi\right)\quad\text{if}\quad \phi>0 \vspace{0.1cm}  \\
V_0\left(1+\frac{A_2}{\Mp^2} \phi\right)\quad\text{if}\quad \phi\leq 0
\end{cases}\, .
\eea 
We work in the regime where the potential is dominated by its constant term, $A_i(\phi)\phi\ll\Mp^2$, where $A_i(\phi)=A_1$ if $\phi>0$ and $A_2$ otherwise.

\subsubsection{Background}
\label{sec:Staro:background}
Inflation first proceeds at $\phi>0$ in the SR regime. Then, if $A_2\ll A_1$, the inflaton's velocity is much larger than the slow-roll attractor when $\phi$ becomes negative, which triggers a phase of transient USR until the late SR attractor is reached again, see \cref{fig:Z_and_eta}. This leads to an enhancement of the curvature perturbation that may result in the formation of primordial black holes, see for instance \Refs{Ivanov:1994pa,Pi:2022zxs}. 

At the background level, the results of \cref{sec:lin:background} apply in each branch of the potential function separately, hence $\bar{\phi}$ and $\bar{\pi}_\phi$ follow \cref{eq:background:sol:linear:pot} (where, when $\phi<0$, $A_1$ needs to be replaced with $A_2$ and $\bar{\pi}_{\phi,\uin}$ with $-A_1$ such that $\bar{\pi}_\phi$ is continuous at the transition time $\eta_\uc$). 

\subsubsection{Linear perturbation theory}
\label{sec:Staro:perturbations}
At the perturbative level, at leading order in $\epsilon_1$, 
one still has $Z''/Z=2/\eta^2$ in each branch, with a sharp feature at the transition~\cite{Briaud:2025hra}. This leads to 
\bea 
u_k(\eta)=\begin{cases}
\frac{e^{-ik\eta}}{\sqrt{2k}}\left(1-\frac{i}{k\eta}\right)\quad\text{for}\quad\eta<\eta_\uc\, ,\\
\alpha_k\frac{e^{-ik\eta}}{\sqrt{2k}}\left(1-\frac{i}{k\eta}\right)+\beta_k\frac{e^{ik\eta}}{\sqrt{2k}}\left(1+\frac{i}{k\eta}\right)\quad\text{for}\quad\eta \geq \eta_\uc\, ,
\end{cases}
\eea
where the Bunch-Davies initial condition has been enforced at $\eta<\eta_\uc$. After the transition, $\alpha_k$ and $\beta_k$ are Bogoliubov coefficients that need to be set by requiring that the induced metric and its extrinsic curvature are continuous on the transition hypersurface. Since $a$, $H$ and $\epsilon_1$ are continuous at the transition, this imposes that the curvature perturbation $\mathcal{R}_k$ and its time-derivative $\mathcal{R}_k'$ are continuous at the transition~\cite{Deruelle:1995kd}. With $\mathcal{R}_k=u_k/Z$, this leads to $u_k$ and $u_k'-aH\epsilon_2u_k/2$ being continuous. Immediately after the transition, one has $\epsilon_2=-6(A_1-A_2)/A_1$, while it vanishes before the transition. Therefore the matching conditions read $u_k(\eta_\uc^+)=u_k(\eta_\uc^-)$ and $u_k'(\eta_\uc^+)=u_k'(\eta_\uc^-)+3 (A_1-A_2)/(A_1 \eta_\uc) u_k(\eta_\uc^-)$, which lead to 
\bea 
\label{eq:Bogo:Staro}
\alpha_k =& 1-\frac{3}{2}i \gamma\left[\left(\frac{k_\uc}{k}\right)^3+\frac{k_\uc}{k}\right]\\
\beta_k =& -\frac{3}{2}i\gamma\left(\frac{k_\uc}{k}\right)^3\left(1-i\frac{k}{k_\uc}\right)^2e^{2i\frac{k}{k_\uc}}
\eea 
where $k_\uc=aH(\eta_\uc)$ is the scale exiting the Hubble radius at the time of the transition, and we have introduced
\bea
\gamma\equiv \frac{A_1-A_2}{A_1}\, .
\eea 

\begin{figure}[t]
    \centering
    \includegraphics[width=0.48\textwidth,trim={0cm 0cm 0cm 0cm}, clip]{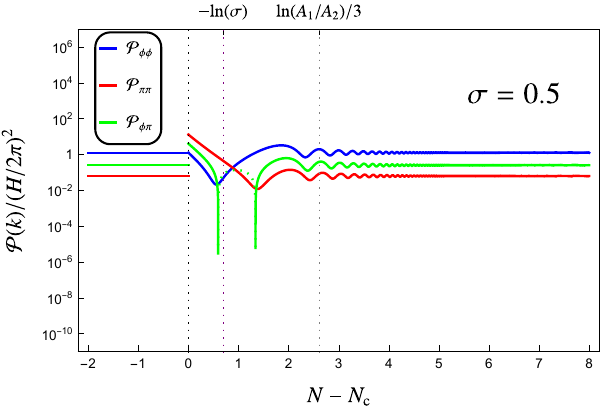}
    \hspace{0.2cm}
    \includegraphics[width=0.48\textwidth,trim={0cm 0cm 0cm 0cm}, clip]{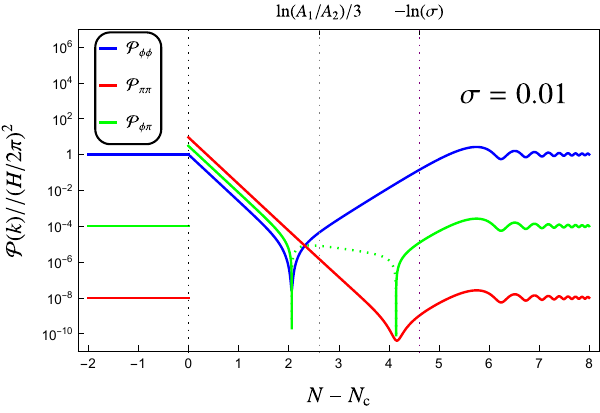}\\
    \caption{Noise correlators in the Starobinsky piecewise-linear potential, as a function of time, for $\sigma=0.5$ (left panel) and $\sigma=0.01$ (right panel). In practice, \cref{eq:noise_amplitudes_Linear} is displayed before the transition, and \cref{eq:noise_amplitudes_Starobinsky} afterwards, with $A_2/A_1=0.0004$. The cross power spectrum ${\cal{P}}_{\phi\pi}$ is shown with a dotted line when it is negative. The vertical dashed lines correspond to three times of interest: the SR-USR transition that occurs at $N=N_\uc$, the USR-SR transition at $N\simeq N_\uc+\ln(A_1/A_2)/3$, and the time $N=N_\uc-\ln(\sigma)$ after which the modes involved in the noise cross out the Hubble radius after the transition (see main text).}
    \label{fig:noiseCov:Starobinsky}
    \end{figure}

When evaluated at the coarse-graining scale $k_\sigma(N)$, the power spectra~\eqref{eq:powerspectra:field} are thus given by \cref{eq:noise_amplitudes_Linear} when $\eta<\eta_\uc$, and by
\bea 
\label{eq:noise_amplitudes_Starobinsky}
{\cal{P}}_{\f \f}\left[k_\sigma(N),N\right]&=\frac{H^2}{4\pi^2}\left|\alpha_{k_\sigma}(1-i\sigma)e^{i\sigma}-\beta_{k_\sigma}(1+i\sigma)e^{-i\sigma}\right|^2\, ,\\
{\cal{P}}_{\f\pi}\left[k_\sigma(N),N\right]&=-\frac{H^2}{4\pi^2}\sigma^2\Rea\left\lbrace\left[\alpha_{k_\sigma}(1-i\sigma)e^{i\sigma}-\beta_{k_\sigma}(1+i\sigma)e^{-i\sigma}\right]\left[\alpha_{k_\sigma}^*e^{-i\sigma}-\beta_{k_\sigma}^*e^{i\sigma}\right] \right\rbrace \, ,\\
{\cal{P}}_{\pi\pi}\left[k_\sigma(N),N\right]&=\frac{H^2}{4\pi^2}\sigma^4 \left|\alpha_{k_\sigma} e^{i\sigma}-\beta_{k_\sigma} e^{-i\sigma}\right|^2\, .
\eea 
when $\eta\geq\eta_\uc$. These expressions are displayed in \cref{fig:noiseCov:Starobinsky}, which shows how the covariance matrix of the noises evolves across the transition. 
Three times of interest can be identified: (i) the SR-USR transition that occurs at $N=N_\uc$, (ii) the end of the USR phase when the system relaxes to the late-SR attractor -- this takes place at $N\simeq N_\uc+N_{\mathrm{USR}}$ where $N_{\mathrm{USR}}=\ln(A_1/A_2)/3$, and (iii) the time $N=N_\uc+\vert\ln(\sigma)\vert$ after which the modes involved in the noise cross out the Hubble radius after the transition. Before the transition, and after the time $N_\uc+\vert\ln(\sigma)\vert$, the covariance matrix is frozen and given by \cref{eq:noise_amplitudes_Linear}. During the transient phase $N_\uc<N<N_\uc+\vert\ln(\sigma)\vert$ however, it strongly depends on time, and during USR the noise is mostly aligned with the momentum direction, $\mathcal{P}_{\phi\phi}\ll \mathcal{P}_{\pi\pi}$, in contrast to pure SR or USR where $\mathcal{P}_{\phi\phi}\gg \mathcal{P}_{\pi\pi}$. Note that, depending on the value used for $\sigma$, two cases need to be distinguished: if $\vert\ln(\sigma)\vert<N_{\mathrm{USR}}$ (left panel in \cref{fig:noiseCov:Starobinsky}), the transient phase ends before USR terminates, while, if $\vert\ln(\sigma)\vert>N_{\mathrm{USR}}$ (right panel in \cref{fig:noiseCov:Starobinsky}), the whole USR phase takes place in the transient regime.

\begin{figure}[t]
    \centering
    \includegraphics[width=0.8\textwidth,trim={0cm 0cm 0cm 0cm}, clip]{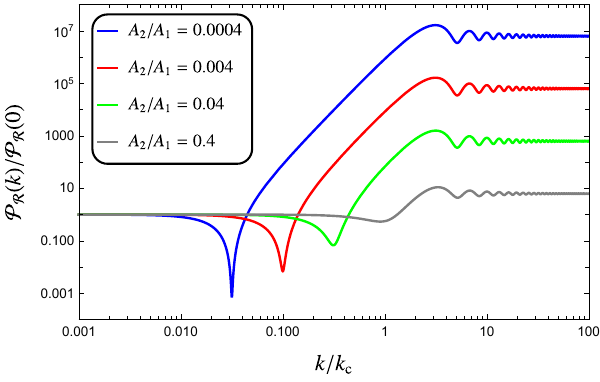}
    \caption{Power spectrum of the curvature perturbation in the Starobinsky's piecewise linear potential.}
    \label{fig:powerspectrum_Starobinsky}
    \end{figure}

For the curvature perturbation $\mathcal{R}=u/(a \bar{\pi}_\phi)$, after the transition one finds
\bea
\label{eq:PR:Staro}
\mathcal{P}_{\mathcal{R}}(k,\eta\geq\eta_\uc) = \frac{H^2}{4\pi^2 A_2^2}\frac{\left\vert \alpha_k(1+ik\eta)e^{-ik\eta}-\beta_k (1-ik\eta)e^{ik\eta} \right\vert^2 }{\left[ 1+\frac{A_1-A_2}{A_2}\left(\frac{\eta}{\eta_\uc}\right)^3\right]^2}\, .
\eea 
At late time, once the inflaton settles down on the slow-roll attractor in the second stage, the curvature perturbation freezes again, $\mathcal{P}_{\mathcal{R}}=H^2\vert \alpha_k-\beta_k\vert^2/(2\pi A_2)^2$, in contrast to the pure USR case discussed in \cref{subsec::pure_USR}. This late-time value is displayed in \cref{fig:powerspectrum_Starobinsky}, where one can check that a larger peak in the power spectrum is obtained at smaller ratio $A_2/A_1$, as announced above.
The integrated power spectrum can be computed by inserting \cref{eq:PR:Staro} into \cref{eq:variance as integrated power spectrum}, but we do not reproduce the corresponding expression here since it is not particularly illuminating.

%

\subsubsection{Stochastic inflation}
\label{sec:Staro:pot:stochastic:inflation}

We now turn to the description of the stochastic dynamics in the Starobinsky's piecewise linear potential. A major difference between this model and the SR and USR cases described in \cref{subsec:Linear_potential,subsec::pure_USR} respectively is that, here, the covariance matrix of the noises depends on time, see \cref{fig:noiseCov:Starobinsky}. This implies that, contrary to what was done before, the noise statistics cannot be pre-computed once for all realisations but rather needs to be integrated along each trajectory as it unfolds, as proposed in \cite{Figueroa:2020jkf,Figueroa:2021zah}. While this is tractable in principle, we defer the implementation of such a program to future work, and below we study the simplified setup where the noise is evaluated along a reference, classical trajectory, on each branch on the potential function. The reason is that our main objective is to verify that gradient interactions are properly accounted for in our improved stochastic formalism, and this is done by comparing its result with linear perturbation theory where gradients are included to all orders. The verification is therefore performed at the perturbative level only, and the above-mentioned approximation is sufficient.

In practice, this implies that once the inflaton crosses the transition point at $\phi=0$, it cannot cross back, and that inflation comprises two distinct phases. The first phase is initiated at $\phi_\uin>0$ on the SR attractor, as in the linear-potential case, and ends when the field crosses $\phi=0$ for the first time. The duration of this first phase is a stochastic quantity denoted $\mathcal{N}_\uc$. The second phases is initiated at $\phi=0$, and the initial velocity is given by the final velocity in the first phase, which is also a stochastic quantity. Then we compute the duration of the second phase, ${\cal{N}}_2$, until the final field value $\phi_\uend$ is reached, and the first-passage time is nothing but $\mathcal{N}=\mathcal{N}_\uc+{\cal{N}}_2$.

From \cref{eq:alpha:beta:def}, at leading order in $\epsilon_1=(\cgmomentum)^2/(2\Mp^2)$, one has $\alpha=-3(A_1-A_2)\Dirac(\cgfield)$ and $\beta=-3$. Since the noises are computed along a reference, classical trajectory, one has $\Dirac(\cgfield) = \Dirac(N-\mathcal{N}_\uc)/\vert\cgmomentum(\mathcal{N}_\uc) \vert =\Dirac(N-\mathcal{N}_\uc)/A_1 $, where we have used the classical equation of motion~\eqref{eq:KG splitted} to evaluate the Jacobian. The source-less equation of motion~\eqref{eq:linearised:sourceless:system} for the local field difference thus reads
\bea
\label{eq:linearised:sourceless:system:Staro}
\frac{ \dd \fieldh}{ \dd N}&=\momentumh_\phi\,,
\\
\frac{ \dd \momentumh_\phi}{ \dd N}&=-3\gamma\Dirac(N-\mathcal{N}_\uc)\fieldh-3 \momentumh_\phi\, .
\eea
Within each phase $N<\mathcal{N}_\uc$ and $N>\mathcal{N}_\uc$ of the inflationary dynamics, this reduces to \cref{eq:linearised:sourceless:system:SR}, hence the same basis of solutions~\eqref{eq:hom:field:fluctuations:sol:SR} can be employed. The only difference is that the solutions are not continuous at the transition, owing to the presence of the Dirac term in \cref{eq:linearised:sourceless:system:Staro}. More precisely, upon integrating \cref{eq:linearised:sourceless:system:Staro} along the infinitesimal interval $[\mathcal{N}_\uc-\epsilon,\mathcal{N}_\uc+\epsilon]$, one finds that $\fieldh$ is continuous, and that 
\bea 
 \momentumh_\phi (\mathcal{N}_\uc+\epsilon) - \momentumh_\phi (\mathcal{N}_\uc-\epsilon) =-3\gamma\fieldh(\mathcal{N}_\uc)\, .
\eea 

By reshuffling the solutions~\eqref{eq:hom:field:fluctuations:sol:SR} by means of this junction condition, one obtains
\bea
\label{eq:hom:sol:field:fluctuations:Staro}
\delta\phi^{\mathrm{IR}}_{(1)}(N)&=1-\gamma\left[1-e^{-3(N-\mathcal{N}_\uc)}\right]\theta\left(N-\mathcal{N}_\uc\right)\,,
\\
\delta\phi^{\mathrm{IR}}_{(2)}(N)&=-\frac{1}{3}e^{-3N}+\frac{1}{3}\gamma e^{-3N_c}\left[1-e^{-3(N-\mathcal{N}_\uc)}\right]\theta\left(N-\mathcal{N}_\uc\right)\,,
\\
\delta\pi^{\mathrm{IR}}_{(1)}(N)&=-3\gamma e^{-3(N-\mathcal{N}_\uc)}\theta\left(N-\mathcal{N}_\uc\right)\,,
\\
\delta\pi^{\mathrm{IR}}_{(2)}(N)&=e^{-3N}+\gamma e^{-3N}
\theta\left(N-\mathcal{N}_\uc\right)\, .
\eea
Inserted into \cref{eq:Langevin gradient term}, these expressions lead to
\bea
\label{eq:Langevin gradient term:Staro:1}
\frac{\mathrm{d}}{\mathrm{d}N}\xi_\Delta^{(1)} &=-5 \xi_\Delta^{(1)} \, ,\\
\frac{\mathrm{d}}{\mathrm{d}N}\xi_\Delta^{(2)} &= -2\xi_\Delta^{(2)} -\sigma^2 \xi_\field\, , \\
\frac{\mathrm{d}}{\mathrm{d}N}\xi_\Delta^{(3)} &= -2\xi_\Delta^{(3)} -\frac{\sigma^2}{3} \xi_{\momentum}\, , \\
\frac{\mathrm{d}}{\mathrm{d}N}\xi_\Delta^{(4)} &= -5\xi_\Delta^{(4)} +\frac{\sigma^2}{3} \xi_{\momentum} \, ,
\eea
when $N<\mathcal{N}_\uc$, and
\bea
\label{eq:Langevin gradient term:Staro:2}
\frac{\mathrm{d}}{\mathrm{d}N}\xi_\Delta^{(1)} =&\left[\frac{3(1+\gamma)}{\gamma e^{3(N-\mathcal{N}_\uc)}-1-\gamma}-2\right] \xi_\Delta^{(1)} - \sigma^2\left[\gamma^2-\gamma(1+\gamma)e^{-3(N-\mathcal{N}_\uc)}\right]\xi_\field  \, ,\\
\frac{\mathrm{d}}{\mathrm{d}N}\xi_\Delta^{(2)} =& \left[\frac{3\gamma}{\left(\gamma-1\right) e^{3(N-\mathcal{N}_\uc)}-\gamma}-2\right]\xi_\Delta^{(2)} +\sigma^2 \left[\gamma^2-1-\gamma(1+\gamma)e^{-3(N-\mathcal{N}_\uc)}\right]\xi_\field\, , \\
\frac{\mathrm{d}}{\mathrm{d}N}\xi_\Delta^{(3)} =& \left[\frac{3\gamma}{\left(\gamma-1\right) e^{3(N-\mathcal{N}_\uc)}-\gamma}-2\right]\xi_\Delta^{(3)} 
\\ & 
-\frac{\sigma^2}{3}\left[1-2\gamma^2+\gamma(\gamma-1)e^{3(N-\mathcal{N}_\uc)}+\gamma(\gamma+1)e^{-3(N-\mathcal{N}_\uc)}\right] \xi_{\momentum}\, , \\
\frac{\mathrm{d}}{\mathrm{d}N}\xi_\Delta^{(4)} =& \left[\frac{3(1+\gamma)}{\gamma e^{3(N-\mathcal{N}_\uc)}-1-\gamma}-2\right]\xi_\Delta^{(4)} 
\\ &
+\frac{\sigma^2}{3}\left[1-2\gamma^2+\gamma(\gamma-1)e^{3(N-\mathcal{N}_\uc)}+\gamma(\gamma+1)e^{-3(N-\mathcal{N}_\uc)}\right] \xi_{\momentum} \, ,
\eea
when $N\geq\mathcal{N}_\uc$. 

In the first phase, the considerations presented in \cref{sec:line:pot:stochastic:inflation} apply:  since the gradient noises must vanish initially, one has $\xi_\Delta^{(1)}=0$, and since $\xi_\Delta^{(3)}$ and $\xi_\Delta^{(4)}$ are sourced by $\xi_{\momentum}$, which is of order $\sigma^2$ according to \cref{eq:noise_amplitudes_Linear}, they can be discarded as well. The only gradient noise that remains is $\xi_\Delta^{(2)}$, and the system of Langevin equations reduces to \cref{eq:Langevin equations linear potential with gradient}. 

In the second phase, the system of Langevin equations satisfied by the gradient noises is seemingly more involved. In particular, $\xi_{\momentum}$ cannot be discarded since it is enhanced during the transient phase after the transition, see \cref{fig:noiseCov:Starobinsky}. The system~\eqref{eq:Langevin gradient term:Staro:2} can however be simplified by introducing
\bea 
\label{eq:noise:rescaling}
\tilde{\xi}_\Delta^{(1)} =& \frac{\xi_\Delta^{(1)}}{1+\gamma-\gamma e^{3(N-\mathcal{N}_\uc)}}\, , \quad\quad
\tilde{\xi}_\Delta^{(2)} = \frac{\xi_\Delta^{(2)}}{1-\gamma+\gamma e^{-3(N-\mathcal{N}_\uc)}}\, ,\\
\tilde{\xi}_\Delta^{(3)} =& \frac{\xi_\Delta^{(3)}}{1-\gamma+\gamma e^{-3(N-\mathcal{N}_\uc)}}
\, , \quad\quad
\tilde{\xi}_\Delta^{(4)} = \frac{\xi_\Delta^{(4)}}{1+\gamma-\gamma e^{3(N-\mathcal{N}_\uc)}}\, .
\eea 
In terms of these rescaled noises, \cref{eq:Langevin gradient term:Staro:2} leads to
\bea 
\label{eq:Langevin gradient term:Staro:2:rescaled}
\frac{\mathrm{d}}{\mathrm{d}N}\tilde\xi_\Delta^{(1)} &=-5 \tilde{\xi}_\Delta^{(1)} +\sigma^2 \gamma e^{-3(N-\mathcal{N}_\uc)}\xi_\field  \, ,\\
\frac{\mathrm{d}}{\mathrm{d}N}\tilde{\xi}_\Delta^{(2)} &= -2\tilde{\xi}_\Delta^{(2)} -(1+\gamma)\sigma^2 \xi_\field\, , \\
\frac{\mathrm{d}}{\mathrm{d}N}\tilde{\xi}_\Delta^{(3)} &= -2\tilde{\xi}_\Delta^{(3)} -\frac{\sigma^2}{3}\left[1+\gamma-\gamma e^{3(N-\mathcal{N}_\uc)}\right] \xi_{\momentum}\, , \\
\frac{\mathrm{d}}{\mathrm{d}N}\tilde\xi_\Delta^{(4)} &=-5 \tilde{\xi}_\Delta^{(4)} +\frac{\sigma^2}{3}\left[1-\gamma+\gamma e^{-3(N-\mathcal{N}_\uc)}\right] \xi_{\momentum}\, .
\eea
This system of Langevin equations is simpler, and has a form closer to that of \cref{eq:Langevin gradient term:Staro:1}, to which it manifestly reduces in the limit $\gamma=0$ (\ie $A_1=A_2$). 

Let us then determine the initial conditions this system needs to be solved with. Across the transition, since the homogenous solutions~\eqref{eq:hom:sol:field:fluctuations:Staro} contain Heaviside distributions, the Langevin equations for the gradient noises contain Dirac distributions arising from the terms $ {\fieldh_{(1)}}'/\fieldh_{(1)}$ and ${\fieldh_{(2)}}'/\fieldh_{(2)}$ in \cref{eq:Langevin gradient term}. In principle, this leads to non-trivial matching conditions for the gradient noises, but in \cref{eq:hom:sol:field:fluctuations:Staro} one can check that the Heaviside terms are always multiplied by factors that vanish at the transition, hence $\fieldh_{(1)}$ and $\fieldh_{(2)}$ are in fact continuous and the gradient noises are also continuous.\footnote{One can check that this is explicitly the case: the non-smooth term in \cref{eq:Langevin gradient term} is of the form $\dd \xi_\Delta^{(1)}/\dd N \ni ({\fieldh_{(1)}}'/\fieldh_{(1)})\xi_\Delta^{(1)} $, which leads to
\bea
\frac{\xi_\Delta^{(1)}(\mathcal{N}_\uc+\epsilon)}{\xi_\Delta^{(1)}(\mathcal{N}_\uc-\epsilon)}= \frac{\fieldh_{(1)}(\mathcal{N}_\uc+\epsilon)}{\fieldh_{(1)}(\mathcal{N}_\uc-\epsilon)}=1
\eea
when integrated across the transition.
}
Since, as explained above, $\xi_\Delta^{(1)}$, $\xi_\Delta^{(3)}$ and $\xi_\Delta^{(4)}$ vanish during the first phase, $\tilde{\xi}_\Delta^{(1)}$, $\tilde{\xi}_\Delta^{(3)}$ and $\tilde{\xi}_\Delta^{(4)}$ should be set to zero at the onset of the second phase, and $\tilde{\xi}_\Delta^{(2)}$ should be initiated to the value acquired by $\xi_\Delta^{(2)}$ at the end of the first phase.

Even though the system~\eqref{eq:Langevin gradient term:Staro:2:rescaled} is more compact than the original one, one may cast the stochastic dynamics in terms of an even simpler system. To do so, one first notices that \cref{eq:Langevin gradient term:Staro:2} can be formally solved as
\bea 
\tilde\xi_\Delta^{(1)} =& \gamma\sigma^2 e^{-5(N-\mathcal{N}_\uc)}\int_{\mathcal{N}_\uc}^N e^{2(N'-\mathcal{N}_\uc)}\xi_\field(N')\dd N'\, ,\\
\tilde\xi_\Delta^{(2)} =&\xi_\Delta^{(2)}(\mathcal{N}_\uc)e^{-2(N-\mathcal{N}_\uc)}-(1+\gamma)\sigma^2 e^{-2(N-\mathcal{N}_\uc)}\int_{\mathcal{N}_\uc}^N e^{2(N'-\mathcal{N}_\uc)}\xi_\field(N')\dd N'\, ,\\
\tilde\xi_\Delta^{(3)} =& \frac{\sigma^2}{3} e^{-2(N-\mathcal{N}_\uc)}\int_{\mathcal{N}_\uc}^N \left[\gamma e^{5(N'-\mathcal{N}_\uc)}-(1+\gamma)e^{2(N'-\mathcal{N}_\uc)}\right]\xi_{\momentum}(N')\dd N'\, ,
\\
\tilde\xi_\Delta^{(4)} =& \frac{\sigma^2}{3} e^{-5(N-\mathcal{N}_\uc)}\int_{\mathcal{N}_\uc}^N \left[\gamma e^{2(N'-\mathcal{N}_\uc)}+(1-\gamma)e^{5(N'-\mathcal{N}_\uc)}\right]\xi_{\momentum}(N')\dd N'\, ,
\eea 
where the initial conditions mentioned above have been imposed. Making use of \cref{eq:noise:rescaling}, various simplifications arise when summing up the four gradient noises, and one finds
\bea 
\sum_{i=1}^4\xi_\Delta^{(i)} =&
\underbrace{
\left\lbrace
\left(1-\gamma\right)\xi_\Delta^{(2)}(\mathcal{N}_\uc)
-\sigma^2 \int_{\mathcal{N}_\uc}^N e^{2(N'-\mathcal{N}_\uc)}\left[\xi_\field(N')+\frac{1}{3}\xi_{\momentum}(N')\right]\dd N' 
 \right\rbrace e^{-2(N-\mathcal{N}_\uc)}}_{\xi_\Delta^{(a)}}
\\ &
+\underbrace{\left[\frac{\sigma^2}{3}\int_{\mathcal{N}_\uc}^N e^{5(N'-\mathcal{N}_\uc)}\xi_{\momentum}(N')\dd N'
+\gamma \xi_\Delta^{(2)}(\mathcal{N}_\uc)\right]e^{-5(N-\mathcal{N}_\uc)}}_{\xi_\Delta^{(b)}}\, .
\eea 
Let us collect the terms arising in the first line of the right-hand side of the above into a quantity denoted $\xi_\Delta^{(a)}$, and the second line into $\xi_\Delta^{(b)}$. These two gradient noises follow the Langevin equations\footnote{The form~\eqref{eq:Langevin:extended:Staro:2:noises} of the Langevin equation for the gradient noises is not only simpler than \cref{eq:Langevin gradient term:Staro:2:rescaled}, it is also more amenable to numerical implementation: while $\xi_\Delta^{(3)}$ and $\xi_\Delta^{(4)}$ grow exponentially at late time, their exponential contributions cancel out from the combinations $\xi_\Delta^{(a)}$ and $\xi_\Delta^{(b)}$.}
\bea 
\label{eq:Langevin:extended:Staro:2:noises}
\frac{\mathrm{d}}{\mathrm{d}N}\xi_\Delta^{(a)} &= -2 \xi_\Delta^{(a)} - \sigma^2 \xi_\field - \frac{\sigma^2}{3}\xi_{\momentum}\, ,\\
\frac{\mathrm{d}}{\mathrm{d}N}\xi_\Delta^{(b)} &= -5 \xi_\Delta^{(b)}+\frac{\sigma^2}{3}\xi_{\momentum}\, ,
\eea 
subject to the initial conditions $\xi_\Delta^{(a)}(\mathcal{N}_\uc)=(1-\gamma)\xi_\Delta^{(2)}(\mathcal{N}_\uc)$ and $\xi_\Delta^{(b)}(\mathcal{N}_\uc)=\gamma \xi_\Delta^{(2)}(\mathcal{N}_\uc)$. Together with \cref{eq:Langevin:extended}, \ie
\bea 
\label{eq:Langevin:extended:Staro:2}
\frac{\dd}{\dd N} {\f}^{\mathrm{IR}}= &{\pi}_\phi^{\mathrm{IR}}+{\xi}_\f\, ,\\
\frac{\dd}{\dd N}{\pi}_\phi^{\mathrm{IR}}= & -3{\pi}_\phi^{\mathrm{IR}}- 3A_2+{\xi}_{\pi_\phi}+\xi_\Delta^{(a)}+\xi_\Delta^{(b)}
\, ,
\eea 
they constitute the system of Langevin equations to be solved in the second phase.

Note that, these equations being linear, $\cgfield$ and $\cgmomentum$ still have Gaussian statistics, the moments of which can be expressed as nested time integrals of the field power spectra, as was done in \cref{sec:lin:stochastic:versus:CPT:field:pert,sec:USR:stochastic:versus:CPT:field:pert} for SR and USR respectively. However, in the present case, the field power spectra~\eqref{eq:noise_amplitudes_Starobinsky} are scale and time dependent, and the corresponding integrals are not straightforward to perform. This is why we now proceed to solve the Langevin equations numerically.

\subsubsection{Numerical implementation}
\label{sec:Staro:pot:numerics}

%
\begin{table}
    \renewcommand{\arraystretch}{1.3}
    \centering
    \begin{tabular}{c @{\quad} c @{\quad} c @{\quad} c @{\quad} c @{\quad} c @{\quad} c @{\quad} c} 
        \hline 
        Parameters  & $H/\Mp$ & $A_1/\Mp$ & $A_2/\Mp$ & $\phi_\uin/\Mp$ &  $\pi_{\phi,\uin}/\Mp$ &  $\phi_\uend/\Mp$ \\ \hline 
        Input & $2.0\times10^{-6}$ &$0.01$ &$0.0004$ &$0.2$ &$-0.01$  &$-0.0034$ \\
        \hline 
    \end{tabular}
    \caption{Parameter values used for numerical simulations in the Starobinsky piecewise model.}
    \label{table:Starobinsky}
\end{table}

The Langevin equations~\eqref{eq:Langevin equations linear potential with gradient} in the first phase
and \cref{eq:Langevin:extended:Staro:2:noises,eq:Langevin:extended:Staro:2} in the second phase are solved numerically for the parameter values listed in \cref{table:Starobinsky}.

\begin{figure}[t]
\centering
\includegraphics[width=0.48\textwidth,trim={0cm 0cm 0cm 0cm}, clip]{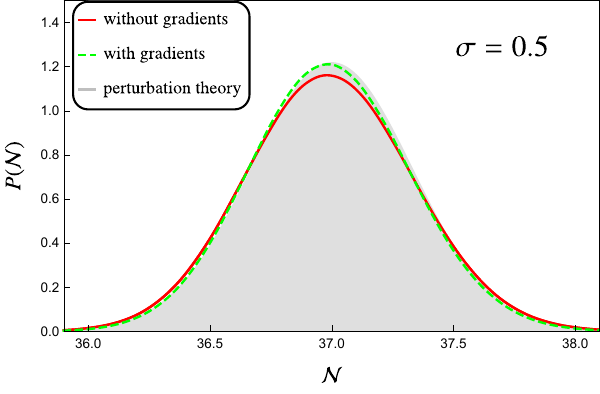}
    \hspace{0.2cm}
    \includegraphics[width=0.48\textwidth,trim={0cm 0cm 0cm 0cm}, clip]{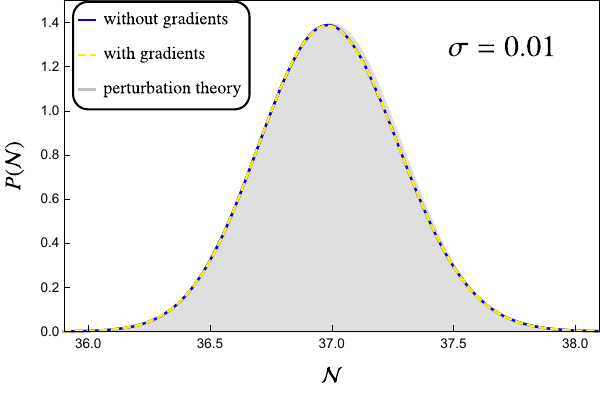}
    \includegraphics[width=0.48\textwidth,trim={0cm 0cm 0cm 0cm}, clip]{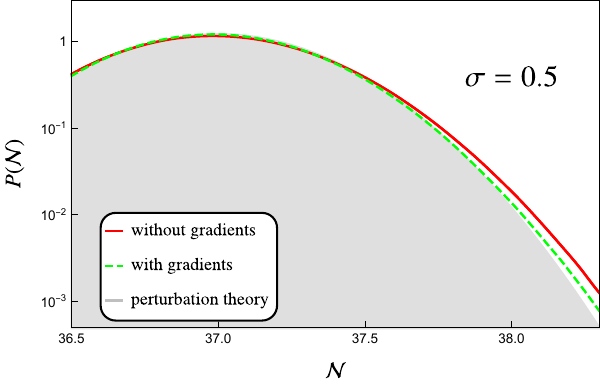}
    \hspace{0.2cm}
    \includegraphics[width=0.48\textwidth,trim={0cm 0cm 0cm 0cm}, clip]{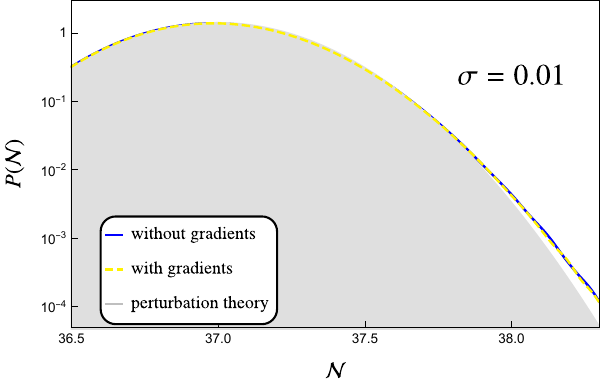}
\caption{First-passage-time distribution in the Starobinsky piecewise-linear model for the parameters listed in \cref{table:Starobinsky}, for $\sigma=0.5$ (left panel) and $\sigma=0.01$ (right panel). Solid lines are obtained from $10^6$ realisations of the Langevin equations~\eqref{eq:Langevin:standard} without gradient interactions, while dashed curves include gradient corrections and are drawn from \cref{eq:Langevin equations linear potential with gradient,eq:Langevin:extended:Staro:2:noises,eq:Langevin:extended:Staro:2}. The grey-shaded curves are Gaussian distributions following the prediction~\eqref{eq:PR:Staro} from linear perturbation theory.}
\label{fig:PDF_Starobinsky}
\end{figure}

\begin{figure}[t]
\centering
    \includegraphics[width=0.48\textwidth,trim={0cm 0cm 0cm 0cm}, clip]{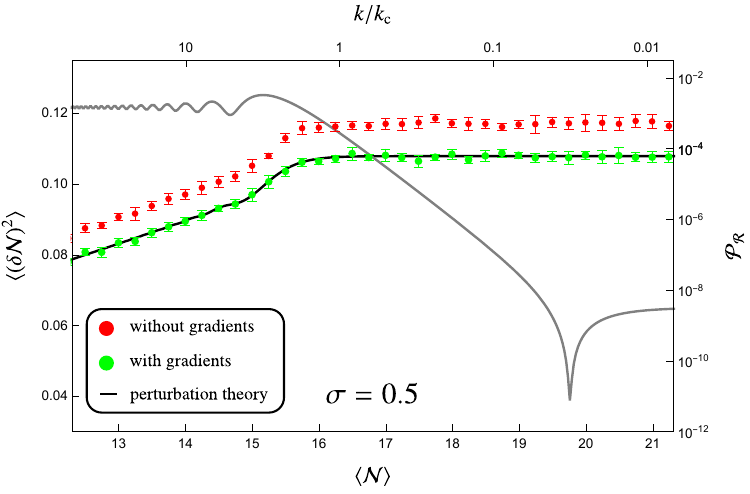}
    \hspace{0.2cm}
    \includegraphics[width=0.48\textwidth,trim={0cm 0cm 0cm 0cm}, clip]{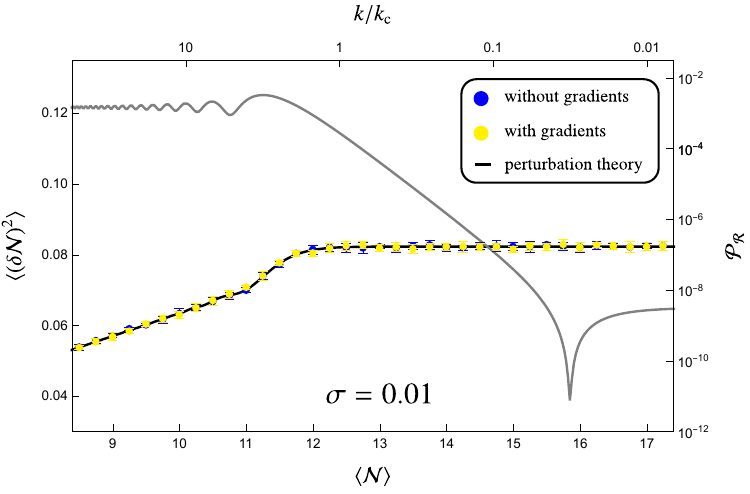}
\caption{Variance of the first-passage-time distribution, $\langle(\delta {\cal{N}})^2\rangle$, for different initial field values, parametrised by their mean number of e-folds $\langle {\cal{N}}\rangle$, in the Starobinsky's piecewise-linear model. The colour code is the same as in \cref{fig:PDF_Starobinsky}, where perturbation theory, obtained by inserting \cref{eq:PR:Staro} into \cref{eq:variance as integrated power spectrum}, is displayed with black solid lines [\cref{eq:PR:Staro} is shown in grey and its value is reported on the rightmost vertical axis for indication]. On the upper horizontal axis, the value of $k=\sigma a_\uend H e^{-\langle \mathcal{N}\rangle}$ is displayed. $10^4$ realisations have been used for each point, whose error bars have been evaluated using Jackknife resampling.}
\label{fig:Integrated_Powerspectrum_Starobinsky}
\end{figure}

In \cref{fig:PDF_Starobinsky}, the distribution function for the first-passage time is displayed for $\sigma=0.5$ (left panel) and $\sigma=0.01$ (right panel), with and without gradient corrections, and compared to the result~\eqref{eq:PR:Staro} from linear perturbation theory. As before, unless $\sigma$ is small enough, only by including gradient interactions in the stochastic formalism can one reproduce the result expected from perturbation theory, where gradient corrections are present to all orders. This is confirmed in \cref{fig:Integrated_Powerspectrum_Starobinsky}, where the second moment of the curvature perturbation is displayed.\footnote{Since $\langle (\delta \mathcal{N})^2\rangle$ is dominated by fluctuations at small scales, where $\mathcal{P}_{\mathcal{R}}$ is larger (see \cref{fig:powerspectrum_Starobinsky}), the contributions from intermediate scales, where gradient effects are more pronounced, are however difficult to resolve in \cref{fig:Integrated_Powerspectrum_Starobinsky}.} As in \cref{subsec:Linear_potential,subsec::pure_USR}, the pullback effect is clearly visible: the amplitude of curvature perturbations is reduced by gradient interactions, both at the level of its variance and along its upper tail.

There are however two main differences with respect to the SR and USR cases studied previously. First, in models with dynamical transitions such as the Starobinsky piecewise-linear potential, gradient interactions do not only constitute a small correction at super-Hubble scales, but they substantially contribute to their dynamics after the transition, as explained in \cref{sec:failure:zeroth:order:in:transitions}. The results of \cref{fig:PDF_Starobinsky,fig:Integrated_Powerspectrum_Starobinsky} thus confirm that our improved stochastic formalism is able to properly account for gradient interactions even when they play a prominent role.

Second, while in SR and USR the condition $\sigma^2\ll 1$ is sufficient to ensure that gradient corrections are negligible, in the Starobinsky piecewise-linear potential that condition rather reads~\cite{Jackson:2023obv}
\bea 
\label{eq:sigma:cond:Staro}
\sigma^2\ll \frac{1-\gamma}{\gamma}\sim e^{-3 N_{\mathrm{USR}}}\, .
\eea 
This guarantees that the scales at which gradient interactions become large are matched to the homogeneous background after the transition, see the discussion below \cref{eq:ratio_k2_correction_transient_USR}. In order to include backreaction at all super-Hubble scales (\textit{i.e.}, taking $\sigma^2$ smaller than one but not necessarily smaller than $e^{-3 N_{\mathrm{USR}}}$), it is therefore necessary to include gradient interactions in the stochastic formalism.\footnote{Using a value of $\sigma$ closer to one has indeed two advantages. First, it allows one to let a wider range of scales  backreact, as we further discuss in \cref{sec:conclusion}. Second, it reduces the duration of the transient phase identified in \cref{fig:noiseCov:Starobinsky}, during which the noise covariance depends on time. This makes the approximation employed here, where the noise covariance is computed on a reference classical trajectory past $\mathcal{N}_\uc$ (as opposed to evaluating it along each stochastic realisation separately), more accurate.}

%

\section{Conclusion}
\label{sec:conclusion}

Most inflationary models studied in the context of PBHs involve a transition between an initial slow-roll attractor phase and a USR period, during which gradient interactions become relevant even at super-Hubble scales. These gradient effects are nonetheless discarded by the separate-universe or $\delta N$ approach, which describes the dynamics of super-Hubble fluctuations non perturbatively in terms of an ensemble of background, homogeneous universes. This is notably the case in the stochastic-inflation formalism, whose standard formulation is therefore not suited to models with such transitions.

In this work, we have shown how the leading gradient interaction, which is the only one that may play a role at super-Hubble scales, can be incorporated in the stochastic-inflation formalism, at the expense of two modifications. 
\begin{itemize}
    \item First, the covariance matrix of the noises, computed using linear perturbation theory within the coarse-graining radius $(\sigma H)^{-1}$, should not be evaluated at leading order in the gradient expansion (\ie at leading order in $\sigma$, as often done) but should retain all terms of order $\sigma^2$. This modification is straightforward, and can be combined with non-linear small-scale corrections, such as loop contributions~\cite{Nassiri-Rad:2025dsa}. 
    \item Second, above the coarse-graining radius, neighbour patches interact via gradients. In principle, this prevents the use of a separate-universe description, where different patches evolve independently. However, we have shown that gradient interactions can be recast as a memory effect within each patch, whose stochastic dynamics becomes non-Markovian, \ie subject to coloured noises. Moreover, this seemingly non-Markovian dynamics can in fact be rephrased in terms of a larger set of Langevin equations with white noises only (the same noises as in the standard formulation). 
\end{itemize}
Our results thus extend the ability to describe long-wavelengths perturbations in  terms  of  separate  universes, and to  rely  on  standard  techniques  for  solving  Markovian stochastic differential equations, even in the presence of gradient interactions.

We have then applied our improved formalism to three different test cases, to check its validity. First, in a pure slow-roll model where the inflaton's potential is linear and gradient effects are expected to be small, we have shown that our formalism correctly captures gradient corrections in the moments of the phase-space fields. This was done by analytically solving the Langevin equations, which are linear in that case. We also reconstructed the distribution function of the curvature perturbations by numerically evolving the Langevin equations and sampling the first-passage time. We found that gradient corrections are again properly reproduced, even for values of $\sigma$ close to one. The same results were obtained in a pure USR regime where gradient effects are also small. In both cases, we have found that gradient interactions make the tail of the first-passage-time distribution less heavy. This is because, when a patch inflates for an anomalously short or long time, it is pulled back by its neighbour patches, which act on it as a spring force. This pullback effect seems to be a systematic consequence of gradient interactions and reduces the probability for large curvature perturbations to occur. 

Finally, we have studied the Starobinsky's piecewise-linear potential, which features a transition between a slow-roll and a USR phase. In this model, gradient interactions play an important role over a finite range of super-Hubble scales after the transition, which is not captured by the standard stochastic-inflation formalism. There again, we have found that gradient interactions are properly accounted for in our improved formalism, and that the pullback effect is responsible for a reduction in the tails of the first-passage-time distributions. The consequences of the pullback effect on the abundance of PBHs and other high-density objects remains to be investigated.

\begin{figure}[t]
\centering
    \includegraphics[width=0.7\textwidth,trim={0cm 0cm 0cm 0cm}, clip]{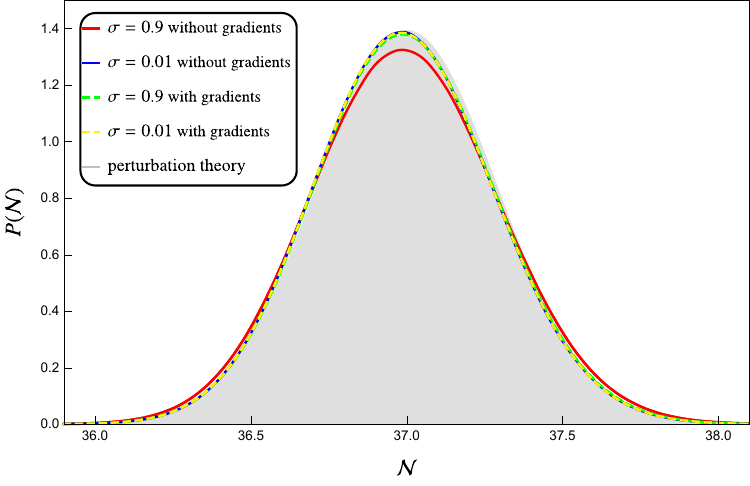}
    \\
    \includegraphics[width=0.7\textwidth,trim={0cm 0cm 0cm 0cm}, clip]{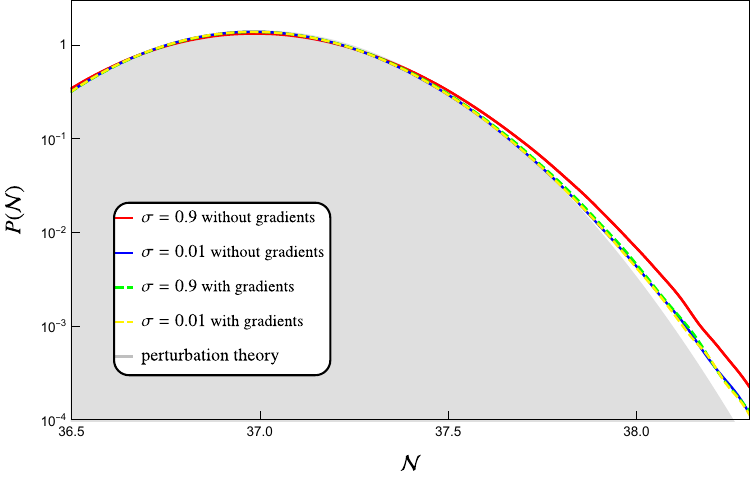}\\
\caption{First-passage-time distribution in the Starobinsky piecewise-linear model for the parameters listed in \cref{table:Starobinsky}, for $\sigma=0.5$ and $\sigma=0.01$, with $\phi_\uin/\Mp=0.2$ and $\phi_\uend/\Mp=-0.0034$ for $\sigma=0.01$ and $\phi_\uin/\Mp=0.23912$ and $\phi_\uend/\Mp=-0.0033844$ for $\sigma=0.9$ (these values are adapted in such a way that the same range of physical scales is included in both cases). Solid lines are obtained from $10^5$ realisations of the Langevin equations~\eqref{eq:Langevin:standard} without gradient interactions, while dashed curves include gradient corrections and are drawn from \cref{eq:Langevin equations linear potential with gradient,eq:Langevin:extended:Staro:2:noises,eq:Langevin:extended:Staro:2}. The grey-shaded curve is the Gaussian distribution following the prediction~\eqref{eq:PR:Staro} from linear perturbation theory.}
\label{fig:PDF_Starobinsky_different_phi_i}
\end{figure}

As mentioned above, one of the advantages of our improved formalism is that $\sigma$ can be set to somewhat larger values than what it should be in the standard stochastic-inflation formalism, allowing one to include a wider range of scales in the modelling of backreaction. To illustrate this effect, in \cref{fig:PDF_Starobinsky_different_phi_i} the first-passage-time distribution in the Starobinsky's piecewise-linear potential is displayed for two different values of $\sigma$, one that satisfies \cref{eq:sigma:cond:Staro} and one that does not. Since the relationship between field value and scale, $k=\sigma a_\uend H e^{-\langle \mathcal{N}\rangle(\phi)}$,  depends on $\sigma$, the values of $\phi_\uin$ and $\phi_\uend$ have been adapted in such a way that the same range of scales is included in the two cases. Therefore, the only difference between the curves with different values of $\sigma$ is that, when $\sigma$ is larger, more scales are accounted for in backreaction. That difference thus only arises at the non-linear level. In the absence of gradient interactions, one can see that the result depends rather substantially on $\sigma$, and that increasing $\sigma$ leads to heavier tails. In contrast, when gradient interactions are included, the dependence on $\sigma$ cannot be resolved.
Our improved formalism thus delivers results that are less sensitive to the detailed choice of the coarse-graining scale, although backreaction in the presence of gradient interactions remains to be investigated. 

Indeed, in this work we chose to investigate a model where gradient corrections are under perturbative control, since our goal was to check that our improved stochastic formalism is able to handle them properly. The downside is that genuine ``gradient-stochastic'' effects were hardly visible, but this does not preclude the existence of other setups where they are more prominent. In hybrid inflation~\cite{Linde:1993cn} for instance, non-perturbative stochastic effects are at play around the saddle point of the potential~\cite{Martin:2011ib, Tada:2023fvd, Murata:2025onc}, and the pullback mechanism might be crucial to determine how trajectory bundles split between different vacua. Another case of interest is multiple-field setups with sharp turns~\cite{Bjorkmo:2019fls}, where the misalignment between the background and noise directions~\cite{Grain:2017dqa} can only be resorbed by increasing $\sigma$, which requires to include gradient effects. In such models we expect our formalism to go beyond existing results, and we plan to investigate them in the future.

Other directions deserve further analyses and we end this article by mentioning a few. In the language of the gradient expansion (see \cref{sec:Importance_gradient}), our formalism incorporates all contributions up to order $k^2$, and this is sufficient in most single-field models even when the power spectrum of curvature perturbations grows faster than $k^4$~\cite{Carrilho:2019oqg,Ragavendra:2020sop,Tasinato:2020vdk,Cole:2022xqc}. This is however not enough close to the dip of the power spectrum, where an accidental cancellation between terms of order $k^0$ and terms of order $k^2$ occurs if the inflaton’s velocity does not flip sign, and the amplitude at the dip is set by higher-order contributions~\cite{Byrnes:2018txb,Ozsoy:2019lyy,Briaud:2025hra,Fujita:2025imc}.
The gradient expansion would also have to be further investigated in multi-field setups ~\cite{Palma:2020ejf,Fumagalli:2020adf,Braglia:2020taf}. 
In these situations our approach may have to be extended to include next-to-next-to-leading contributions in gradients, which should be possible following similar lines as those presented here.
Let us also note that the validity of the gradient expansion when truncated at order $k^2$ has been established within perturbation theory only. Incorporating higher gradient contributions would allow one to test the gradient expansion at the non-perturbative level, and determine for instance whether or not the pullback effect is correctly captured by leading gradients. The extension of our approach beyond the decoupling limit also remains to be investigated.

Finally, our work share common objectives with \Refa{Artigas:2024ajh}, which recently proposed to include gradient effects in the separate-universe approach by allowing the separate patches to be spatially curved (see \Refa{Raveendran:2025pnz} for possible limitations of this approach). In this way, the super-Hubble evolution of the curvature perturbation at super-Hubble scales is properly accounted for even in the presence of SR-USR transitions, and this was verified at the linear level. It would be interesting to investigate whether or not a stochastic formalism can be derived from this setup, and how it would compare with our approach.

\acknowledgments
We thank Danilo Artigas, Parth Bhargava, Andrew Gow, Kazuya Koyama and David Wands for very interesting discussions. 
RK is supported by JSPS KAKENHI grants 24KJ2108 and JSPS Overseas Challenge Program for Young Researchers.
RK thanks LPENS for hospitality where most of this work was conducted.

\appendix

\crefalias{section}{appendix}

\section{Smoothing function $I$}
\label{app:I}

In this appendix, we show how the smoothing function $I$ appearing in \cref{sec:continuous expression} can be obtained using properties of the Fourier transform. First, introducing the Fourier modes $\cgfield_{\bm{k}}(N)$ of the coarse-grained field $\cgfield(\bm{x},N)$, which are stochastic quantities, the gradient term is given by
\bea
\label{eq:A1}
\Delta\cgfield(\bm{x},N)= -\int\frac{ \dd ^3\bm{k}}{(2\pi)^{3/2}}k^2\cgfield_{\bm{k}}(N)e^{-i\bm{k}\cdot\bm{x}}\,.
\eea
Upon inverse Fourier transforming the mode functions $\cgfield_{\bm{k}}$, we obtain
\bea
\label{eq:expression gradient term}
\Delta\cgfield(\bm{x},N)= \int \dd ^3\bm{y} \cgfield(\bm{y},N)I\left(|\bm{x}-\bm{y}|\right)\,,
\eea
where the function $I$ is defined as
\bea
\label{eq:I:int:representation}
I\left(|\bm{z}|\right) = -\frac{1}{2\pi^2}\int_0^{+\infty}\mathrm{d}k k^4\sinc\left(k|\bm{z}|\right)\, .
\eea
This expression, which formally diverges, is nothing but the integral representation of the distribution 
\bea 
\label{eq:I:distrib:app}
I\left(|\bm{x}-\bm{y}|\right)=\Delta_{\bm{y}}\Dirac(\bm{y}-\bm{x})
\eea
introduced in \cref{sec:continuous expression}. 

Then, in terms of the un-coarse-grained field, \cref{eq:A1} reads
\bea
\Delta\cgfield(\bm{x},N)= -\int\frac{ \dd ^3\bm{k}}{(2\pi)^{3/2}}k^2\field_{\bm{k}}(N)W\left(\frac{k}{\sigma a H}\right)e^{-i\bm{k}\cdot\bm{x}}\, ,
\eea
which can again be inverse Fourier transformed as 
\bea
\label{eq:expression gradient term:2}
\Delta\cgfield(\bm{x},N)= \int \dd ^3\bm{y} \field(\bm{y},N)\mathcal{I}\left(|\bm{x}-\bm{y}|\right)
\eea
with 
\bea
\label{eq:calI:int:representation}
\mathcal{I}\left(|\bm{z}|\right) = -\frac{1}{2\pi^2}\int_0^{+\infty}\mathrm{d}k W\left(\frac{k}{\sigma a H}\right) k^4\sinc\left(k|\bm{z}|\right)\, .
\eea
If $W$ is taken as a Heaviside step function, as we assumed throughout this work, the integral can be performed explicitly, and the result is
\bea
\label{eq:I:explicit:app}
\mathcal{I}\left(|\bm{z}|\right)=-\frac{k_\sigma^5}{2\pi^2} \left[ \frac{5(6-|k_\sigma \bm{z}|^2)\cos(|k_\sigma\bm{z}|)}{|k_\sigma\bm{z}|^4}+\frac{15(-2+|k_\sigma\bm{z}|^2)\sin(|k_\sigma\bm{z}|)}{|k_\sigma\bm{z}|^5} \right]\, ,
\eea
where we recall that $k_\sigma=\sigma a H$. This makes it clear that the function $\mathcal{I}$ decays when $\vert\bm{z}\vert \gg (\sigma a H)^{-1} $, hence that it selects patches that are close enough to $\bm{x}$ and that contribute to the gradient interaction, in agreement with the discussion around \cref{fig:memory effect}. 

Importantly, although $\int\dd^3\bm{z}I(\vert\bm{z}\vert)$ formally diverges when \cref{eq:I:int:representation} is employed, one has to recall that $I$ needs to be understood as a distribution and \cref{eq:I:distrib:app} trivially leads to
\bea
\label{eq:vanishing integral I}
\int \dd ^3 \bm{z}  \hspace{0.05cm} I(|\bm{z}|)=0\, .
\eea
This shows that the homogeneous part of the field does not contribute to the gradient noise, in agreement with the discretised expression~\eqref{eq:discretised Laplace}.

\section{Solution for the local field difference}
\label{app:sol:local:field:difference}

In this appendix, we check that \cref{eq:general solution field difference} is  a solution of the linear system~\eqref{eq:Langevin deltaphi}. First, let us rewrite \cref{eq:general solution field difference} as 
\bea
\label{eq:general solution field difference:app}
& \delta\cgfield\left(\bm{x},\bm{y},N\right) = 
\\ &\quad \int^N_{\Ni}\mathrm{d}N'\left\lbrace G_{\phi\pi}(N,N') \left[\xi_\field\left(\bm{y},N'\right)-\xi_\field\left(\bm{x},N'\right)\right]
+G_{\phi\phi}(N,N')\left[\xi_{\momentum}\left(\bm{y},N'\right)-\xi_{\momentum}\left(\bm{x},N'\right)\right]\right\rbrace\,,
\eea
where 
\bea 
\label{eq:G:def}
G_{\phi\pi}(N,N') =& \fieldh_{(2)}\left(N\right)\Momentumh_{(1)}\left(N'\right)-\fieldh_{(1)}\left(N\right)\Momentumh_{(2)}\left(N'\right) \\
= & \frac{\fieldh_{(2)}\left(N\right)\momentumh_{(1)}\left(N'\right)-\fieldh_{(1)}\left(N\right)\momentumh_{(2)}\left(N'\right)}{\fieldh_{(2)}\left(N'\right)\momentumh_{(1)}\left(N'\right)-\fieldh_{(1)}\left(N'\right)\momentumh_{(2)}\left(N'\right)}\, ,
\\
G_{\phi\phi}(N,N')=&\fieldh_{(1)}\left(N\right)\Fieldh_{(2)}\left(N'\right)-\fieldh_{(2)}\left(N\right)\Fieldh_{(1)}\left(N'\right)\\
= & \frac{\fieldh_{(1)}\left(N\right)\fieldh_{(2)}\left(N'\right)-\fieldh_{(2)}\left(N\right)\fieldh_{(1)}\left(N'\right)}{\fieldh_{(2)}\left(N'\right)\momentumh_{(1)}\left(N'\right)-\fieldh_{(1)}\left(N'\right)\momentumh_{(2)}\left(N'\right)}\, .
\eea 
From these expressions it is clear that, in the coincident limit $N'\to N$, $G_{\phi\pi}(N,N)=1$ and $G_{\phi\phi}(N,N)=0$. Therefore, when differentiating \cref{eq:general solution field difference:app} with respect to time, one finds
\bea 
& \frac{\dd\delta\cgfield}{\dd N}\left(\bm{x},\bm{y},N\right) = 
\xi_\field\left(\bm{y},N\right)-\xi_\field\left(\bm{x},N\right)\\
& +
\int^N_{\Ni}\mathrm{d}N'\left\lbrace \partial_N G_{\phi\pi}(N,N') \left[\xi_\field\left(\bm{y},N'\right)-\xi_\field\left(\bm{x},N'\right)\right]
+\partial_N G_{\phi\phi}(N,N')\left[\xi_{\momentum}\left(\bm{y},N'\right)-\xi_{\momentum}\left(\bm{x},N'\right)\right]\right\rbrace .
\eea 
From the first entry of \cref{eq:Langevin deltaphi}, one can thus identify the second line of the above with $\delta\cgmomentum$, \ie
\bea
\label{eq:deltapi:lin:app}
&\delta\cgmomentum\left(\bm{x},\bm{y},N\right) = \\
& \int^N_{\Ni}\mathrm{d}N'\left\lbrace \partial_N G_{\phi\pi}(N,N') \left[\xi_\field\left(\bm{y},N'\right)-\xi_\field\left(\bm{x},N'\right)\right]
+\partial_N G_{\phi\phi}(N,N')\left[\xi_{\momentum}\left(\bm{y},N'\right)-\xi_{\momentum}\left(\bm{x},N'\right)\right]\right\rbrace .
\eea 
Moreover, from the homogeneous equations~\eqref{eq:linearised:sourceless:system}, one finds
\bea 
\label{eq:dN:Green}
\partial_N G_{\phi\pi}(N,N')= & \frac{\momentumh_{(2)}\left(N\right)\momentumh_{(1)}\left(N'\right)-\momentumh_{(1)}\left(N\right)\momentumh_{(2)}\left(N'\right)}{\fieldh_{(2)}\left(N'\right)\momentumh_{(1)}\left(N'\right)-\fieldh_{(1)}\left(N'\right)\momentumh_{(2)}\left(N'\right)}\, ,\\
\partial_N G_{\phi\phi}(N,N')= &
\frac{\momentumh_{(1)}\left(N\right)\fieldh_{(2)}\left(N'\right)-\momentumh_{(2)}\left(N\right)\fieldh_{(1)}\left(N'\right)}{\fieldh_{(2)}\left(N'\right)\momentumh_{(1)}\left(N'\right)-\fieldh_{(1)}\left(N'\right)\momentumh_{(2)}\left(N'\right)}\, .
\eea 
This makes it clear that, in the coincident limit, $\lim_{N'\to N}\partial_N G_{\phi\pi}(N,N')=0$ and $\lim_{N'\to N} \partial_N G_{\phi\phi}(N,N')=1$. By time differentiating \cref{eq:deltapi:lin:app} one thus obtains
\bea 
\label{eq:dphiIR:dN:lin:app}
&\frac{\dd\delta\cgmomentum}{\dd N}\left(\bm{x},\bm{y},N\right) = 
\xi_{\momentum}\left(\bm{y},N\right)-\xi_{\momentum}\left(\bm{x},N\right)\\
& +\int^N_{\Ni}\mathrm{d}N'\left\lbrace \partial^2_N G_{\phi\pi}(N,N') \left[\xi_\field\left(\bm{y},N'\right)-\xi_\field\left(\bm{x},N'\right)\right]
+\partial^2_N G_{\phi\phi}(N,N')\left[\xi_{\momentum}\left(\bm{y},N'\right)-\xi_{\momentum}\left(\bm{x},N'\right)\right]\right\rbrace .
\eea 
Differentiating \cref{eq:dN:Green} one more time with respect to $N$, and using the homogeneous system~\eqref{eq:linearised:sourceless:system}, one has
\bea 
\partial^2_N G_{\phi\pi}(N,N')= &\alpha G_{\phi\pi}(N,N')+\beta \partial_N G_{\phi\pi}(N,N')\, , \\
\partial^2_N G_{\phi\phi}(N,N')=& \alpha G_{\phi\phi}(N,N') +  \beta \partial_N G_{\phi\phi}(N,N') \, .
\eea 
Upon replacing in \cref{eq:dphiIR:dN:lin:app} one thus finds
\bea 
\label{eq:dphiIR:dN:lin:app:2}
&\frac{\dd\delta\cgmomentum}{\dd N}\left(\bm{x},\bm{y},N\right) = 
\xi_{\momentum}\left(\bm{y},N'\right)-\xi_{\momentum}\left(\bm{x},N'\right)\\
 & + \alpha \int^N_{\Ni}\mathrm{d}N'\left\lbrace  G_{\phi\pi}(N,N') \left[\xi_\field\left(\bm{y},N'\right)-\xi_\field\left(\bm{x},N'\right)\right]
+ G_{\phi\phi}(N,N')\left[\xi_{\momentum}\left(\bm{y},N'\right)-\xi_{\momentum}\left(\bm{x},N'\right)\right]\right\rbrace \\
 & + \beta \int^N_{\Ni}\mathrm{d}N'\left\lbrace \partial_N G_{\phi\pi}(N,N') \left[\xi_\field\left(\bm{y},N'\right)-\xi_\field\left(\bm{x},N'\right)\right]
+\partial_N G_{\phi\phi}(N,N')\left[\xi_{\momentum}\left(\bm{y},N'\right)-\xi_{\momentum}\left(\bm{x},N'\right)\right]\right\rbrace\, .
\eea 
In the second line, one recognises $\delta\cgfield(\bm{x},\bm{y},N)$ given in \cref{eq:general solution field difference:app}, while in the third line, one recognises $\delta\cgmomentum(\bm{x},\bm{y},N)$ given in \cref{eq:deltapi:lin:app}. We have thus shown that
\bea 
\label{eq:dphiIR:dN:lin:app:3}
&\frac{\dd\delta\cgmomentum}{\dd N}\left(\bm{x},\bm{y},N\right) = 
\xi_{\momentum}\left(\bm{y},N\right)-\xi_{\momentum}\left(\bm{x},N\right)
+ \alpha \delta\cgfield\left(\bm{x},\bm{y},N\right)
 + \beta \delta\cgmomentum\left(\bm{x},\bm{y},N\right)\, ,
\eea 
which indeed corresponds to the second entry of \cref{eq:Langevin deltaphi}.

\section{Field moments in the linear potential}
\label{app:moments:linear}

In this appendix, we complement the calculation performed in \cref{eq:deltaphi^2:lin} by considering the two other field moments, namely $\langle \delta\field \delta\momentum\rangle$ and $\langle (\delta\momentum)^2\rangle$, in the linear potential studied in \cref{subsec:Linear_potential}.

Let us start by $\langle \delta\field \delta\momentum\rangle$. Using \cref{eq:sol:lin:SR}, one has
\bea 
\langle \delta\field \delta\momentum\rangle =&
\int_{N_\uin}^N \dd N_1 \int_{N_\uin}^N \dd N_3 \left\langle \xi_\f(N_1)\left[\xi_{\momentum} (N_3)+\xi_\Delta^{(2)}(N_3)\right] \right\rangle e^{3(N_3-N)}
\\ & \hspace{-2cm} +
\int_{N_\uin}^N\dd N_1\int_{N_\uin}^{N_1}\dd N_2\int_{N_\uin}^N\dd N_3 \left\langle \left[\xi_\Delta^{(2)}(N_2)+\xi_{\momentum}(N_2)\right]\left[\xi_\Delta^{(2)}(N_3)+\xi_{\momentum}(N_3)\right] \right\rangle e^{3(N_2-N_1+N_3-N)}\, .
\eea 
In addition to \cref{eq:linear:noise:correlators}, one needs to compute
\bea 
\label{eq:linear:noise:correlators:2}
\left\langle \xi_{\momentum}(N_1)\xi_\Delta^{(2)}(N_2) \right\rangle =&
-\sigma^2 e^{2(N_1-N_2)}\mathcal{P}_{\phi\pi}\left[k_\sigma(N_1),N_1\right]\theta(N_2-N_1)\, ,\\
\left\langle\xi_\Delta^{(2)}(N_1)\xi_\Delta^{(2)}(N_2) \right\rangle =& \sigma^4\int_{N_\uin}^{\mathrm{min}(N_1,N_2)}\dd N' e^{4N'-2N_1-2N_2}\mathcal{P}_{\phi\phi}\left[k_\sigma(N'),N'\right]\, ,
\eea 
and this leads to 
\bea 
& \langle \delta\field \delta\momentum\rangle =
 \int_{N_\uin}^N \dd N_1 e^{3(N_1-N)} \mathcal{P}_{\phi\pi}\left[k_{\sigma}(N_1),N_1\right] 
\\ &
 -\sigma^2\int_{N_\uin}^N \dd N_1 \int_{N_\uin}^N \dd N_3  e^{2N_1+N_3-3N}\mathcal{P}_{\phi\phi}\left[k_{\sigma}(N_1),N_1\right]\theta(N_3-N_1)
 \\ & 
+\sigma^4 \int_{N_\uin}^N\dd N_1\int_{N_\uin}^{N_1}\dd N_2\int_{N_\uin}^N\dd N_3\int_{N_\uin}^{\mathrm{min}(N_2,N_3)}\dd N' e^{N_2-3N_1+N_3-3N+4N'}\mathcal{P}_{\phi\phi}\left[k_{\sigma}(N'),N'\right]
 \\ & 
-\sigma^2 \int_{N_\uin}^N\dd N_1\int_{N_\uin}^{N_1}\dd N_2\int_{N_\uin}^N\dd N_3 e^{N_2-3N_1+5N_3-3N}\mathcal{P}_{\phi\pi}\left[k_{\sigma}(N_3),N_3\right]\theta(N_2-N_3)
 \\ & 
-\sigma^2 \int_{N_\uin}^N\dd N_1\int_{N_\uin}^{N_1}\dd N_2\int_{N_\uin}^N\dd N_3 e^{5N_2-3N_1+N_3-3N}\mathcal{P}_{\phi\pi}\left[k_{\sigma}(N_2),N_2\right]\theta(N_3-N_2)
 \\ & 
 +\int_{N_\uin}^N\dd N_1 \int_{N_\uin}^{N_1}\dd N_2 e^{3(2N_2-N_1-N)}\mathcal{P}_{\pi\pi}\left[k_{\sigma}(N_2),N_2\right]\, .
\eea 
The power spectra being constant, the above integrals can be readily performed, and one finds
\bea 
 \langle \delta\field \delta\momentum\rangle = &
\frac{\mathcal{P}_{\phi\pi}}{3}\left[1-e^{3\left(N_\uin-N\right)}\right]
-\frac{\sigma^2}{6}\mathcal{P}_{\phi\phi}\left(1-e^{N_\uin-N}\right)^2\left(1+2 e^{N_\uin-N}\right)
\\ &
+\frac{\sigma^4}{72} \mathcal{P}_{\phi\phi}\left(1-e^{N_\uin-N}\right)^4\left(1+2 e^{N_\uin-N}\right)^2
\\ &
-\frac{\sigma^2}{90}\mathcal{P}_{\phi\pi}\left(1-e^{N_\uin-N}\right)^3\left[1+5e^{3\left(N_\uin-N\right)}+6e^{2\left(N_\uin-N\right)}+3e^{N_\uin-N}\right]\\ &
-\frac{\sigma^2}{90}\mathcal{P}_{\phi\pi}\left(1-e^{N_\uin-N}\right)^3\left[4+5e^{3\left(N_\uin-N\right)}+9e^{2\left(N_\uin-N\right)}+12e^{N_\uin-N}\right]
\\ &
+\frac{\mathcal{P}_{\pi\pi}}{18}\left[1-e^{3\left(N_\uin-N\right)}\right]^2\, .
\eea 
By replacing the power spectra by their expressions given in \cref{eq:noise_amplitudes_Linear}, one finally obtains
\bea 
 \langle \delta\field \delta\momentum\rangle = &
\left(\frac{H}{2\pi}\right)^2\left\lbrace 
- \frac{\sigma^2}{2}\left[1-e^{2(N_\uin-N)}\right]
- \frac{\sigma^4}{24}\left(1-e^{N_\uin-N}\right)^3\left(1+3e^{N_\uin-N}\right)
\right. \\ & \left.
+ \frac{\sigma^6}{72}\left(1-e^{N_\uin-N}\right)^4\left(1+2e^{N_\uin-N}\right)^2
\right\rbrace\, .
\eea 
This needs to be compared with the prediction of perturbation theory, namely
\bea
\int_{k_\sigma(N_\uin)}^{k_\sigma(N)} \dd\ln k\, \mathcal{P}_{\phi\pi}\left(k,N\right) = - \left(\frac{H}{2\pi}\right)^2\frac{\sigma^2}{2}\left[1-e^{2(N_\uin-N)}\right]\, ,
\eea
where we have used that $\mathcal{P}_{\phi\pi}=-(k\eta^2)[H/(2\pi)]^2$. At order $\sigma^2$, the two expressions above coincide.

Let us now consider  $\langle (\delta\momentum)^2\rangle$.  From \cref{eq:sol:lin:SR}, one has
\bea 
\label{eq:deltapi:squared:interm}
\langle(\delta\momentum)^2\rangle = \int_{N_\uin}^N \dd N_1\int_{N_\uin}^N \dd N_2 e^{3(N_1+N_2-2N)}
\left\langle
\left[\xi_{\momentum}(N_1)+\xi_\Delta^{(2)}(N_1)\right] 
\left[\xi_{\momentum}(N_2)+\xi_\Delta^{(2)}(N_2)\right] 
\right\rangle\, .
\eea 
Making use of \cref{eq:linear:noise:correlators:2}, and since the power spectra do not depend on time, this reduces to
\bea
\langle(\delta\momentum)^2\rangle =& \frac{\mathcal{P}_{\pi\pi}}{6}\left[1-e^{6(N_\uin-N)}\right]
-\frac{\sigma^2}{15}\mathcal{P}_{\phi\pi}\left[1-6e^{5(N_\uin-N)}+5e^{6(N_\uin-N)}\right]
\\ &
+\frac{\sigma^4}{60}\mathcal{P}_{\phi\phi}\left[1-15e^{4(N_\uin-N)}-10e^{6(N_\uin-N)}+24e^{5(N_\uin-N)}\right]\, .
\eea 
Upon replacing the power spectra by their expressions given in \cref{eq:noise_amplitudes_Linear}, most terms cancel out, and one is left with
\bea
\label{eq:deltapi:squared:linear}
\langle(\delta\momentum)^2\rangle =& \frac{H^2}{16\pi^2}\sigma^4\left[1-e^{4(N_\uin-N)}\right]+\mathcal{O}(\sigma^6)\, .
\eea 
In cosmological perturbation theory, the power spectrum of the momentum fluctuation reads $\mathcal{P}_{\pi\pi}(k,N)=(k\eta)^4 H^2/(4\pi^2)$, which leads to 
\bea
\int_{k_\sigma(N_\uin)}^{k_\sigma(N)} \dd\ln k\, \mathcal{P}_{\pi\pi}\left(k,N\right) =\frac{H^2}{16\pi^2}\sigma^4\left[1-e^{4(N_\uin-N)}\right]
\, .
\eea
This expression coincides with the one found for $\langle(\delta\momentum)^2\rangle$ at order $\sigma^4$.

\bibliographystyle{JHEP}
\bibliography{biblio}

\providecommand{\href}[2]{#2}\begingroup\raggedright\begin{thebibliography}{100}

\bibitem{Hawking:1971ei}
S.~Hawking, \emph{{Gravitationally collapsed objects of very low mass}},
  {\emph{Mon. Not. Roy. Astron. Soc.} {\bfseries 152} (1971) 75}.

\bibitem{Carr:1974nx}
B.J.~Carr and S.W.~Hawking, \emph{{Black holes in the early Universe}},
  {\emph{Mon. Not. Roy. Astron. Soc.} {\bfseries 168} (1974) 399}.

\bibitem{Carr:1975qj}
B.J.~Carr, \emph{{The Primordial black hole mass spectrum}},
  \href{https://doi.org/10.1086/153853}{\emph{Astrophys. J.} {\bfseries 201}
  (1975) 1}.

\bibitem{Carr:2009jm}
B.J.~Carr, K.~Kohri, Y.~Sendouda and J.~Yokoyama, \emph{{New cosmological
  constraints on primordial black holes}},
  \href{https://doi.org/10.1103/PhysRevD.81.104019}{\emph{Phys. Rev. D}
  {\bfseries 81} (2010) 104019}
  [\href{https://arxiv.org/abs/0912.5297}{{\ttfamily 0912.5297}}].

\bibitem{Niikura:2017zjd}
H.~Niikura et~al., \emph{{Microlensing constraints on primordial black holes
  with Subaru/HSC Andromeda observations}},
  \href{https://doi.org/10.1038/s41550-019-0723-1}{\emph{Nature Astron.}
  {\bfseries 3} (2019) 524} [\href{https://arxiv.org/abs/1701.02151}{{\ttfamily
  1701.02151}}].

\bibitem{Katz:2018zrn}
A.~Katz, J.~Kopp, S.~Sibiryakov and W.~Xue, \emph{{Femtolensing by Dark Matter
  Revisited}}, \href{https://doi.org/10.1088/1475-7516/2018/12/005}{\emph{JCAP}
  {\bfseries 12} (2018) 005}
  [\href{https://arxiv.org/abs/1807.11495}{{\ttfamily 1807.11495}}].

\bibitem{Montero-Camacho:2019jte}
P.~Montero-Camacho, X.~Fang, G.~Vasquez, M.~Silva and C.M.~Hirata,
  \emph{{Revisiting constraints on asteroid-mass primordial black holes as dark
  matter candidates}},
  \href{https://doi.org/10.1088/1475-7516/2019/08/031}{\emph{JCAP} {\bfseries
  08} (2019) 031} [\href{https://arxiv.org/abs/1906.05950}{{\ttfamily
  1906.05950}}].

\bibitem{Carr:2020gox}
B.~Carr, K.~Kohri, Y.~Sendouda and J.~Yokoyama, \emph{{Constraints on
  primordial black holes}},
  \href{https://doi.org/10.1088/1361-6633/ac1e31}{\emph{Rept. Prog. Phys.}
  {\bfseries 84} (2021) 116902}
  [\href{https://arxiv.org/abs/2002.12778}{{\ttfamily 2002.12778}}].

\bibitem{1975A&A....38....5M}
P.~Meszaros, \emph{{Primeval black holes and galaxy formation}}, {\emph{Astron.
  Astrophys.} {\bfseries 38} (1975) 5}.

\bibitem{Duechting:2004dk}
N.~Duechting, \emph{{Supermassive black holes from primordial black hole
  seeds}}, \href{https://doi.org/10.1103/PhysRevD.70.064015}{\emph{Phys. Rev.
  D} {\bfseries 70} (2004) 064015}
  [\href{https://arxiv.org/abs/astro-ph/0406260}{{\ttfamily
  astro-ph/0406260}}].

\bibitem{Kawasaki:2012kn}
M.~Kawasaki, A.~Kusenko and T.T.~Yanagida, \emph{{Primordial seeds of
  supermassive black holes}},
  \href{https://doi.org/10.1016/j.physletb.2012.03.056}{\emph{Phys. Lett. B}
  {\bfseries 711} (2012) 1} [\href{https://arxiv.org/abs/1202.3848}{{\ttfamily
  1202.3848}}].

\bibitem{Clesse:2015wea}
S.~Clesse and J.~Garc\'\i{}a-Bellido, \emph{{Massive Primordial Black Holes
  from Hybrid Inflation as Dark Matter and the seeds of Galaxies}},
  \href{https://doi.org/10.1103/PhysRevD.92.023524}{\emph{Phys. Rev. D}
  {\bfseries 92} (2015) 023524}
  [\href{https://arxiv.org/abs/1501.07565}{{\ttfamily 1501.07565}}].

\bibitem{Carr:2018rid}
B.~Carr and J.~Silk, \emph{{Primordial Black Holes as Generators of Cosmic
  Structures}}, \href{https://doi.org/10.1093/mnras/sty1204}{\emph{Mon. Not.
  Roy. Astron. Soc.} {\bfseries 478} (2018) 3756}
  [\href{https://arxiv.org/abs/1801.00672}{{\ttfamily 1801.00672}}].

\bibitem{Liu:2022bvr}
B.~Liu and V.~Bromm, \emph{{Accelerating Early Massive Galaxy Formation with
  Primordial Black Holes}},
  \href{https://doi.org/10.3847/2041-8213/ac927f}{\emph{Astrophys. J. Lett.}
  {\bfseries 937} (2022) L30}
  [\href{https://arxiv.org/abs/2208.13178}{{\ttfamily 2208.13178}}].

\bibitem{Hutsi:2022fzw}
G.~H\"utsi, M.~Raidal, J.~Urrutia, V.~Vaskonen and H.~Veerm\"ae, \emph{{Did
  JWST observe imprints of axion miniclusters or primordial black holes?}},
  \href{https://doi.org/10.1103/PhysRevD.107.043502}{\emph{Phys. Rev. D}
  {\bfseries 107} (2023) 043502}
  [\href{https://arxiv.org/abs/2211.02651}{{\ttfamily 2211.02651}}].

\bibitem{Starobinsky:1982ee}
A.A.~Starobinsky, \emph{{Dynamics of Phase Transition in the New Inflationary
  Universe Scenario and Generation of Perturbations}},
  \href{https://doi.org/10.1016/0370-2693(82)90541-X}{\emph{Phys. Lett. B}
  {\bfseries 117} (1982) 175}.

\bibitem{Starobinsky:1985ibc}
A.A.~Starobinsky, \emph{{Multicomponent de Sitter (Inflationary) Stages and the
  Generation of Perturbations}}, {\emph{JETP Lett.} {\bfseries 42} (1985) 152}.

\bibitem{Sasaki:1995aw}
M.~Sasaki and E.D.~Stewart, \emph{{A General analytic formula for the spectral
  index of the density perturbations produced during inflation}},
  \href{https://doi.org/10.1143/PTP.95.71}{\emph{Prog. Theor. Phys.} {\bfseries
  95} (1996) 71} [\href{https://arxiv.org/abs/astro-ph/9507001}{{\ttfamily
  astro-ph/9507001}}].

\bibitem{Sasaki:1998ug}
M.~Sasaki and T.~Tanaka, \emph{{Superhorizon scale dynamics of multiscalar
  inflation}}, \href{https://doi.org/10.1143/PTP.99.763}{\emph{Prog. Theor.
  Phys.} {\bfseries 99} (1998) 763}
  [\href{https://arxiv.org/abs/gr-qc/9801017}{{\ttfamily gr-qc/9801017}}].

\bibitem{Wands:2000dp}
D.~Wands, K.A.~Malik, D.H.~Lyth and A.R.~Liddle, \emph{{A New approach to the
  evolution of cosmological perturbations on large scales}},
  \href{https://doi.org/10.1103/PhysRevD.62.043527}{\emph{Phys. Rev. D}
  {\bfseries 62} (2000) 043527}
  [\href{https://arxiv.org/abs/astro-ph/0003278}{{\ttfamily
  astro-ph/0003278}}].

\bibitem{Salopek:1990jq}
D.S.~Salopek and J.R.~Bond, \emph{{Nonlinear evolution of long wavelength
  metric fluctuations in inflationary models}},
  \href{https://doi.org/10.1103/PhysRevD.42.3936}{\emph{Phys. Rev.} {\bfseries
  D42} (1990) 3936}.

\bibitem{Rigopoulos:2003ak}
G.I.~Rigopoulos and E.P.S.~Shellard, \emph{{The separate universe approach and
  the evolution of nonlinear superhorizon cosmological perturbations}},
  \href{https://doi.org/10.1103/PhysRevD.68.123518}{\emph{Phys. Rev. D}
  {\bfseries 68} (2003) 123518}
  [\href{https://arxiv.org/abs/astro-ph/0306620}{{\ttfamily
  astro-ph/0306620}}].

\bibitem{Tanaka:2007gh}
Y.~Tanaka and M.~Sasaki, \emph{{Gradient expansion approach to nonlinear
  superhorizon perturbations. II. A Single scalar field}},
  \href{https://doi.org/10.1143/PTP.118.455}{\emph{Prog. Theor. Phys.}
  {\bfseries 118} (2007) 455}
  [\href{https://arxiv.org/abs/0706.0678}{{\ttfamily 0706.0678}}].

\bibitem{Lyth:2003im}
D.H.~Lyth and D.~Wands, \emph{{Conserved cosmological perturbations}},
  \href{https://doi.org/10.1103/PhysRevD.68.103515}{\emph{Phys. Rev. D}
  {\bfseries 68} (2003) 103515}
  [\href{https://arxiv.org/abs/astro-ph/0306498}{{\ttfamily
  astro-ph/0306498}}].

\bibitem{Lyth:2004gb}
D.H.~Lyth, K.A.~Malik and M.~Sasaki, \emph{{A General proof of the conservation
  of the curvature perturbation}},
  \href{https://doi.org/10.1088/1475-7516/2005/05/004}{\emph{JCAP} {\bfseries
  05} (2005) 004} [\href{https://arxiv.org/abs/astro-ph/0411220}{{\ttfamily
  astro-ph/0411220}}].

\bibitem{Starobinsky:1986fx}
A.A.~Starobinsky, \emph{{Stochastic De Sitter (inflationary) stage in the early
  universe}}, \href{https://doi.org/10.1007/3-540-16452-9_6}{\emph{Lect. Notes
  Phys.} {\bfseries 246} (1986) 107}.

\bibitem{Enqvist:2008kt}
K.~Enqvist, S.~Nurmi, D.~Podolsky and G.~Rigopoulos, \emph{{On the divergences
  of inflationary superhorizon perturbations}},
  \href{https://doi.org/10.1088/1475-7516/2008/04/025}{\emph{JCAP} {\bfseries
  04} (2008) 025} [\href{https://arxiv.org/abs/0802.0395}{{\ttfamily
  0802.0395}}].

\bibitem{Fujita:2013cna}
T.~Fujita, M.~Kawasaki, Y.~Tada and T.~Takesako, \emph{{A new algorithm for
  calculating the curvature perturbations in stochastic inflation}},
  \href{https://doi.org/10.1088/1475-7516/2013/12/036}{\emph{JCAP} {\bfseries
  12} (2013) 036} [\href{https://arxiv.org/abs/1308.4754}{{\ttfamily
  1308.4754}}].

\bibitem{Fujita:2014tja}
T.~Fujita, M.~Kawasaki and Y.~Tada, \emph{{Non-perturbative approach for
  curvature perturbations in stochastic $\delta N$ formalism}},
  \href{https://doi.org/10.1088/1475-7516/2014/10/030}{\emph{JCAP} {\bfseries
  10} (2014) 030} [\href{https://arxiv.org/abs/1405.2187}{{\ttfamily
  1405.2187}}].

\bibitem{Vennin:2015hra}
V.~Vennin and A.A.~Starobinsky, \emph{{Correlation Functions in Stochastic
  Inflation}}, \href{https://doi.org/10.1140/epjc/s10052-015-3643-y}{\emph{Eur.
  Phys. J. C} {\bfseries 75} (2015) 413}
  [\href{https://arxiv.org/abs/1506.04732}{{\ttfamily 1506.04732}}].

\bibitem{Pattison:2017mbe}
C.~Pattison, V.~Vennin, H.~Assadullahi and D.~Wands, \emph{{Quantum diffusion
  during inflation and primordial black holes}},
  \href{https://doi.org/10.1088/1475-7516/2017/10/046}{\emph{JCAP} {\bfseries
  10} (2017) 046} [\href{https://arxiv.org/abs/1707.00537}{{\ttfamily
  1707.00537}}].

\bibitem{Biagetti:2018pjj}
M.~Biagetti, G.~Franciolini, A.~Kehagias and A.~Riotto, \emph{{Primordial Black
  Holes from Inflation and Quantum Diffusion}},
  \href{https://doi.org/10.1088/1475-7516/2018/07/032}{\emph{JCAP} {\bfseries
  07} (2018) 032} [\href{https://arxiv.org/abs/1804.07124}{{\ttfamily
  1804.07124}}].

\bibitem{Panagopoulos:2019ail}
G.~Panagopoulos and E.~Silverstein, \emph{{Primordial Black Holes from
  non-Gaussian tails}},  \href{https://arxiv.org/abs/1906.02827}{{\ttfamily
  1906.02827}}.

\bibitem{Figueroa:2020jkf}
D.G.~Figueroa, S.~Raatikainen, S.~Rasanen and E.~Tomberg, \emph{{Non-Gaussian
  Tail of the Curvature Perturbation in Stochastic Ultraslow-Roll Inflation:
  Implications for Primordial Black Hole Production}},
  \href{https://doi.org/10.1103/PhysRevLett.127.101302}{\emph{Phys. Rev. Lett.}
  {\bfseries 127} (2021) 101302}
  [\href{https://arxiv.org/abs/2012.06551}{{\ttfamily 2012.06551}}].

\bibitem{Pattison:2021oen}
C.~Pattison, V.~Vennin, D.~Wands and H.~Assadullahi, \emph{{Ultra-slow-roll
  inflation with quantum diffusion}},
  \href{https://doi.org/10.1088/1475-7516/2021/04/080}{\emph{JCAP} {\bfseries
  04} (2021) 080} [\href{https://arxiv.org/abs/2101.05741}{{\ttfamily
  2101.05741}}].

\bibitem{Ezquiaga:2019ftu}
J.M.~Ezquiaga, J.~Garc\'\i{}a-Bellido and V.~Vennin, \emph{{The exponential
  tail of inflationary fluctuations: consequences for primordial black holes}},
  \href{https://doi.org/10.1088/1475-7516/2020/03/029}{\emph{JCAP} {\bfseries
  03} (2020) 029} [\href{https://arxiv.org/abs/1912.05399}{{\ttfamily
  1912.05399}}].

\bibitem{Vennin:2020kng}
V.~Vennin, \emph{{Stochastic inflation and primordial black holes}},
  {\emph{Habilitation thesis} (2020) }
  [\href{https://arxiv.org/abs/2009.08715}{{\ttfamily 2009.08715}}].

\bibitem{Ando:2020fjm}
K.~Ando and V.~Vennin, \emph{{Power spectrum in stochastic inflation}},
  \href{https://doi.org/10.1088/1475-7516/2021/04/057}{\emph{JCAP} {\bfseries
  04} (2021) 057} [\href{https://arxiv.org/abs/2012.02031}{{\ttfamily
  2012.02031}}].

\bibitem{Tada:2021zzj}
Y.~Tada and V.~Vennin, \emph{{Statistics of coarse-grained cosmological fields
  in stochastic inflation}},
  \href{https://doi.org/10.1088/1475-7516/2022/02/021}{\emph{JCAP} {\bfseries
  02} (2022) 021} [\href{https://arxiv.org/abs/2111.15280}{{\ttfamily
  2111.15280}}].

\bibitem{Achucarro:2021pdh}
A.~Achucarro, S.~Cespedes, A.-C.~Davis and G.A.~Palma, \emph{{The hand-made
  tail: non-perturbative tails from multifield inflation}},
  \href{https://doi.org/10.1007/JHEP05(2022)052}{\emph{JHEP} {\bfseries 05}
  (2022) 052} [\href{https://arxiv.org/abs/2112.14712}{{\ttfamily
  2112.14712}}].

\bibitem{Kitajima:2021fpq}
N.~Kitajima, Y.~Tada, S.~Yokoyama and C.-M.~Yoo, \emph{{Primordial black holes
  in peak theory with a non-Gaussian tail}},
  \href{https://doi.org/10.1088/1475-7516/2021/10/053}{\emph{JCAP} {\bfseries
  10} (2021) 053} [\href{https://arxiv.org/abs/2109.00791}{{\ttfamily
  2109.00791}}].

\bibitem{Hooshangi:2021ubn}
S.~Hooshangi, M.H.~Namjoo and M.~Noorbala, \emph{{Rare events are
  nonperturbative: Primordial black holes from heavy-tailed distributions}},
  \href{https://doi.org/10.1016/j.physletb.2022.137400}{\emph{Phys. Lett. B}
  {\bfseries 834} (2022) 137400}
  [\href{https://arxiv.org/abs/2112.04520}{{\ttfamily 2112.04520}}].

\bibitem{Gow:2022jfb}
A.D.~Gow, H.~Assadullahi, J.H.P.~Jackson, K.~Koyama, V.~Vennin and D.~Wands,
  \emph{{Non-perturbative non-Gaussianity and primordial black holes}},
  \href{https://doi.org/10.1209/0295-5075/acd417}{\emph{EPL} {\bfseries 142}
  (2023) 49001} [\href{https://arxiv.org/abs/2211.08348}{{\ttfamily
  2211.08348}}].

\bibitem{Cai:2022erk}
Y.-F.~Cai, X.-H.~Ma, M.~Sasaki, D.-G.~Wang and Z.~Zhou, \emph{{Highly
  non-Gaussian tails and primordial black holes from single-field inflation}},
  \href{https://doi.org/10.1088/1475-7516/2022/12/034}{\emph{JCAP} {\bfseries
  12} (2022) 034} [\href{https://arxiv.org/abs/2207.11910}{{\ttfamily
  2207.11910}}].

\bibitem{Animali:2022otk}
C.~Animali and V.~Vennin, \emph{{Primordial black holes from stochastic
  tunnelling}},  \href{https://arxiv.org/abs/2210.03812}{{\ttfamily
  2210.03812}}.

\bibitem{Jackson:2022unc}
J.H.P.~Jackson, H.~Assadullahi, K.~Koyama, V.~Vennin and D.~Wands,
  \emph{{Numerical simulations of stochastic inflation using importance
  sampling}}, \href{https://doi.org/10.1088/1475-7516/2022/10/067}{\emph{JCAP}
  {\bfseries 10} (2022) 067}
  [\href{https://arxiv.org/abs/2206.11234}{{\ttfamily 2206.11234}}].

\bibitem{Ezquiaga:2022qpw}
J.M.~Ezquiaga, J.~Garc{\'\i}a-Bellido and V.~Vennin, \emph{{Massive Galaxy
  Clusters Like El Gordo Hint at Primordial Quantum Diffusion}},
  \href{https://doi.org/10.1103/PhysRevLett.130.121003}{\emph{Phys. Rev. Lett.}
  {\bfseries 130} (2023) 121003}
  [\href{https://arxiv.org/abs/2207.06317}{{\ttfamily 2207.06317}}].

\bibitem{Rigopoulos:2022gso}
G.~Rigopoulos and A.~Wilkins, \emph{{Computing first-passage times with the
  functional renormalisation group}},
  \href{https://doi.org/10.1088/1475-7516/2023/04/046}{\emph{JCAP} {\bfseries
  04} (2023) 046} [\href{https://arxiv.org/abs/2211.09649}{{\ttfamily
  2211.09649}}].

\bibitem{Briaud:2023eae}
V.~Briaud and V.~Vennin, \emph{{Uphill inflation}},
  \href{https://doi.org/10.1088/1475-7516/2023/06/029}{\emph{JCAP} {\bfseries
  06} (2023) 029} [\href{https://arxiv.org/abs/2301.09336}{{\ttfamily
  2301.09336}}].

\bibitem{Hooshangi:2023kss}
S.~Hooshangi, M.H.~Namjoo and M.~Noorbala, \emph{{Tail diversity from
  inflation}}, \href{https://doi.org/10.1088/1475-7516/2023/09/023}{\emph{JCAP}
  {\bfseries 09} (2023) 023}
  [\href{https://arxiv.org/abs/2305.19257}{{\ttfamily 2305.19257}}].

\bibitem{Kawaguchi:2023mgk}
R.~Kawaguchi, T.~Fujita and M.~Sasaki, \emph{{Highly asymmetric probability
  distribution from a finite-width upward step during inflation}},
  \href{https://doi.org/10.1088/1475-7516/2023/11/021}{\emph{JCAP} {\bfseries
  11} (2023) 021} [\href{https://arxiv.org/abs/2305.18140}{{\ttfamily
  2305.18140}}].

\bibitem{Raatikainen:2023bzk}
S.~Raatikainen, S.~R{\"a}s{\"a}nen and E.~Tomberg, \emph{{Primordial Black Hole
  Compaction Function from Stochastic Fluctuations in Ultraslow-Roll
  Inflation}},
  \href{https://doi.org/10.1103/PhysRevLett.133.121403}{\emph{Phys. Rev. Lett.}
  {\bfseries 133} (2024) 121403}
  [\href{https://arxiv.org/abs/2312.12911}{{\ttfamily 2312.12911}}].

\bibitem{Launay:2024qsm}
Y.L.~Launay, G.I.~Rigopoulos and E.P.S.~Shellard, \emph{{Stochastic inflation
  in general relativity}},
  \href{https://doi.org/10.1103/PhysRevD.109.123523}{\emph{Phys. Rev. D}
  {\bfseries 109} (2024) 123523}
  [\href{https://arxiv.org/abs/2401.08530}{{\ttfamily 2401.08530}}].

\bibitem{Vennin:2024yzl}
V.~Vennin and D.~Wands, \emph{{Quantum Diffusion and~Large Primordial
  Perturbations from~Inflation}},  (2025),
  \href{https://doi.org/10.1007/978-981-97-8887-3\_8}{DOI}
  [\href{https://arxiv.org/abs/2402.12672}{{\ttfamily 2402.12672}}].

\bibitem{Inui:2024sce}
R.~Inui, H.~Motohashi, S.~Pi, Y.~Tada and S.~Yokoyama, \emph{{Constant roll and
  non-Gaussian tail in light of logarithmic duality}},
  \href{https://doi.org/10.1088/1475-7516/2025/02/042}{\emph{JCAP} {\bfseries
  02} (2025) 042} [\href{https://arxiv.org/abs/2409.13500}{{\ttfamily
  2409.13500}}].

\bibitem{Animali:2024jiz}
C.~Animali and V.~Vennin, \emph{{Clustering of primordial black holes from
  quantum diffusion during inflation}},
  \href{https://doi.org/10.1088/1475-7516/2024/08/026}{\emph{JCAP} {\bfseries
  08} (2024) 026} [\href{https://arxiv.org/abs/2402.08642}{{\ttfamily
  2402.08642}}].

\bibitem{Mizuguchi:2024kbl}
Y.~Mizuguchi, T.~Murata and Y.~Tada, \emph{{STOLAS: STOchastic LAttice
  Simulation of cosmic inflation}},
  \href{https://doi.org/10.1088/1475-7516/2024/12/050}{\emph{JCAP} {\bfseries
  12} (2024) 050} [\href{https://arxiv.org/abs/2405.10692}{{\ttfamily
  2405.10692}}].

\bibitem{Choudhury:2025kxg}
S.~Choudhury, \emph{{Stochastic origin of primordial fluctuations in the Sky}},
   \href{https://arxiv.org/abs/2503.17635}{{\ttfamily 2503.17635}}.

\bibitem{Murata:2025onc}
T.~Murata and Y.~Tada, \emph{{Stochastic-tail of the curvature perturbation in
  hybrid inflation}},  \href{https://arxiv.org/abs/2507.22439}{{\ttfamily
  2507.22439}}.

\bibitem{Kuroda:2025coa}
T.~Kuroda, A.~Naruko, V.~Vennin and M.~Yamaguchi, \emph{{Primordial black holes
  from a curvaton: the role of bimodal distributions}},
  \href{https://doi.org/10.1088/1475-7516/2025/07/052}{\emph{JCAP} {\bfseries
  07} (2025) 052} [\href{https://arxiv.org/abs/2504.09548}{{\ttfamily
  2504.09548}}].

\bibitem{Animali:2025pyf}
C.~Animali, P.~Auclair, B.~Blachier and V.~Vennin, \emph{{Harvesting primordial
  black holes from stochastic trees with FOREST}},
  \href{https://doi.org/10.1088/1475-7516/2025/05/019}{\emph{JCAP} {\bfseries
  05} (2025) 019} [\href{https://arxiv.org/abs/2501.05371}{{\ttfamily
  2501.05371}}].

\bibitem{Cruces:2025typ}
D.~Cruces, S.~Pi and M.~Sasaki, \emph{{$\delta n$ formalism: A new formulation
  for the probability density of the curvature perturbation}},
  \href{https://arxiv.org/abs/2505.24590}{{\ttfamily 2505.24590}}.

\bibitem{Miyamoto:2025qqm}
K.~Miyamoto and Y.~Tada, \emph{{Calculating the power spectrum in stochastic
  inflation by Monte Carlo simulation and least squares curve fitting}},
  \href{https://arxiv.org/abs/2508.17654}{{\ttfamily 2508.17654}}.

\bibitem{Pattison:2019hef}
C.~Pattison, V.~Vennin, H.~Assadullahi and D.~Wands, \emph{{Stochastic
  inflation beyond slow roll}},
  \href{https://doi.org/10.1088/1475-7516/2019/07/031}{\emph{JCAP} {\bfseries
  07} (2019) 031} [\href{https://arxiv.org/abs/1905.06300}{{\ttfamily
  1905.06300}}].

\bibitem{Firouzjahi:2020jrj}
H.~Firouzjahi, A.~Nassiri-Rad and M.~Noorbala, \emph{{Stochastic nonattractor
  inflation}}, \href{https://doi.org/10.1103/PhysRevD.102.123504}{\emph{Phys.
  Rev. D} {\bfseries 102} (2020) 123504}
  [\href{https://arxiv.org/abs/2009.04680}{{\ttfamily 2009.04680}}].

\bibitem{Mishra:2023lhe}
S.S.~Mishra, E.J.~Copeland and A.M.~Green, \emph{{Primordial black holes and
  stochastic inflation beyond slow roll. Part I. Noise matrix elements}},
  \href{https://doi.org/10.1088/1475-7516/2023/09/005}{\emph{JCAP} {\bfseries
  09} (2023) 005} [\href{https://arxiv.org/abs/2303.17375}{{\ttfamily
  2303.17375}}].

\bibitem{Jackson:2023obv}
J.H.P.~Jackson, H.~Assadullahi, A.D.~Gow, K.~Koyama, V.~Vennin and D.~Wands,
  \emph{{The separate-universe approach and sudden transitions during
  inflation}}, \href{https://doi.org/10.1088/1475-7516/2024/05/053}{\emph{JCAP}
  {\bfseries 05} (2024) 053}
  [\href{https://arxiv.org/abs/2311.03281}{{\ttfamily 2311.03281}}].

\bibitem{Artigas:2024ajh}
D.~Artigas, S.~Pi and T.~Tanaka, \emph{{Extended {\ensuremath{\delta}}N
  Formalism: Nonspatially Flat Separate-Universe Approach}},
  \href{https://doi.org/10.1103/PhysRevLett.134.221001}{\emph{Phys. Rev. Lett.}
  {\bfseries 134} (2025) 221001}
  [\href{https://arxiv.org/abs/2408.09964}{{\ttfamily 2408.09964}}].

\bibitem{Raveendran:2025pnz}
R.N.~Raveendran, \emph{{On the validity of separate-universe approach in
  transient ultra-slow-roll inflation}},
  \href{https://arxiv.org/abs/2506.23571}{{\ttfamily 2506.23571}}.

\bibitem{Garcia-Bellido:2017mdw}
J.~Garcia-Bellido and E.~Ruiz~Morales, \emph{{Primordial black holes from
  single field models of inflation}},
  \href{https://doi.org/10.1016/j.dark.2017.09.007}{\emph{Phys. Dark Univ.}
  {\bfseries 18} (2017) 47} [\href{https://arxiv.org/abs/1702.03901}{{\ttfamily
  1702.03901}}].

\bibitem{Ballesteros:2017fsr}
G.~Ballesteros and M.~Taoso, \emph{{Primordial black hole dark matter from
  single field inflation}},
  \href{https://doi.org/10.1103/PhysRevD.97.023501}{\emph{Phys. Rev. D}
  {\bfseries 97} (2018) 023501}
  [\href{https://arxiv.org/abs/1709.05565}{{\ttfamily 1709.05565}}].

\bibitem{Karam:2022nym}
A.~Karam, N.~Koivunen, E.~Tomberg, V.~Vaskonen and H.~Veerm{\"a}e,
  \emph{{Anatomy of single-field inflationary models for primordial black
  holes}}, \href{https://doi.org/10.1088/1475-7516/2023/03/013}{\emph{JCAP}
  {\bfseries 03} (2023) 013}
  [\href{https://arxiv.org/abs/2205.13540}{{\ttfamily 2205.13540}}].

\bibitem{Mukhanov:1981xt}
V.F.~Mukhanov and G.V.~Chibisov, \emph{{Quantum Fluctuations and a Nonsingular
  Universe}}, {\emph{JETP Lett.} {\bfseries 33} (1981) 532}.

\bibitem{Sasaki:1986hm}
M.~Sasaki, \emph{{Large Scale Quantum Fluctuations in the Inflationary
  Universe}}, \href{https://doi.org/10.1143/PTP.76.1036}{\emph{Prog. Theor.
  Phys.} {\bfseries 76} (1986) 1036}.

\bibitem{Bunch:1978yq}
T.S.~Bunch and P.C.W.~Davies, \emph{{Quantum Field Theory in de Sitter Space:
  Renormalization by Point Splitting}},
  \href{https://doi.org/10.1098/rspa.1978.0060}{\emph{Proc. Roy. Soc. Lond. A}
  {\bfseries 360} (1978) 117}.

\bibitem{Artigas:2021zdk}
D.~Artigas, J.~Grain and V.~Vennin, \emph{{Hamiltonian formalism for
  cosmological perturbations: the~separate-universe approach}},
  \href{https://doi.org/10.1088/1475-7516/2022/02/001}{\emph{JCAP} {\bfseries
  02} (2022) 001} [\href{https://arxiv.org/abs/2110.11720}{{\ttfamily
  2110.11720}}].

\bibitem{Artigas:2023kyo}
D.~Artigas, J.~Grain and V.~Vennin, \emph{{Hamiltonian formalism for
  cosmological perturbations: fixing the gauge}},
  \href{https://doi.org/10.1088/1475-7516/2025/01/083}{\emph{JCAP} {\bfseries
  01} (2025) 083} [\href{https://arxiv.org/abs/2309.17184}{{\ttfamily
  2309.17184}}].

\bibitem{Nambu:1987ef}
Y.~Nambu and M.~Sasaki, \emph{{Stochastic Stage of an Inflationary Universe
  Model}}, \href{https://doi.org/10.1016/0370-2693(88)90974-4}{\emph{Phys.
  Lett. B} {\bfseries 205} (1988) 441}.

\bibitem{Nambu:1988je}
Y.~Nambu and M.~Sasaki, \emph{{Stochastic approach to chaotic inflation and the
  distribution of universes}},
  \href{https://doi.org/10.1016/0370-2693(89)90385-7}{\emph{Phys. Lett. B}
  {\bfseries 219} (1989) 240}.

\bibitem{Kandrup:1988sc}
H.E.~Kandrup, \emph{{STOCHASTIC INFLATION AS A TIME DEPENDENT RANDOM WALK}},
  \href{https://doi.org/10.1103/PhysRevD.39.2245}{\emph{Phys. Rev. D}
  {\bfseries 39} (1989) 2245}.

\bibitem{Nakao:1988yi}
K.-i.~Nakao, Y.~Nambu and M.~Sasaki, \emph{{Stochastic Dynamics of New
  Inflation}}, \href{https://doi.org/10.1143/PTP.80.1041}{\emph{Prog. Theor.
  Phys.} {\bfseries 80} (1988) 1041}.

\bibitem{Nambu:1989uf}
Y.~Nambu, \emph{{Stochastic Dynamics of an Inflationary Model and Initial
  Distribution of Universes}},
  \href{https://doi.org/10.1143/PTP.81.1037}{\emph{Prog. Theor. Phys.}
  {\bfseries 81} (1989) 1037}.

\bibitem{Mollerach:1990zf}
S.~Mollerach, S.~Matarrese, A.~Ortolan and F.~Lucchin, \emph{{Stochastic
  inflation in a simple two field model}},
  \href{https://doi.org/10.1103/PhysRevD.44.1670}{\emph{Phys. Rev. D}
  {\bfseries 44} (1991) 1670}.

\bibitem{Linde:1993xx}
A.D.~Linde, D.A.~Linde and A.~Mezhlumian, \emph{{From the Big Bang theory to
  the theory of a stationary universe}},
  \href{https://doi.org/10.1103/PhysRevD.49.1783}{\emph{Phys. Rev. D}
  {\bfseries 49} (1994) 1783}
  [\href{https://arxiv.org/abs/gr-qc/9306035}{{\ttfamily gr-qc/9306035}}].

\bibitem{Starobinsky_1994}
A.A.~Starobinsky and J.~Yokoyama, \emph{Equilibrium state of a self-interacting
  scalar field in the de sitter background},
  \href{https://doi.org/10.1103/physrevd.50.6357}{\emph{Physical Review D}
  {\bfseries 50} (1994) 6357–6368}.

\bibitem{Finelli:2008zg}
F.~Finelli, G.~Marozzi, A.~Starobinsky, G.~Vacca and G.~Venturi,
  \emph{{Generation of fluctuations during inflation: Comparison of stochastic
  and field-theoretic approaches}},
  \href{https://doi.org/10.1103/PhysRevD.79.044007}{\emph{Phys. Rev. D}
  {\bfseries 79} (2009) 044007}
  [\href{https://arxiv.org/abs/0808.1786}{{\ttfamily 0808.1786}}].

\bibitem{Finelli:2010sh}
F.~Finelli, G.~Marozzi, A.A.~Starobinsky, G.P.~Vacca and G.~Venturi,
  \emph{{Stochastic growth of quantum fluctuations during slow-roll
  inflation}}, \href{https://doi.org/10.1103/PhysRevD.82.064020}{\emph{Phys.
  Rev. D} {\bfseries 82} (2010) 064020}
  [\href{https://arxiv.org/abs/1003.1327}{{\ttfamily 1003.1327}}].

\bibitem{Grain:2017dqa}
J.~Grain and V.~Vennin, \emph{{Stochastic inflation in phase space: Is slow
  roll a stochastic attractor?}},
  \href{https://doi.org/10.1088/1475-7516/2017/05/045}{\emph{JCAP} {\bfseries
  05} (2017) 045} [\href{https://arxiv.org/abs/1703.00447}{{\ttfamily
  1703.00447}}].

\bibitem{Cheung:2007st}
C.~Cheung, P.~Creminelli, A.L.~Fitzpatrick, J.~Kaplan and L.~Senatore,
  \emph{{The Effective Field Theory of Inflation}},
  \href{https://doi.org/10.1088/1126-6708/2008/03/014}{\emph{JHEP} {\bfseries
  03} (2008) 014} [\href{https://arxiv.org/abs/0709.0293}{{\ttfamily
  0709.0293}}].

\bibitem{Leach:2001zf}
S.M.~Leach, M.~Sasaki, D.~Wands and A.R.~Liddle, \emph{{Enhancement of
  superhorizon scale inflationary curvature perturbations}},
  \href{https://doi.org/10.1103/PhysRevD.64.023512}{\emph{Phys. Rev. D}
  {\bfseries 64} (2001) 023512}
  [\href{https://arxiv.org/abs/astro-ph/0101406}{{\ttfamily
  astro-ph/0101406}}].

\bibitem{Lifshitz:1960}
E.M.~Lifshitz and I.M.~Khalatnikov, \emph{{About singularities of cosmological
  solutions of the gravitational equations. I}}, {\emph{ZhETF} {\bfseries 39}
  (1960) 149}.

\bibitem{Comer:1994np}
G.~Comer, N.~Deruelle, D.~Langlois and J.~Parry, \emph{{Growth or decay of
  cosmological inhomogeneities as a function of their equation of state}},
  \href{https://doi.org/10.1103/PhysRevD.49.2759}{\emph{Phys. Rev. D}
  {\bfseries 49} (1994) 2759}.

\bibitem{Lyth:2005fi}
D.H.~Lyth and Y.~Rodriguez, \emph{{The Inflationary prediction for primordial
  non-Gaussianity}},
  \href{https://doi.org/10.1103/PhysRevLett.95.121302}{\emph{Phys. Rev. Lett.}
  {\bfseries 95} (2005) 121302}
  [\href{https://arxiv.org/abs/astro-ph/0504045}{{\ttfamily
  astro-ph/0504045}}].

\bibitem{Cai:2018dkf}
Y.-F.~Cai, X.~Chen, M.H.~Namjoo, M.~Sasaki, D.-G.~Wang and Z.~Wang,
  \emph{{Revisiting non-Gaussianity from non-attractor inflation models}},
  \href{https://doi.org/10.1088/1475-7516/2018/05/012}{\emph{JCAP} {\bfseries
  05} (2018) 012} [\href{https://arxiv.org/abs/1712.09998}{{\ttfamily
  1712.09998}}].

\bibitem{Pi:2022ysn}
S.~Pi and M.~Sasaki, \emph{{Logarithmic Duality of the Curvature
  Perturbation}},  \href{https://arxiv.org/abs/2211.13932}{{\ttfamily
  2211.13932}}.

\bibitem{Starobinsky:1992ts}
A.A.~Starobinsky, \emph{{Spectrum of adiabatic perturbations in the universe
  when there are singularities in the inflation potential}}, {\emph{JETP Lett.}
  {\bfseries 55} (1992) 489}.

\bibitem{Artigas:2025nbm}
D.~Artigas, E.~Frion, T.~Miranda, V.~Vennin and D.~Wands, \emph{{On the
  Hamilton-Jacobi approach to inflation beyond slow roll}},
  \href{https://doi.org/10.1088/1475-7516/2025/08/032}{\emph{JCAP} {\bfseries
  08} (2025) 032} [\href{https://arxiv.org/abs/2504.05937}{{\ttfamily
  2504.05937}}].

\bibitem{Van_Kampen1989-gg}
N.G.~van Kampen, \emph{Langevin-like equation with colored noise}, {\emph{J.
  Stat. Phys.} {\bfseries 54} (1989) 1289}.

\bibitem{gardiner2004handbook}
C.W.~Gardiner, \emph{Handbook of stochastic methods for physics, chemistry and
  the natural sciences}, vol.~13 of \emph{Springer Series in Synergetics},
  Springer-Verlag, third~ed. (2004).

\bibitem{Starobinsky:1986fxa}
A.A.~Starobinsky, \emph{{Multicomponent de Sitter (Inflationary) Stages and the
  Generation of Perturbations}}, {\emph{JETP Lett.} {\bfseries 42} (1985) 152}.

\bibitem{Sasaki1996}
M.~Sasaki and E.D.~Stewart, \emph{{A General Analytic Formula for the Spectral
  Index of the Density Perturbations Produced during Inflation}},
  \href{https://doi.org/10.1143/PTP.95.71}{\emph{Progress of Theoretical
  Physics} {\bfseries 95} (1996) 71}
  [\href{https://arxiv.org/abs/astro-ph/9507001}{{\ttfamily
  astro-ph/9507001}}].

\bibitem{Clough:2016ymm}
K.~Clough, E.A.~Lim, B.S.~DiNunno, W.~Fischler, R.~Flauger and S.~Paban,
  \emph{{Robustness of Inflation to Inhomogeneous Initial Conditions}},
  \href{https://doi.org/10.1088/1475-7516/2017/09/025}{\emph{JCAP} {\bfseries
  09} (2017) 025} [\href{https://arxiv.org/abs/1608.04408}{{\ttfamily
  1608.04408}}].

\bibitem{Caravano:2024tlp}
A.~Caravano, K.~Inomata and S.~Renaux-Petel, \emph{{Inflationary Butterfly
  Effect: Nonperturbative Dynamics from Small-Scale Features}},
  \href{https://doi.org/10.1103/PhysRevLett.133.151001}{\emph{Phys. Rev. Lett.}
  {\bfseries 133} (2024) 151001}
  [\href{https://arxiv.org/abs/2403.12811}{{\ttfamily 2403.12811}}].

\bibitem{Caravano:2024moy}
A.~Caravano, G.~Franciolini and S.~Renaux-Petel, \emph{{Ultraslow-roll
  inflation on the lattice: Backreaction and nonlinear effects}},
  \href{https://doi.org/10.1103/PhysRevD.111.063518}{\emph{Phys. Rev. D}
  {\bfseries 111} (2025) 063518}
  [\href{https://arxiv.org/abs/2410.23942}{{\ttfamily 2410.23942}}].

\bibitem{Inoue:2001zt}
S.~Inoue and J.~Yokoyama, \emph{{Curvature perturbation at the local extremum
  of the inflaton's potential}},
  \href{https://doi.org/10.1016/S0370-2693(01)01369-7}{\emph{Phys. Lett. B}
  {\bfseries 524} (2002) 15}
  [\href{https://arxiv.org/abs/hep-ph/0104083}{{\ttfamily hep-ph/0104083}}].

\bibitem{Kinney:2005vj}
W.H.~Kinney, \emph{{Horizon crossing and inflation with large eta}},
  \href{https://doi.org/10.1103/PhysRevD.72.023515}{\emph{Phys. Rev. D}
  {\bfseries 72} (2005) 023515}
  [\href{https://arxiv.org/abs/gr-qc/0503017}{{\ttfamily gr-qc/0503017}}].

\bibitem{Dimopoulos:2017ged}
K.~Dimopoulos, \emph{{Ultra slow-roll inflation demystified}},
  \href{https://doi.org/10.1016/j.physletb.2017.10.066}{\emph{Phys. Lett. B}
  {\bfseries 775} (2017) 262}
  [\href{https://arxiv.org/abs/1707.05644}{{\ttfamily 1707.05644}}].

\bibitem{Pattison:2018bct}
C.~Pattison, V.~Vennin, H.~Assadullahi and D.~Wands, \emph{{The attractive
  behaviour of ultra-slow-roll inflation}},
  \href{https://doi.org/10.1088/1475-7516/2018/08/048}{\emph{JCAP} {\bfseries
  08} (2018) 048} [\href{https://arxiv.org/abs/1806.09553}{{\ttfamily
  1806.09553}}].

\bibitem{Ivanov:1994pa}
P.~Ivanov, P.~Naselsky and I.~Novikov, \emph{{Inflation and primordial black
  holes as dark matter}},
  \href{https://doi.org/10.1103/PhysRevD.50.7173}{\emph{Phys. Rev. D}
  {\bfseries 50} (1994) 7173}.

\bibitem{Pi:2022zxs}
S.~Pi and J.~Wang, \emph{{Primordial black hole formation in Starobinsky's
  linear potential model}},
  \href{https://doi.org/10.1088/1475-7516/2023/06/018}{\emph{JCAP} {\bfseries
  06} (2023) 018} [\href{https://arxiv.org/abs/2209.14183}{{\ttfamily
  2209.14183}}].

\bibitem{Briaud:2025hra}
V.~Briaud, A.~Karam, N.~Koivunen, E.~Tomberg, H.~Veerm{\"a}e and V.~Vennin,
  \emph{{How deep is the dip and how tall are the wiggles in inflationary power
  spectra?}}, \href{https://doi.org/10.1088/1475-7516/2025/05/097}{\emph{JCAP}
  {\bfseries 05} (2025) 097}
  [\href{https://arxiv.org/abs/2501.14681}{{\ttfamily 2501.14681}}].

\bibitem{Deruelle:1995kd}
N.~Deruelle and V.F.~Mukhanov, \emph{{On matching conditions for cosmological
  perturbations}}, \href{https://doi.org/10.1103/PhysRevD.52.5549}{\emph{Phys.
  Rev. D} {\bfseries 52} (1995) 5549}
  [\href{https://arxiv.org/abs/gr-qc/9503050}{{\ttfamily gr-qc/9503050}}].

\bibitem{Figueroa:2021zah}
D.G.~Figueroa, S.~Raatikainen, S.~Rasanen and E.~Tomberg, \emph{{Implications
  of stochastic effects for primordial black hole production in ultra-slow-roll
  inflation}}, \href{https://doi.org/10.1088/1475-7516/2022/05/027}{\emph{JCAP}
  {\bfseries 05} (2022) 027}
  [\href{https://arxiv.org/abs/2111.07437}{{\ttfamily 2111.07437}}].

\bibitem{Nassiri-Rad:2025dsa}
A.~Nassiri-Rad, H.~Sheikhahmadi and H.~Firouzjahi, \emph{{Stochastic Inflation
  with Interacting Noises}},
  \href{https://arxiv.org/abs/2508.09946}{{\ttfamily 2508.09946}}.

\bibitem{Linde:1993cn}
A.D.~Linde, \emph{{Hybrid inflation}},
  \href{https://doi.org/10.1103/PhysRevD.49.748}{\emph{Phys. Rev. D} {\bfseries
  49} (1994) 748} [\href{https://arxiv.org/abs/astro-ph/9307002}{{\ttfamily
  astro-ph/9307002}}].

\bibitem{Martin:2011ib}
J.~Martin and V.~Vennin, \emph{{Stochastic Effects in Hybrid Inflation}},
  \href{https://doi.org/10.1103/PhysRevD.85.043525}{\emph{Phys. Rev. D}
  {\bfseries 85} (2012) 043525}
  [\href{https://arxiv.org/abs/1110.2070}{{\ttfamily 1110.2070}}].

\bibitem{Tada:2023fvd}
Y.~Tada and M.~Yamada, \emph{{Stochastic dynamics of multi-waterfall hybrid
  inflation and formation of primordial black holes}},
  \href{https://doi.org/10.1088/1475-7516/2023/11/089}{\emph{JCAP} {\bfseries
  11} (2023) 089} [\href{https://arxiv.org/abs/2306.07324}{{\ttfamily
  2306.07324}}].

\bibitem{Bjorkmo:2019fls}
T.~Bjorkmo, \emph{{Rapid-Turn Inflationary Attractors}},
  \href{https://doi.org/10.1103/PhysRevLett.122.251301}{\emph{Phys. Rev. Lett.}
  {\bfseries 122} (2019) 251301}
  [\href{https://arxiv.org/abs/1902.10529}{{\ttfamily 1902.10529}}].

\bibitem{Carrilho:2019oqg}
P.~Carrilho, K.A.~Malik and D.J.~Mulryne, \emph{{Dissecting the growth of the
  power spectrum for primordial black holes}},
  \href{https://doi.org/10.1103/PhysRevD.100.103529}{\emph{Phys. Rev. D}
  {\bfseries 100} (2019) 103529}
  [\href{https://arxiv.org/abs/1907.05237}{{\ttfamily 1907.05237}}].

\bibitem{Ragavendra:2020sop}
H.V.~Ragavendra, P.~Saha, L.~Sriramkumar and J.~Silk, \emph{{Primordial black
  holes and secondary gravitational waves from ultraslow roll and punctuated
  inflation}}, \href{https://doi.org/10.1103/PhysRevD.103.083510}{\emph{Phys.
  Rev. D} {\bfseries 103} (2021) 083510}
  [\href{https://arxiv.org/abs/2008.12202}{{\ttfamily 2008.12202}}].

\bibitem{Tasinato:2020vdk}
G.~Tasinato, \emph{{An analytic approach to non-slow-roll inflation}},
  \href{https://doi.org/10.1103/PhysRevD.103.023535}{\emph{Phys. Rev. D}
  {\bfseries 103} (2021) 023535}
  [\href{https://arxiv.org/abs/2012.02518}{{\ttfamily 2012.02518}}].

\bibitem{Cole:2022xqc}
P.S.~Cole, A.D.~Gow, C.T.~Byrnes and S.P.~Patil, \emph{{Smooth vs instant
  inflationary transitions: steepest growth re-examined and primordial black
  holes}}, \href{https://doi.org/10.1088/1475-7516/2024/05/022}{\emph{JCAP}
  {\bfseries 05} (2024) 022}
  [\href{https://arxiv.org/abs/2204.07573}{{\ttfamily 2204.07573}}].

\bibitem{Byrnes:2018txb}
C.T.~Byrnes, P.S.~Cole and S.P.~Patil, \emph{{Steepest growth of the power
  spectrum and primordial black holes}},
  \href{https://doi.org/10.1088/1475-7516/2019/06/028}{\emph{JCAP} {\bfseries
  06} (2019) 028} [\href{https://arxiv.org/abs/1811.11158}{{\ttfamily
  1811.11158}}].

\bibitem{Ozsoy:2019lyy}
O.~{\"O}zsoy and G.~Tasinato, \emph{{On the slope of the curvature power
  spectrum in non-attractor inflation}},
  \href{https://doi.org/10.1088/1475-7516/2020/04/048}{\emph{JCAP} {\bfseries
  04} (2020) 048} [\href{https://arxiv.org/abs/1912.01061}{{\ttfamily
  1912.01061}}].

\bibitem{Fujita:2025imc}
T.~Fujita, R.~Kawaguchi, M.~Sasaki and Y.~Tada, \emph{{Dip and non-linearity in
  the curvature perturbation from inflation with a transient non-slow-roll
  stage}},  \href{https://arxiv.org/abs/2503.19744}{{\ttfamily 2503.19744}}.

\bibitem{Palma:2020ejf}
G.A.~Palma, S.~Sypsas and C.~Zenteno, \emph{{Seeding primordial black holes in
  multifield inflation}},
  \href{https://doi.org/10.1103/PhysRevLett.125.121301}{\emph{Phys. Rev. Lett.}
  {\bfseries 125} (2020) 121301}
  [\href{https://arxiv.org/abs/2004.06106}{{\ttfamily 2004.06106}}].

\bibitem{Fumagalli:2020adf}
J.~Fumagalli, S.~Renaux-Petel, J.W.~Ronayne and L.T.~Witkowski, \emph{{Turning
  in the landscape: A new mechanism for generating primordial black holes}},
  \href{https://doi.org/10.1016/j.physletb.2023.137921}{\emph{Phys. Lett. B}
  {\bfseries 841} (2023) 137921}
  [\href{https://arxiv.org/abs/2004.08369}{{\ttfamily 2004.08369}}].

\bibitem{Braglia:2020taf}
M.~Braglia, X.~Chen and D.K.~Hazra, \emph{{Probing Primordial Features with the
  Stochastic Gravitational Wave Background}},
  \href{https://doi.org/10.1088/1475-7516/2021/03/005}{\emph{JCAP} {\bfseries
  03} (2021) 005} [\href{https://arxiv.org/abs/2012.05821}{{\ttfamily
  2012.05821}}].

\end{thebibliography}\endgroup
\end{document}